\documentclass[aps,pra,10pt]{revtex4-2}

\pdfoutput=1
\usepackage[T1]{fontenc}
\usepackage{fullpage}
\usepackage{amsfonts}
\usepackage{amsmath}
\usepackage{slashed}
\usepackage{graphicx}
\usepackage{latexsym}
\usepackage[dvipsnames]{xcolor}
\usepackage[breaklinks]{hyperref}
\hypersetup{colorlinks=true,citecolor=blue,linkcolor=blue,urlcolor=NavyBlue}

\begin{document}

\title{Revealing the Origin of Neutrino Masses through\\ Displaced Shower Searches in the CMS Muon System}

\author{Wei Liu}
\email{wei.liu@njust.edu.cn}
\affiliation{Department of Applied Physics and MIIT Key Laboratory of Semiconductor Microstructure\\ and Quantum Sensing, Nanjing University of Science and Technology, Nanjing 210094, China}

\author{Suchita Kulkarni}
\email{suchita.kulkarni@uni-graz.at}
\affiliation{Institute of Physics, NAWI Graz, University of Graz, Universit\"atsplatz 5, A-8010 Graz, Austria}

\author{Frank F. Deppisch}
\email{f.deppisch@ucl.ac.uk}
\affiliation{University College London, Gower Street, London WC1E 6BT, United Kingdom}

\begin{abstract}
We study the potential to probe the origin of neutrino masses, by searching for long-lived right-handed neutrinos (RHNs) $N$ in the $B-L$ model and in the RHN-extended Standard Model (SM) Effective Field Theory (EFT). Despite the small active-sterile mixing $|V_{\ell N}|^2$, RHNs are produced abundantly via SM and exotic Higgs production, as long as the Higgs mixing or EFT operator coefficient is sufficiently large. We reinterpret a search for displaced showers in the CMS muon system and we find that it is sensitive to parameter space at and below the seesaw floor, $|V_{\ell N}|^2 \approx 10^{-12}$ ($\ell = e$, $\tau$) for $m_N \approx 40$ GeV. With existing data constraining such well-motivated scenarios of neutrino mass generation, we determine the projected sensitivity at the HL-LHC, motivating dedicated searches for long-lived RHNs with decay lengths $\approx 10$~m.
\end{abstract}

\maketitle

\section{Introduction}
The Standard Model (SM) does not incorporate the observed neutrino masses confirmed in neutrino oscillations~\cite{Davis:1994jw, Super-Kamiokande:1998kpq, KamLAND:2002uet,SNO:2002tuh}, with the most popular explanation being the seesaw mechanism~\cite{Minkowski:1977sc} where heavy right-handed neutrinos~(RHNs) are introduced. If RHN masses $m_N$ are of electroweak~(EW) scale, the active-sterile mixing strength $V_{\ell N}$ required to generate the neutrino mass $m_\nu$ is $|V_{\ell N}|^2 \sim m_\nu / m_N \lesssim 10^{-11} \times(40~\text{GeV}/m_N)$. While the absolute light neutrino mass scale has not been determined yet, the KATRIN experiment limits the effective $\beta$ decay mass to $m_\beta < 0.45$~eV at 90~\% confidence level~(CL)~\cite{Katrin:2024tvg}, whereas solar neutrino oscillations point toward a smallest scale $m_\nu > \sqrt{\Delta m^2_\text{sol}} = 9\times 10^{-3}$~eV~\cite{Esteban:2020cvm}. This makes RHNs long-lived and given the displaced signature, RHNs have received increased attention in phenomenological studies \cite{Kaneta:2016vkq, Bondarenko:2018ptm, Bryman:2019bjg, Balaji:2019fxd, Bolton:2019pcu, Liu:2021akf, Abdullahi:2022jlv, Zhang:2023nxy, Barducci:2023hzo, Liu:2023klu}, and numerous collider searches have been carried out \cite{CMS:2012wqj, LHCb:2014osd, ATLAS:2015gtp, CMS:2015qur, CMS:2016aro, NA62:2017qcd, CMS:2018iaf, CMS:2018jxx, ATLAS:2019kpx, LHCb:2020wxx, CMS:2022fut, ATLAS:2023tkz, CMS:2024ake}.

While suggestive, the seesaw mechanism does not explain the origin of RHN Majorana masses. This is addressed in models where lepton number is broken spontaneously, generating the RHN masses. This includes, for example, the $U(1)_{B-L}$ gauge model and extensions \cite{Davidson:1978pm, Mohapatra:1980qe, Debnath:2023akj, FileviezPerez:2024fzc}, models based on Left-Right \cite{Mohapatra:1974gc, Senjanovic:1975rk, Mohapatra:1979ia} and Pati-Salam \cite{Pati:1974yy} symmetry as well as certain Two-Higgs-Doublet models \cite{Batell:2024ddo}. Such scenarios generically predict exotic gauge and scalar bosons that mix with the corresponding SM states, enabling new portals of RHN production. We here consider the $B-L$ (baryon - lepton number) model \cite{Davidson:1978pm, Mohapatra:1980qe} as a prominent and minimal example, where masses are generated through spontaneous breaking of a $U(1)_{B-L}$ gauge symmetry. The model predicts three RHNs $N_i$, an exotic gauge boson $Z'$ and a $B-L$-breaking Higgs~$\Phi$. The RHNs may also explain the baryon asymmetry via leptogenesis \cite{Fukugita:1986hr, Luty:1992un, Davidson:2008bu, Liu:2023klu} and the $B-L$ model can be conformally invariant, not only explaining why $B-L$ is broken near the TeV scale and inducing EW symmetry breaking \cite{Iso:2009ss, Iso:2009nw}, but also predicting primordial black holes as dark matter~\cite{Baldes:2023rqv}. The low energy effects of this and similar models, at SM scales and below, can also be described within the RHN-extended extended SM Effective Field Theory~($\nu_R$SMEFT) framework~\cite{Graesser:2007pc, Graesser:2007yj, delAguila:2008ir, Aparici:2009fh, Liao:2016qyd}.

RHNs in the $B-L$ model not only interact with the SM particles via the active-sterile neutrino mixing $V_{\ell N}$, but also through the $B-L$ gauge and Higgs portals. Thus, RHNs can be produced via the SM Higgs, $B-L$ Higgs, SM $Z$ and $B-L$ $Z'$. It is well known that enabling these portals allows testing small active-sterile mixing strengths \cite{Deppisch:2013cya, Batell:2016zod, Deppisch:2019kvs, Accomando:2017qcs, Das:2019fee, Cheung:2021utb, Chiang:2019ajm, FileviezPerez:2020cgn, Das:2018tbd, Liu:2022kid,Han:2021pun, Das:2017nvm, Graesser:2007yj, Maiezza:2015lza, Deppisch:2018eth, Mason:2019okp, Accomando:2016rpc, Gao:2019tio, Gago:2015vma, Jones-Perez:2019plk, Liu:2022ugx, Li:2023dbs, Deppisch:2023sga, Bernal:2023coo}, as the RHNs still decay as $N \to \nu_\ell h, \nu_\ell Z^*, \ell W^*$, controlled by $|V_{\ell N}|$ ($\ell = e,\mu,\tau$). This leads to macroscopic proper decay lengths ($m_N < m_{W,Z}$)~\cite{Atre:2009rg},
\begin{align}
\label{eq:length}
    L_N^0 \approx 3~\text{m} \times 
    \left(\frac{10^{-12}}{|V_{\ell N}|^2}\right) 
    \left(\frac{40~\text{GeV}}{m_N}\right)^5.
\end{align}
Long-lived particles (LLPs), such as RHNs, can be searched for at Large Hadron Collider (LHC) detectors. We here focus on a search for displaced showers in the CMS muon endcap \cite{CMS:2021juv} for a scalar LLP~$S$, pair-produced via the Higgs, $pp\to h\to SS$, and decaying to quarks or $\tau$ leptons. It provides the most stringent limit for LLP proper decay lengths $6 - 40$~m and LLP masses $7 - 40$~GeV. ATLAS searched for hadronic LLP decays \cite{ATLAS:2018tup, ATLAS:2019jcm, ATLAS:2023oti} but with weaker limits. The CMS muon endcap search has been considered in~\cite{Cottin:2022nwp} for sterile neutrinos but we reinterpret it in the $B-L$ model and $\nu_R$SMEFT, focusing on Higgs production where we find that it is sensitive to the seesaw floor, i.e., for active-sterile mixing required to generate light neutrino masses. Our analysis shows for the first time that existing data from the LHC probes the origin of neutrino masses in well motivated scenarios. 

\section{Right-handed Neutrino Production and Decay}
The $B-L$ model extends the SM by an Abelian gauge symmetry $U(1)_{B-L}$. It is spontaneously broken by the vacuum expectation value (VEV) of a SM-singlet Higgs $\Phi$, resulting in a heavy $Z'$ gauge boson and heavy Majorana RHNs. The relevant features of and constraints on the $B-L$ model are summarized in Appendix~\ref{sec:model}. For simplicity, we assume that a single RHN $N$ is light enough to be produced and that it mixes with a single lepton flavour $\ell$ at a time through $V_{\ell N}$. We concentrate on $\ell = e, \tau$ due to the search discussed in the following section: First, RHN LLPs decaying to muons rarely produce a particle shower required by the search and second, the search vetoes events with muons too close to a jet~\cite{CMS:2021juv}. We therefore expect no sensitivity to muon flavour.

We first compare the four resonant RHN production channels at the LHC, namely via the $Z'$, SM $Z$, SM-like Higgs $h$ and $(B-L)$-like Higgs $\Phi$. We omit production via the SM charged and neutral currents as they are suppressed by the small active-sterile mixing $|V_{\ell N}|$. The RHNs instead decay to SM particles via these currents with decay lengths $L_N^0 \approx 10$~m at the seesaw floor, see Eq.~\eqref{eq:length}. Produced from electroweak-scale or heavier states, the RHNs are boosted, increasing the observed decay length to $L_N = \beta\gamma L_N^0$, with an average boost factor $\langle\beta\gamma\rangle \approx 1-10$. Such long decay lengths are best probed in detectors far away from the interaction point (IP), namely the muon system, forming the outer layer of the CMS detector at a distance of $\approx 8 - 13$~m from the IP.

\subsection{Gauge Portal}
With quarks and RHNs charged under $B-L$, RHNs are produced at the LHC through the $Z'$ gauge portal, $pp\to Z'\to NN$. As a benchmark, we use $m_{Z'} = 500$~GeV and the cross section at the 13~TeV LHC is $\sigma(pp\to Z'\to NN) \approx \sigma(pp\to Z') \times \text{Br}(Z'\to NN)$, with $\sigma(pp\to Z') \approx 0.4~\text{fb} \times (g_{B-L} / 10^{-3})^2$. The branching ratio is Br$(Z'\to NN) \approx 1/13$ for $m_N \ll m_{Z'}$, leading to
\begin{align}
    \sigma(pp \to Z'(500) \to NN) 
    \approx 0.75~\text{fb} \!\times\! 
    \left(\frac{g_{B-L}}{5\times 10^{-3}}\right)^2\!\!.
\end{align}
The most stringent limit from ATLAS resonance searches \cite{ATLAS:2019erb} is $g_{B-L} \lesssim 5\times 10^{-3}$ for $m_{Z'} = 500$~GeV.

The SM $Z$ can mix with the $Z'$, inducing the SM gauge portal $pp\to Z\to NN$, controlled by the effective coupling $\beta = g_{B-L}\sin\theta_{Z Z^\prime}$. The SM cross section at the 13~TeV LHC is $\sigma(pp\to Z) \approx 5.8\times 10^7$~fb~\cite{ATLAS:2016oxs}, and the branching ratio is $\text{Br}(Z\to NN) \approx 0.48\times\beta^2$ for $m_N \ll m_Z$. This results in
\begin{align}
    \sigma(pp\to Z\to NN) \approx 28~\text{fb} \times \left(\frac{\beta}{10^{-3}}\right)^2,
\end{align}
consistent with atomic parity violation limits~\cite{Dev:2021otb}.

\subsection{Higgs Portal}
\label{sec:higgs-cs}
The SM Higgs and the $B-L$ Higgs also mix, through the Higgs mixing parameter $\sin\alpha$. RHNs can then be produced via the SM Higgs, $pp\to h\to NN$, where we only consider gluon-gluon fusion; while vector boson fusion and associated Higgs production are also possible, they contribute at most 10\% to the overall cross section~\cite{CERNYellowReportPageAt13TeV} and the resulting final states are too soft to pass the missing transverse energy requirement $E_T^\text{miss} > 200$~GeV, discussed in Sec.~\ref{sec:dis-shower}. The $pp\to h\to NN$ production cross section is $\sigma(pp \to h \to NN) \approx \cos^2\alpha \times \sigma(pp \to h)_\text{SM} \times \text{Br}(h \to NN)$ \cite{Accomando:2016rpc, Deppisch:2018eth}. Here, $\sigma(pp\to h)_\text{SM} = 44\pm 4$~pb ($50\pm 7$~pb) is the SM Higgs production cross section at the 13~TeV (14~TeV) LHC via gluon-gluon fusion~\cite{CERNYellowReportPageAt13TeV, LHCHiggsWG}. The branching ratio can be approximated as~\cite{Accomando:2016rpc} 
\begin{align}
    \text{Br}(h &\to NN)
    \approx \frac{\tan^2\alpha}{16\pi}
    \frac{m_N^2}{\langle\Phi\rangle^2}
    \frac{m_h}{\Gamma_h^\text{SM}}
    \left(1 - \frac{4m_N^2}{m_h^2}\right)^{3/2},
\end{align}
with the SM Higgs width $\Gamma_h^\text{SM} \approx 4.1$~MeV~\cite{LHCHiggsCrossSectionWorkingGroup:2016ypw}. It is suppressed by $\sin\alpha$ and the RHN Yukawa coupling with the $B-L$ Higgs $\Phi$, $y_N =  m_N / \langle\Phi\rangle = 2 g_{B-L} m_N / m_{Z'}$. The cross section is maximized for $m_N = m_h/\sqrt{10} \approx 40$~GeV,
\begin{align}
\label{eq:higgs-cs}
    \sigma(pp \to h \to NN) 
    \lesssim 7.8~\text{fb} \!\times\! 
    \left(\frac{\sin\alpha}{0.1}\right)^2
    \!\left(\frac{5~\text{TeV}}{\langle\Phi\rangle}\right)^2  
    \!\!.
\end{align}
Both the Higgs mixing $\sin\alpha$ and the $B-L$ VEV $\langle\Phi\rangle$ are experimentally constrained through direct and indirect Higgs probes and $Z'$ searches. In our analysis, we choose $\langle\Phi\rangle = 3.75$~TeV, satisfying the current limit from EW precision tests, $\langle\Phi\rangle \gtrsim 3.5$~TeV~\cite{Cacciapaglia:2006pk, ALEPH:2006bhb}.

Similarly, RHNs can be produced via $\Phi$. The SM quarks and gluons couple to $\Phi$ with a suppression as $\sigma(pp\to\Phi\to NN) \approx \sin^2\alpha\times\sigma(pp\to\Phi)_\text{SM} \times \text{Br}(\Phi\to NN)$, where $\sigma(pp\to\Phi)_\text{SM}$ is the cross section treating $\Phi$ as the SM Higgs but with mass $m_\Phi$. $\Phi$ decays to heavy SM fermions and boson pairs via mixing to the SM Higgs, and RHN pairs via the Yukawa coupling $y_N$. For $m_\Phi < 2 m_W$, the branching ratio is
\begin{align}
    \text{Br}(\Phi \to NN)
    = \frac{\Gamma(\Phi\to NN)}{\sin^2\alpha \, \Gamma_\Phi^\text{SM} + \Gamma(\Phi\to NN)},
\end{align}
where $\Gamma_\Phi^\text{SM}$ is the decay width treating $\Phi$ as the SM Higgs but with mass $m_\Phi$, and the partial width is $\Gamma(\Phi\to NN) = \cos^2\alpha/(16\pi) m_N^2 m_\Phi / \langle\Phi\rangle^2$ for $m_N \ll m_\Phi$ \cite{Liu:2022ugx}. This results in the 13~TeV LHC cross section
\begin{align}
    \label{eq:phi-cs}
    \sigma(pp \to \Phi(150) \to NN) 
    \lesssim 100~\text{fb} \times  
    \left(\frac{5~\text{TeV}}{\langle\Phi\rangle}\right)^2,
\end{align}
for $m_\Phi < 2 m_W \approx 150$~GeV. For $m_\Phi \gtrsim 2 m_{W,Z} \approx 200$~GeV, the decays $\Phi\to WW, ZZ$ dominate,
\begin{align}
    \sigma(pp \to \Phi(200) \to NN) 
    \lesssim 1~\text{fb} \times 
    \left(\frac{5~\text{TeV}}{\langle\Phi\rangle}\right)^2.
\end{align}
The cross section decreases rapidly with larger $m_\Phi$ and we use $m_\Phi = 150$~GeV as benchmark.

\section{Displaced Showers in the CMS Muon System}
\label{sec:dis-shower}
The overall rates above motivate a detailed simulation to match the CMS endcap search. The model is implemented in Universal {\tt FeynRules} Output (UFO)~\cite{Degrande:2011ua}, following \cite{Deppisch:2018eth, Deppisch:2019kvs}. The event generator {\tt MadGraph5aMC@NLO}~v3.4.1 \cite{Alwall:2014hca} is then used for a parton level simulation of the events with the shower jets and matrix element jets matched \cite{Mangano:2006rw, Hoeche:2005vzu}. Events are fed to {\tt PYTHIA}~v8.235 \cite{Sjostrand:2014zea} for parton showering, hadronization and heavy hadron decays. Clustering of events is performed by {\tt FastJet}~v3.2.1 \cite{Cacciari:2011ma} and detector effects by {\tt Delphes}~v3.5.1 \cite{deFavereau:2013fsa}. 

The CMS search~\cite{CMS:2021juv}, which uses the muon detector as a sampling calorimeter, identifies LLP showers. The updated search \cite{CMS:2024bvl} might improve the detector efficiency but there is no detailed reinterpretation information available yet. There is also a similar search for vector-like leptons~\cite{CMS:2024nua}. We do not expect this to have a large impact as the triggers are largely the same. LLP decays produce photons, electrons, tau or quarks inside the muon detector creating collimated hadronic and electromagnetic showers and giving rise to a large number of hits in a small detector region. These hits are identified by cathode strip chambers~(CSCs), which detect charged particles and they are clustered (called CSC clusters) by identifying high-density regions. As the search strategy relies on identification and reconstruction of original information using CSC clusters, to aid reinterpretation, CMS provides an updated card and modules for {\tt Delphes} \cite{delphes_pr, hepdata.104408.v2, CMS:2021juv}. We implement these to determine the reconstruction efficiency and kinematic information of the CSC cluster events, and apply the selection criteria of the CMS search:

The \emph{missing transverse energy}, defined as the negative vector sum of visible $p_T$ energy deposited in the tracker and calorimeter, is $E_T^\text{miss} > 200$~GeV, as trigger requirement. If the signal contained only LLPs decaying beyond the calorimeter, there is nothing to trigger on. We thus simulate the signal up to two initial state jets, which are prompt and deposit energy in the calorimeter. 

\emph{No electron~(muon)} with transverse momentum $p_T > 35~(25)$~GeV and pseudorapidity $|\eta| < 2.5~(2.4)$, to remove $W$ and top background.
    
\emph{At least one CSC cluster} with $|\Delta\phi| < 0.75$ to ensure that it originated from the LLP decay. Here, $\Delta\phi(\mathbf{x}_\text{CSC}, \mathbf{p}_T^\text{miss})$ is defined as the azimuthal angle between the missing transverse momentum and the cluster location from the IP. The LLP is then close to the missing momentum, which points opposite to the vector sum of the visible $p_T$. 

Events with \emph{clusters too close to a jet~(muon) are removed}, for $\Delta R = \sqrt{(\Delta\eta)^2 + (\Delta\phi)^2} < 0.4$. The CSC cluster is then not matched to jets (muons) with $p_T > 10~(20)$~GeV and it is not created by LLPs inside the jet, e.g. $K_L$, or muon bremsstrahlung. 
    
The \emph{average time of detector hits} in the CSC cluster, relative to the collision is $-5~\text{ns} < \langle\Delta t_\text{CSC}\rangle < 12.5$~ns, to reject pileup clusters.

\begin{figure}[t!]
    \centering
    \includegraphics[width=0.5\columnwidth]{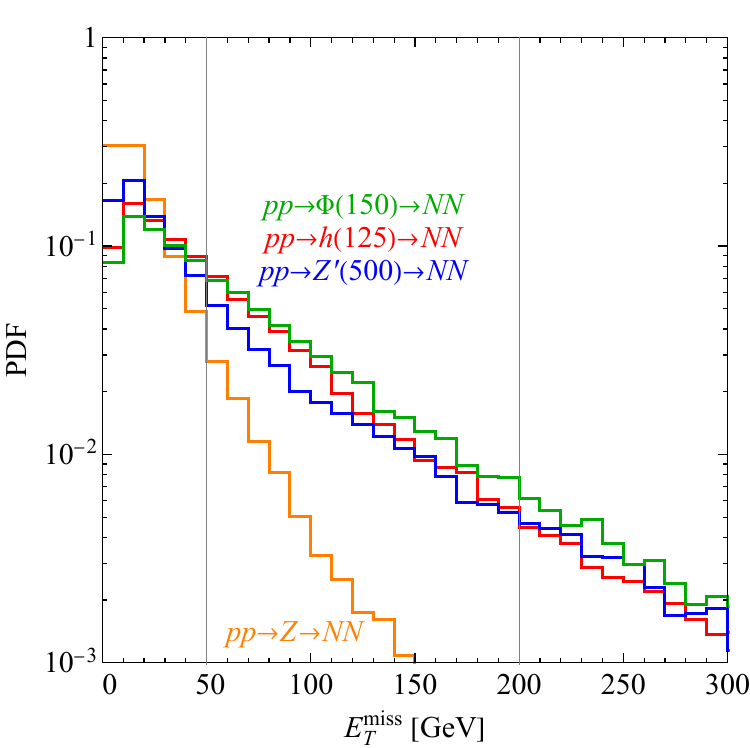}
    \caption{Missing transverse energy $E_T^\text{miss}$ for $pp \to Z' \to NN$ ($m_{Z'} = 500$~GeV, blue), $Z$~(green), $h$~(red) and $\Phi$ ($m_\Phi = 150$~GeV, orange), with $m_N = 30$~GeV and $|V_{\ell N}|^2 = 10^{-12}$. The vertical lines indicate the thresholds $E_T^\text{miss} > 200$~GeV and $> 50$~GeV in the CMS search and the soft trigger strategy, respectively.} 
\label{fig:MET}
\end{figure}
Due to $E_T^\text{miss} > 200$~GeV, only a small number of events passes the selection criteria, see Fig.~\ref{fig:MET}. The rate drops rapidly for larger $E_T^\text{miss}$, especially for SM $Z$ production, due to the fact that missing transverse energy either originates from initial state radiation, which is enhanced for Higgs gluon-gluon fusion, or from heavier resonant particle masses, as in the $Z'$ and $\Phi$ modes. Nevertheless, only $\approx 1\%$ of events satisfy the trigger requirement. In the SM $Z$ mode, only $\lesssim 0.1\%$ pass the trigger and we thus expect no appreciable sensitivity in this case.

To calculate the dominant gluon-gluon fusion Higgs production, we use the effective gluon-gluon-Higgs coupling at leading order~\cite{Alwall:2011uj}. We have checked that this is accurate, giving a conservative determination; while the transverse momentum of the Higgs can change when including higher order corrections \cite{Jones:2018hbb, Becker:2020rjp}, by comparing the leading order effective treatment with the next-to-leading order fully resolved top quark mass-dependent vertex model \cite{Maltoni:2014eza}, the effective model leads to a softer Higgs for $p_T < 200$~GeV, while the distributions are comparable for $200~\text{GeV} < p_T < 400$~GeV. For $p_T > 400$~GeV, the effective model overestimates by more than a factor of two. Nevertheless, as the cross section is much smaller than for $200~\text{GeV} < p_T < 400$~GeV, we consider our calculation accurate for $p_T > 200$~GeV.

Given that new, dedicated Level-1 and high level triggers targeting our signature have been collecting data during the LHC Run-3~\cite{CMS-DP-2022-062}, Ref.~\cite{Cottin:2022nwp} argues that the requirement on $E_T^\text{miss}$ can be relaxed to $E_T^\text{miss} > 50$~GeV, resulting in an improved signal rate, cf. Fig.~\ref{fig:MET}, while the background can still be controlled by requiring larger CSC clusters. We refer to this option as the `soft trigger strategy'. The softened $E_T^\text{miss}$ threshold means that we are more susceptible to Higgs production corrections as we are probing softer Higgs~$p_T$. Our results remain conservative as we underestimate the Higgs $p_T$ distribution by about a factor of two.

We simulate the displaced shower signature at the 13~TeV LHC with 137~fb$^{-1}$ integrated luminosity and the 14~TeV HL-LHC with 3000~fb$^{-1}$. The background mainly comes from punch-through jets and muon bremsstrahlung, controlled by the above selection criteria and determined from CMS data: The number of expected background events is $b = 2.0\pm 1.0$ and the number of observed events is $N = 3$, excluding signal events $s > 6.1$ at 95\% CL \cite{CMS:2021juv}. At the HL-LHC, after taking into account a 20\% (60\%) signal loss due to pileup, and requiring more hits to push the background to be negligible, $s > 56~(3.0)$ is required at 95~\%~CL with the CMS (soft) trigger strategy, where we scale the number of background events according to luminosity, $b\approx 40~(0.0)$ at the HL-LHC with the CMS (soft) trigger~\cite{Cottin:2022nwp}.

\section{Probing the Neutrino Mass Generation Mechanism}
Using the above strategy, we reinterpret the CMS search \cite{CMS:2021juv} in the $B-L$ model and the $\nu_R$SMEFT, specifically to probe the RHN mass $m_N$ and the active-sterile mixing $|V_{\ell N}|$ for successful neutrino mass generation. In our scenarios, with a single RHN $N$, a light neutrino acquires the Majorana mass $m_\nu = |V_{\ell N}|^2 m_N$. As indicated in the introduction, we take the range $9\times 10^{-3}~\text{eV} < |V_{\ell N}|^2 m_N < 0.45$~eV as a target to probe the seesaw floor. We only consider active-sterile mixing to $\ell = e, \tau$ flavour as mentioned. 

\subsection{Gauge Portal}
We start with the $Z'$ gauge portal $pp\to Z'\to NN$ where we use the benchmark values $m_{Z'} = 500$~GeV and $g_{B-L} = 5\times 10^{-3}$. No appreciable sensitivity is obtained in this case using the CMS data with either the original or the soft trigger strategy, since the requirement on $E_T^\text{miss}$ removes most of the signal. At the HL-LHC with 3000~fb$^{-1}$, our reinterpretation also only predicts a sensitivity for the soft trigger strategy. If the $B-L$ gauge coupling is larger than $g_{B-L} \gtrsim 0.0035$, RHNs can be probed in this case, within the mass range $10~\text{GeV} \lesssim m_N \lesssim 80$~GeV.

Prospects for the SM gauge portal $pp\to Z\to NN$ are similar. For a $ZZ'$ mixing $\beta = 10^{-3}$, no sensitivity at 95\%~CL is achieved using current data. It can only be probed at the HL-LHC using the soft trigger strategy, for $m_N \lesssim 40$~GeV. While the total cross section for $Z$ is sizeable, Fig.~\ref{fig:MET} demonstrates that a large number of events are removed due to small $E_T^\text{miss}$.

\subsection{Higgs Portal}
\begin{figure}[t!]
    \centering
    \includegraphics[width=0.49\textwidth]{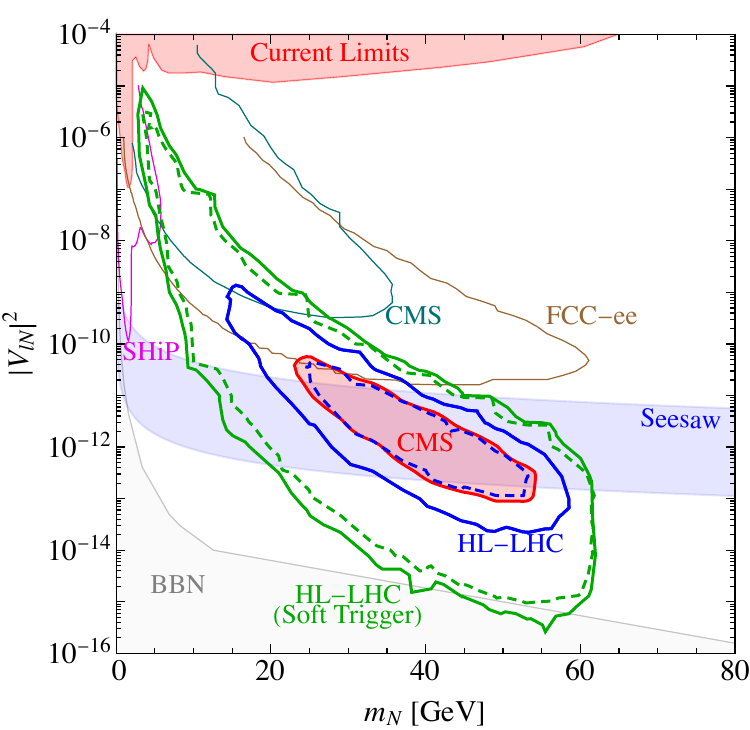}
    \caption{Sensitivity of $pp\to h\to NN$ at 95\%~CL on the RHN mass $m_N$ and the active-sterile mixing $|V_{\ell N}|^2$ ($\ell = e, \tau$). The scenarios are for reinterpreting CMS data (red region), as well as at the 14~TeV HL-LHC with 3000~fb$^{-1}$ using the CMS (blue contour) and soft (green contour) trigger strategies. The Higgs mixing is $\sin\alpha = 0.24$ (red region and solid contours) and $0.14$ (dashed), and the $B-L$ Higgs VEV is $\langle\Phi\rangle = 3.75$~TeV. The region labelled `Current Limits' is excluded by sterile neutrino searches ($\ell = e$) while the contours labelled SHiP, CMS and FCC-ee give the projected sensitivity of planned experiments \cite{Bolton:2019pcu, Bolton:2022pyf}. The band labelled `Seesaw' indicates $9\times 10^{-3}~\text{eV} < |V_{\ell N}|^2 m_N < 0.45$~eV.}
    \label{fig:higgslinear}
\end{figure}
Prospects for the Higgs portals are much more promising. Reinterpreting the CMS search for $pp\to h\to NN$, we find that existing CMS data constrains parameter space of light neutrino mass generation in the $B-L$ model. This is shown in Fig.~\ref{fig:higgslinear}, where the red region labelled `CMS' is excluded at 95\%~CL, for a Higgs mixing $\sin\alpha = 0.24$ and $B-L$ VEV $\langle\Phi\rangle = 3.75$~TeV. As anticipated in Sec.~\ref{sec:higgs-cs}, the sensitivity is maximal for $m_N\approx m_h/\sqrt{10}\approx 40$~GeV. For a displaced search at distances $\mathcal{O}(1~\text{m})$ in the CMS muon system, this corresponds to the well-motivated seesaw prediction $|V_{\ell N}|^2 \approx 10^{-12}$.

Fig.~\ref{fig:higgslinear}~ also shows the projected sensitivity at the HL-LHC using the CMS (solid blue contour) and soft trigger (solid red contour) strategies. This extends the coverage over $10~\text{GeV} \lesssim m_N \lesssim 60$~GeV. The HL-LHC will improve the sensitivity to the Higgs mixing, to values $\sin\alpha \gtrsim 0.14$ \cite{Cepeda:2019klc, ATLAS:2019mfr}. Using this value, i.e., assuming the HL-LHC will see no sign of the Higgs mixing with an exotic state, the sensitivity decreases to the dashed contours in Fig.~\ref{fig:higgslinear}. In the three scenarios considered, the smallest $\sin\alpha$ for which part of the seesaw region can be probed at 95\%~CL is $\sin\alpha > 0.17$ (CMS reinterpretation), $0.1$ (HL-LHC with CMS trigger) and $0.017$ (HL-LHC with soft trigger).

\begin{figure}[t!]
    \centering
    \includegraphics[width=0.49\textwidth]{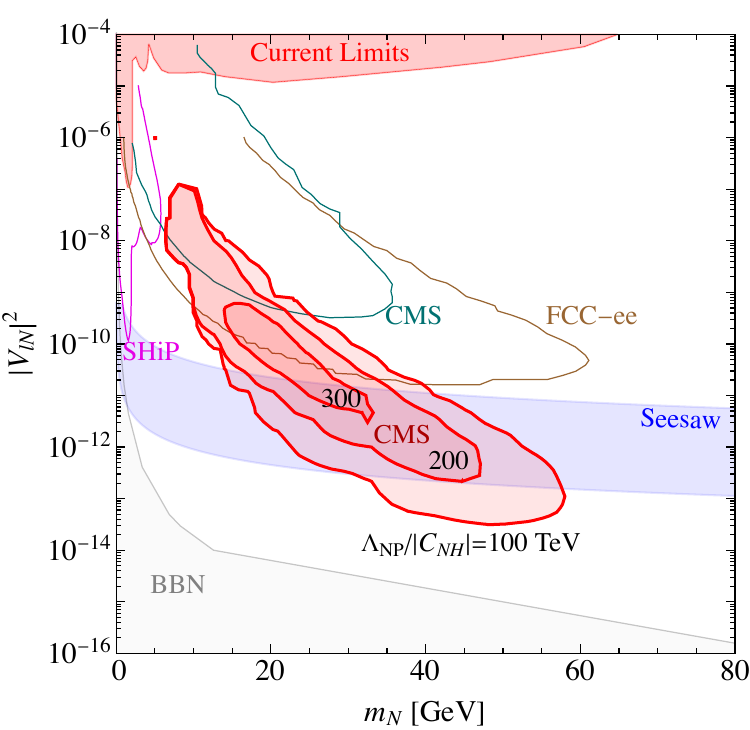}
    \caption{As Fig.~\ref{fig:higgslinear}, but showing the sensitivity of the CMS search to the scale $\Lambda_\text{NP}/|\mathcal{C}_{NH}|$ of the $\nu_R$SMEFT operator~$\mathcal{O}_{NH}$.}
    \label{fig:higgsEFT}
\end{figure}
Our result can also be interpreted in terms of the $\nu_R$SMEFT operator $\mathcal{O}_{NH} = (\bar N^c N)(H^\dagger\cdot H)$. It is the RHN-equivalent of the Weinberg operator, appearing at the same mass dimension-5 and it leads to the vertex $(v\mathcal{C}_{NH}/\Lambda_\text{NP}) hNN$,  enabling the decay $h\to NN$. Here, $\Lambda_\text{NP}$ is the New Physics scale in which the $\nu_R$SMEFT is expanded and $\mathcal{C}_{NH}$ is the dimensionless Wilson coefficient of $\mathcal{O}_{NH}$, see Appendix~\ref{sec:nursmeft} for a brief summary of the $\nu_R$SMEFT. Matching the vertex factor with that in the $B-L$ model, $v\mathcal{C}_{NH}/\Lambda_\text{NP} = y_N\sin\alpha/2 = m_N\sin\alpha/(2\langle\Phi\rangle)$, we can recast the above sensitivities to the $\mathcal{O}_{NH}$ operator scale. As can be seen in Fig.~\ref{fig:higgsEFT}, operator scales as high as $\Lambda_\text{NP}/|\mathcal{C}_{NH}| \approx 300$~TeV are already being probed using the CMS data. At the HL-LHC using the CMS and soft trigger strategies, scales up to $\Lambda_\text{NP}/|\mathcal{C}_{NH}| \approx 500$~TeV and $2000$~TeV can be tested, respectively. These are already comparable to prospects at future lepton colliders~\cite{Barducci:2020icf}.

As discussed in Sec.~\ref{sec:higgs-cs}, the exotic Higgs channel $pp\to\Phi\to NN$ can have a large cross section and is nearly unsuppressed by the Higgs mixing $\sin\alpha$ for $m_h < m_\Phi \lesssim 150$~GeV. The current constraint and future sensitivities are qualitatively similar but better than for SM Higgs production. With the same parameters and scenarios considered above and an exotic Higgs mass of $m_\Phi = 150$~GeV, the smallest Higgs mixing that still results in a 95\%~CL signal is $\sin\alpha \approx 0.07$ (CMS reinterpretation), $0.02$ (HL-LHC with CMS trigger) and $0.002$ (HL-LHC with soft trigger).

\section{Conclusion}
The origin of neutrino masses remains an open issue in particle physics. Although the seesaw mechanism provides an elegant solution, the RHNs it predicts are difficult to probe at colliders due to the small active-sterile mixing required to generate light neutrino masses. We instead consider that RHNs are produced via Higgs or gauge portals in New Physics scenarios, with the $B-L$ gauge model as a prototype. As the SM Higgs portal is the most promising channel, we also interpret our results in the $\nu_R$SMEFT. In such scenarios, RHNs can be produced abundantly at colliders but remain long-lived due to the small active-sterile mixing strengths.

We have reinterpreted the CMS analysis \cite{CMS:2021juv} as a search for RHNs, produced via $pp\to Z', Z, h, \Phi\to NN$ and decaying at displaced vertices in the CMS endcap. We found that the Higgs portals $h$ and $\Phi$ can be used to probe RHNs with masses $m_N \approx 40$~GeV and active-sterile mixing strengths $|V_{\ell N}|^2 \approx 10^{-12}$ using existing CMS data, and for $B-L$ model and $\nu_R$SMEFT parameters satisfying current constraints. The gauge portals $Z$ and $Z'$ offer less promise in the near future.

Our results illustrate that searches for displaced showers in the CMS muon system are powerful probes to reveal the origin of neutrino masses. Existing CMS data is already excluding yet unexplored parameter space in the well-motivated $B-L$ model that covers the seesaw floor, i.e., where a light neutrino mass scale of the order $9\times 10^{-3}~\text{eV} \lesssim m_\nu \lesssim 0.45$ eV is generated. This can be further generalized in the broad context of the $\nu_R$SMEFT, where New Physics scales $\Lambda_\text{NP} \approx 200$~TeV are being tested, which is about one order of magnitude better than the current limit from the Higgs signal strength~\cite{Fernandez-Martinez:2022stj, Fernandez-Martinez:2023phj}, and is expected to improve to $\Lambda_\text{NP} \approx 3000$~TeV at the HL-LHC. Our work motivates a dedicated search for displaced showers in the muon system with potential optimizations applied, such as for 3-body decay modes relevant for our RHN signatures.
 
\begin{acknowledgments}
W. L. is supported by National Natural Science Foundation of China (Grant No.12205153). S. K. is supported by the FWF research group grant FG1 and FWF project number P 36947-N. F. F. D. acknowledges support from the UK Science and Technology Facilities Council (STFC) via the Consolidated Grants ST/P00072X/1 and ST/T000880/1.
\end{acknowledgments}

\appendix
\renewcommand{\theequation}{A.\arabic{equation}}
\setcounter{equation}{0}
\section{The Minimal $B-L$ Gauge Model}
\label{sec:model}

We here briefly summarize the minimal $B-L$ gauge model, and the constraints on relevant model parameters. The $B-L$ model extends the SM gauge group with an additional Abelian gauge symmetry associated with the $B-L$ quantum number, $SU(3)_c\times SU(2)_L \times U(1)_Y \times U(1)_{B-L}$. The particle content is also extended, by including the $B-L$ gauge boson $Z'$, three Weyl RHNs $\nu_R^i$ and the $B-L$ Higgs $\Phi$. The relevant Lagrangian in the interaction eigenstates is 
\begin{align}
\label{LB-L}
    \mathcal{L}_{B-L} &= 
     -\frac{1}{4}Z'_{\mu\nu} Z'^{\mu\nu} 
     + D_\mu\Phi^* D^\mu\Phi + \sum_i \bar\nu_R^i i\slashed{D} \nu_R^i
     - \frac{\epsilon}{2}B^{\mu\nu} Z'_{\mu\nu} \,\,
       \bigg\{+ m^2_{ZZ'} B^\mu Z'_\mu\bigg\} \nonumber\\
    &\quad- \frac{1}{2} \sum_{i,j} 
    \left(\lambda_N^{ij}\bar\nu_R^{i,c}\Phi\nu_R^j + \text{h.c.} \right)
    - \sum_{\alpha,j}\left(\lambda_D^{\alpha j} 
    \bar L^\alpha\cdot\tilde H\nu_R^j + \text{h.c.}\right)
    -{\cal V}(H,\Phi),
\end{align}
with the field strength tensor of the $B-L$ gauge group, $Z^\prime_{\mu\nu} = \partial_\mu Z^\prime_\nu - \partial_\nu Z^\prime_\mu$, hypercharge, $B_{\mu\nu} = \partial_\mu B_\nu - \partial_\nu B_\mu$, and the covariant derivative $D_\mu = D_\mu^\text{SM} - ig_{B-L} Q_{B-L} Z_\mu'$, including the $B-L$ contribution, with $g_{B-L}$ and $Q_{B-L}$ being the $B-L$ gauge coupling and charge, respectively. In Eq.~\eqref{LB-L}, $H$ is the SM Higgs doublet, with $\tilde{H} = i\sigma^2 H^*$ and $L^\alpha$ are the SM lepton doublets. The SM particles have their canonical $Q_{B-L}$ charges while that of the exotic particles take the values $Q_{B-L}(Z') = 0$, $Q_{B-L}(\nu_{Ri}) = -1$ and $Q_{B-L}(\Phi) = 2$. The exotic particles are singlets under the SM gauge symmetries.

\subsection{Higgs Sector}
In Eq.~\eqref{LB-L}, ${\cal V}(H,\Phi)$ is the scalar potential
\begin{align}
\label{VHX}
    {\cal V}(H,\Phi) = 
      m^2(H^\dagger\cdot H) 
    + \mu^2 |\Phi|^2 
    + \lambda_1 (H^\dagger\cdot H)^2
    + \lambda_2 |\Phi|^4 + \lambda_3 (H^\dagger\cdot H) |\Phi|^2.
\end{align}
Both the SM and $B-L$ Higgs acquire vacuum expectation values, $v = \langle H^0\rangle$, $v_{B-L} = \langle\Phi\rangle$, leading to a breaking of the model's gauge group to $SU(3)_c \times U(1)_\text{EM}$. In addition, the scalar potential in Eq.~\eqref{VHX} gives rise to the mass matrix of the Higgs fields $(H^0, \Phi)$ at tree level~\cite{Robens:2015gla},
\begin{align}
\label{mass}
	M_H^2 = \begin{pmatrix}
		2\lambda_1 v^2 & \lambda_3 v_{B-L} v \\
		\lambda_3 v_{B-L} v & 2\lambda_2 v_{B-L}^2
	\end{pmatrix}.
\end{align}
The Higgs masses at tree level are 
\begin{align}
	\label{Higgsmass}
	M^2_{h(\Phi)} 
	&= \lambda_1 v^2 + \lambda_2 v_{B-L}^2 
	-(+) \sqrt{(\lambda_1 v^2 - \lambda_2 v_{B-L}^2)^2 + (\lambda_3 v_{B-L}v)^2},
\end{align} 
with the mixing given by
\begin{align}
\label{Higgs mixing}
	\begin{pmatrix}
		h \\ \Phi'
	\end{pmatrix} = 
	\begin{pmatrix}
		\cos\alpha & -\sin\alpha \\
		\sin\alpha &  \cos\alpha
	\end{pmatrix}
	\begin{pmatrix}
		H^0 \\ \Phi
	\end{pmatrix},
\end{align} 
where the mixing angle $\alpha$ is determined by
\begin{align}
\label{lambda}
	\tan(2\alpha) = \frac{\lambda_3 v_{B-L}v}{\lambda_2 v_{B-L}^2 - \lambda_1 v^2}.
\end{align}
We take $m_h \approx 125~\text{GeV} < m_\Phi$ where we drop the prime on the mass eigenstate \footnote{The case where $m_\Phi < m_h$ is considered in \cite{Liu:2022ugx}.}.

\begin{figure}[t!]
    \centering
    \includegraphics[width=0.5\columnwidth]{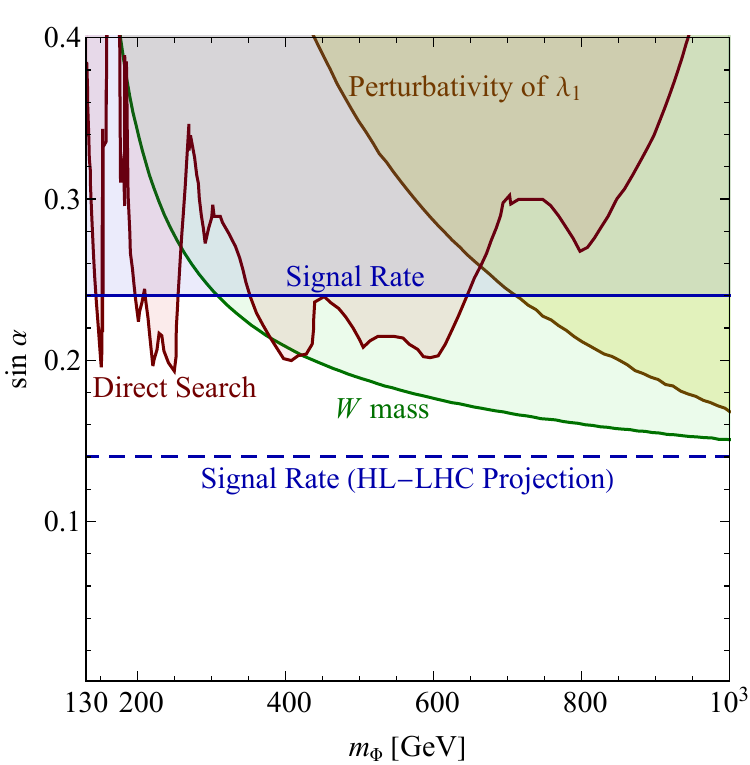}
    \caption{Current limits on the Higgs mixing strength $\sin\alpha$, adapted from \cite{Robens:2022cun}. The bound from measuring the Higgs signal rate has been updated using the LHC Run-2 results~\cite{Robens:2022cun, Papaefstathiou:2022oyi, ATLAS:2021vrm}. Their projection at the HL-LHC with 3000~fb$^{-1}$ integrated luminosity, taking the uncertainty $\delta\mu = |1-\sigma(pp\to h) / \sigma(pp\to h)_\text{SM}| = 0.02$, is shown for comparison~\cite{Cepeda:2019klc, ATLAS:2019mfr}.}
\label{fig:higgslimits}
\end{figure}
A summary of recent experimental limits on the Higgs mixing with an exotic singlet scalar can be found in \cite{Robens:2022cun}. The Higgs mixing can be probed by direct searches for a heavy scalar and a measurement of the Higgs signal rate \cite{Robens:2015gla, Robens:2021rkl, Robens:2022cun}. The existence of an additional scalar also introduces a shift to the $W$ boson mass \cite{Lopez-Val:2014jva, Papaefstathiou:2022oyi}, from which a limit is derived by comparing the experimental Particle Data Group value, $m_W^\text{exp} = 80.379\pm 0.012$~GeV and the SM predicted value, $m_W^\text{SM} = 80.356$~GeV \cite{ParticleDataGroup:2022pth}. Lastly, requiring the scalar coupling $\lambda_1$ to remain perturbative can also be used to infer a limit on the mixing. From Refs.~\cite{Robens:2022cun, Papaefstathiou:2022oyi}, using LHC Run-2 results~\cite{ATLAS:2021vrm}, an upper limit from the signal rate can be obtained, $\sin\alpha \lesssim 0.24$. At the 14~TeV HL-LHC with 3000~fb$^{-1}$ integrated luminosity, a measurement of the Higgs signal rate is expected to achieve an uncertainty of $\delta\mu = |1-\sigma(pp\to h)/\sigma(pp\to h)_\text{SM}| = 0.02$~\cite{Cepeda:2019klc, ATLAS:2019mfr}, with a projected sensitivity on the Higgs mixing of $\sin\alpha \approx 0.14$. The current constraints and the future HL-LHC sensitivity are shown in Fig.~\ref{fig:higgslimits}.

\begin{figure}[t!]
	\centering
	\includegraphics[width=0.5\columnwidth]{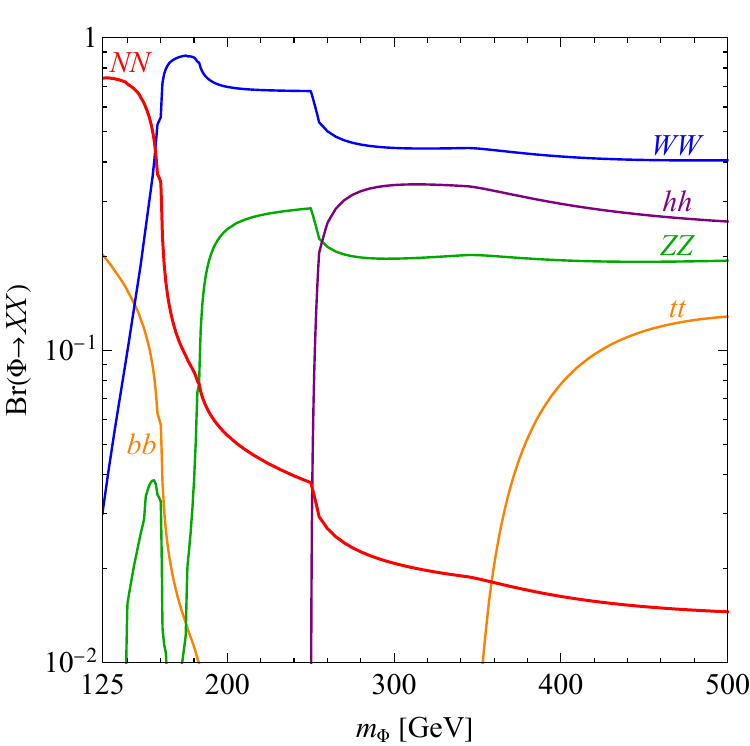}
	\caption{Decay branching ratios of the $B-L$ Higgs $\Phi$ as a function of its mass $m_\Phi$. A single RHN with mass $m_N = 0.3\times m_\Phi$ is assumed and the Higgs mixing strength is taken as $\sin\alpha = 0.08$. The $B-L$ Higgs VEV is fixed at $\langle\Phi\rangle = 3.75$~TeV.} 
	\label{fig:brphi}
\end{figure}
Besides SM Higgs production at the LHC, we also consider the production of the $B-L$ Higgs~$\Phi$ with subsequent decay to RHNs. The branching ratios of $\Phi$ to relevant decay products are shown in Fig.~\ref{fig:brphi} as a function of its mass, calculated in MadGraph for the $B-L$ gauge model. The $B-L$ Higgs decays to SM final states as well as a pair of RHNs, $\Phi \to NN$, where we take the RHN mass $m_N = 0.3\times m_\Phi$ which approximately maximizes the partial decay rate. The decays to SM particles are suppressed by the Higgs mixing which is taken at a small value $\sin\alpha = 0.08$ in the figure to illustrate the potential dominance of the decay $\Phi\to NN$ for $m_\Phi \lesssim 160$~GeV. The $B-L$ Higgs VEV is fixed at $\langle\Phi\rangle = 3.75$~TeV. As noted in the main text, as long as $\Phi\to NN$ is not dominant, the cross section $\sigma(pp\to \Phi \to NN)$ is largely independent of the Higgs mixing as it cancels out between production and decay. As can be seen, Br$(\Phi\to NN) \approx 0.5 - 0.85$ is large for $m_\Phi \lesssim 2m_W \approx 160$~GeV.

\subsection{Gauge Sector}
\begin{figure}[t!]
    \centering
    \includegraphics[width=0.495\textwidth]{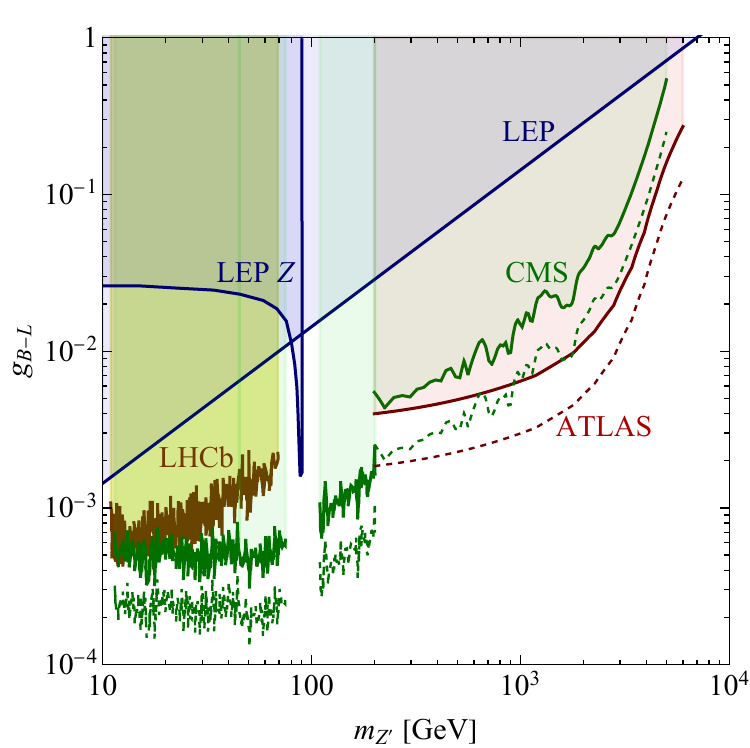}
    \includegraphics[width=0.485\textwidth]{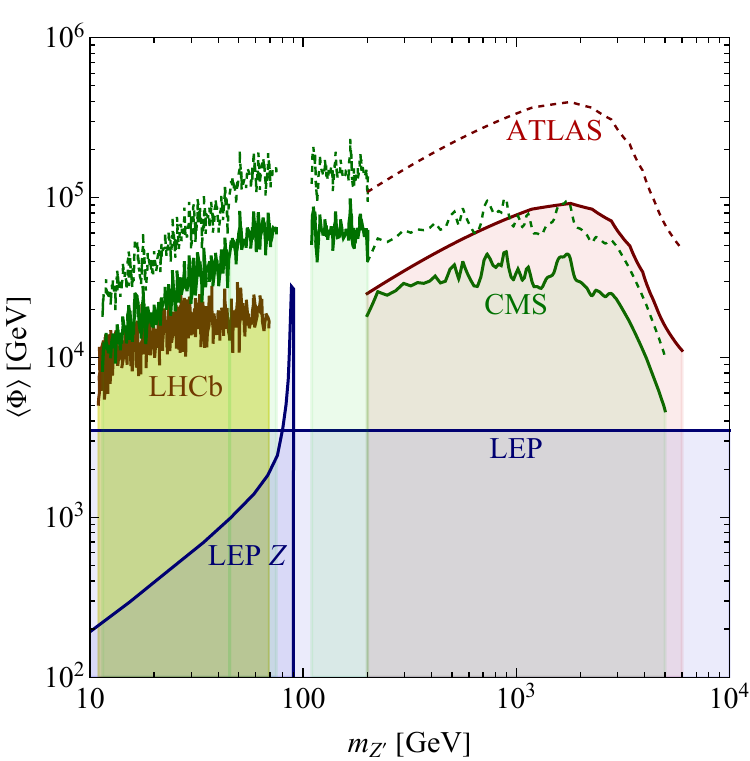}
    \caption{Current limits on the $B-L$ gauge coupling $g_{B-L}$ (left) and the $B-L$ VEV $\langle\Phi\rangle = m_{Z^\prime}/(2g_{B-L})$ (right) as a function of the $Z'$ mass $m_{Z'}$ from resonance searches at CMS~\cite{CMS:2019buh, CMS:2021ctt, Chiang:2019ajm, Liu:2022kid}, ATLAS~\cite{ATLAS:2019erb, Chiang:2019ajm}, LHCb~\cite{LHCb:2019vmc}, electroweak precision tests~\cite{Cacciapaglia:2006pk, ALEPH:2006bhb} and searches for invisible $Z'$ final states \cite{Fox:2011fx} (LEP). Also shown are the projected sensitivities at ATLAS and CMS for the 3000~fb$^{-1}$ HL-LHC.}
\label{fig:limits}
\end{figure}
Arguably the most direct way to probe the $B-L$ model is through the extra gauge boson with mass $m_{Z'} = 2 g_{B-L} v_{B-L}$. In Fig.~\ref{fig:limits}~(left), the current limits and projected sensitivities on the gauge coupling $g_{B-L}$ as a function of $m_{Z'}$ are shown. For the parameter space of our interest, $10~\text{GeV} < m_{Z'} < 10$~TeV, the existing limits mainly arise from colliders through searches at CMS, ATLAS and LHCb, as well as EW precision tests and searches of invisible final states of the $Z'$. Limits from the LHCb were originally presented for a dark photon but can be reinterpreted in the $B-L$ model, following \cite{Ilten:2018crw}.  The limits from EW precision tests~(LEP) are effectively on the vacuum expectation value of the $B-L$ Higgs with $v_{B-L} = m_{Z'}/(2 g_{B-L}) \gtrsim 3.5$~TeV~ \cite{Cacciapaglia:2006pk, ALEPH:2006bhb}. For light $Z'$ with $10~\text{GeV} < m_{Z'} < m_W$, limits are from resonance searches for di-leptons final states at LHCb~\cite{LHCb:2019vmc} and CMS~\cite{CMS:2019buh}. The current best limits are $g_{B-L} \lesssim 6\times 10^{-4}$ for $m_{Z'} \lesssim 70$~GeV. When $m_{Z'} \sim m_{W,Z}$, there are large backgrounds from the SM gauge bosons, hence the limit from LEP becomes dominant, except for $m_{Z'} \lesssim m_Z$ when the search of the invisible decay of $Z'$ at LEP~\cite{Fox:2011fx} is more stringent. The CMS search \cite{CMS:2019buh} is effective for $110~\text{GeV} < m_{Z'} < 200$~GeV, with an upper limit of $g_{B-L} \lesssim 10^{-3}$. For heavier $Z'$, the limits from high mass resonance searches at CMS \cite{CMS:2021ctt} and ATLAS \cite{ATLAS:2019erb}, as well as LEP apply. Among them, the limits from ATLAS are the most stringent for $m_{Z'} \lesssim 6$~TeV. Due to the large mass, the background from the SM gauge bosons is suppressed and the limits become less stringent as $Z'$ becomes heavier, reaching $g_{B-L} \lesssim 0.2$ when $m_{Z'} \approx 6$~TeV. For even heavier $Z'$, one can only rely on the limits from LEP. We also show the estimated sensitivities of ATLAS and CMS at the HL-LHC with 3000~fb$^{-1}$, by scaling the current bounds with luminosity as $g_{B-L} \propto \text{luminosity}^{-1/4}$. The same constraints are displayed in Fig.~\ref{fig:limits}~(right), but interpreted with respect to the $B-L$ VEV $v_{B-L} = \langle\Phi\rangle = m_{Z'} / (2 g_{B-L})$.

In the minimal $B-L$ model, no kinetic or mass mixing between the SM hypercharge and $B-L$ is considered. Kinetic mixing is incorporated by including the $\epsilon$ term in Eq.~\eqref{LB-L} \cite{Dev:2021otb}. The mass mixing term $m_{ZZ'}^2$ violates the model's gauge symmetry but may still be present if the $B-L$ and electroweak symmetry breaking are connected in a broader framework. For our purposes, we take $\epsilon$ and $m_{ZZ'}^2$ as effective, independent parameters at the electroweak scale. The mass mixing induces a field mixing angle $\gamma$ between the two gauge bosons,
\begin{align}
	\tan\gamma = \frac{m^2_{ZZ'}}{m_{Z'}^2 - m_Z^2},
\end{align}
and both kinetic and mass mixing lead to a coupling of the SM-like gauge boson to the $B-L$ current $J_{B-L}^\mu$ \cite{Lindner:2018kjo}, $-\sin\theta_{ZZ'} Z_\mu J_{B-L}^\mu$,
with the overall mixing angle
\begin{align}
	\sin\theta_{ZZ'} \approx 
	\tan\gamma + \frac{\epsilon\sin\theta_W}{m_{Z'}^2 / m_Z^2 - 1},
\end{align}
for $\epsilon$ and $\gamma$ sufficiently small. Here, $\theta_W$ is the electroweak mixing angle. Gauge boson mixing induces an additional parity-violating asymmetry and is thus constrained by precise measurements of atomic parity violation, with $\beta = g_{B-L}\sin\theta_{ZZ'}\lesssim 10^{-3}$~\cite{Dev:2021otb}. This leads to a potential change in the $Z$ production cross section by a factor of $\beta^2 \sim 10^{-6}$, still allowed at the LHC since the uncertainty of the $Z$ production cross section is about 4$\%$~\cite{ATLAS:2016fij, ATLAS:2023sjw}.

\subsection{Neutrino Sector}
\label{sec:model-neutrino}
The Yukawa matrix $y_N$ in Eq.~\eqref{LB-L} gives rise to the RHN masses, generated in breaking the $B-L$ symmetry, with the mass matrix given by $m_N =  y_N \langle\Phi\rangle$. Likewise, the active neutrinos mix with the RHNs via the Dirac mass matrix $m_D = y_\nu v/\sqrt{2}$. The complete mass matrix in the $(\nu_L^c, \nu_R)$ basis is then
\begin{align}
	\label{MD}
	{\cal M} = 
	\begin{pmatrix}
		0   & m_D \\
		m^T_D & M_N
	\end{pmatrix}.
\end{align} 
In the seesaw limit, $M_N \gg m_D$, the light neutrino masses are 
\begin{align}
	\label{seesaw}
	m_\nu = - m_D M^{-1}_N m^T_D,
\end{align}
and the flavour and mass eigenstates of the light and heavy neutrinos are related as  
\begin{align}
	\label{Neutrino}
	\begin{pmatrix}
		\nu_L^c \\ \nu_R
	\end{pmatrix} = 
	\begin{pmatrix}
		U & V_{\ell N} \\
		V_{N\ell} & U_N
	\end{pmatrix}
	\begin{pmatrix}
		\nu^c \\ N
	\end{pmatrix}.
\end{align} 
The mixing and the light neutrino masses are constrained by oscillation experiments, namely, the charged current lepton mixing $U \approx U_\text{PMNS}$, apart from small non-unitarity corrections. For simplicity, we assume the presence of a single RHN $N$ mixing with a single lepton flavour at a time. This relates the light neutrino mass scale $m_\nu$ with the RHN mass $m_N$ as $m_\nu = |V_{\ell N}|^2 m_N$. While this simplification does not allow describing the full light neutrino phenomenology, the mass should not exceed limits on the absolute neutrino mass scale currently set by the KATRIN experiment as $m_\nu < 0.45$~eV at 90~\% CL~\cite{Katrin:2024tvg}. While not a strict lower limit, observations of the solar neutrino oscillation length indicate a smallest non-vanishing neutrino mass scale of $\sqrt{\Delta m^2_\text{sol}} = 9\times 10^{-3}$~eV~\cite{Esteban:2020cvm}. We thus take the range $9\times 10^{-3}~\text{eV} < |V_{\ell N}|^2 m_N < 0.45$~eV as a target to probe the canonical seesaw floor of neutrino mass generation.

RHNs can be searched for through their mixing $V_{\ell N}$ with the active neutrinos, via their resulting participation in SM neutral and charged-current interactions, irrespective of the presence of a $B-L$ gauge interaction. This leads to a wide range of constraints at colliders, in beam dump experiments, meson decay searches, etc.. In the figures in the main text and below, we display current constraints compiled in \cite{Bolton:2019pcu} for comparison with our production mechanisms. The focus of future efforts is on the RHN lifetime frontier and we display the projected sensitivities of SHiP \cite{SHiP:2018xqw}, CMS \cite{Drewes:2019fou} and FCC-ee \cite{Blondel:2022qqo} as representative examples in our parameter space of interest.

\subsection{Sensitivity of the CMS Displaced Shower Search}
\label{sec:model-constraints}
\begin{figure}[t!]
    \centering
    \includegraphics[width=0.49\textwidth]{gghlinear}
    \includegraphics[width=0.49\textwidth]{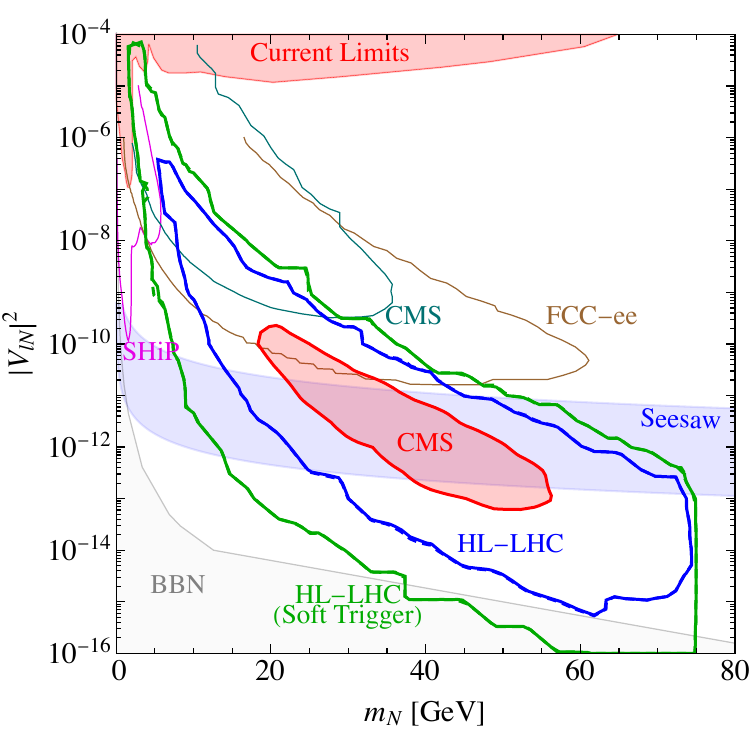} 
    \caption{Sensitivity of $pp\to h\to NN$ (left) and $pp\to\Phi\to NN$ ($m_\Phi = 150$~GeV, right) at 95\%~CL on the RHN mass $m_N$ and the active-sterile mixing $|V_{\ell N}|^2$ ($\ell = e, \tau$). The scenarios are for reinterpreting CMS data (red region), as well as at the 14~TeV HL-LHC with 3000~fb$^{-1}$ using the CMS (blue contour) and soft (green contour) trigger strategies. The Higgs mixing is $\sin\alpha = 0.24$ (red region and solid contours) and $0.14$ (dashed), and the $B-L$ Higgs VEV is $\langle\Phi\rangle = 3.75$~TeV. The region labelled `Current Limits' is excluded by sterile neutrino searches ($\ell = e$) while the contours labelled SHiP, CMS and FCC-ee give the projected sensitivity of planned experiments \cite{Bolton:2019pcu, Bolton:2022pyf}. The band labelled `Seesaw' indicates $9\times 10^{-3}~\text{eV} < |V_{\ell N}|^2 m_N < 0.45$~eV.}
    \label{fig:higgslinearcombined}
\end{figure}
As discussed in the main text, the CMS search for displaced showers in the muon endcap can be interpreted in terms of the $B-L$ gauge model via the processes $pp\to h \to NN$ and $pp\to\Phi\to NN$ with RHNs decaying in the muon system. For completeness, we show the resulting sensitivity in Fig.~\ref{fig:higgslinearcombined}, with the left panel being identical to Fig.~2 in the main text.

\section{Interpretation in the RHN-extended SM Effective Field Theory}
\label{sec:nursmeft}

Instead of choosing a specific ultraviolet-complete scenario such as the $U(1)_{B-L}$ gauge model described above, the effects of heavy New Physics at SM scales and below can be categorized and interpreted within an Effective Field Theory (EFT) approach. Assuming that the SM particle content is extended by a single, sterile and relatively light RHN ($m_N \lesssim \Lambda_\text{EW}$) with no other exotic states present, the so-called RHN-extended SM EFT~($\nu_R$SMEFT)~\cite{Graesser:2007pc, Graesser:2007yj, delAguila:2008ir, Aparici:2009fh,Liao:2016qyd} applies, and the Lagrangian can be expressed as
\begin{align}
\label{eq:lag_nusmeft}
    {\cal L} = {\cal L}_\text{SM}  
             + \bar \nu_R \slashed\partial \nu_R 
             - \frac{1}{2} m_N \bar \nu_R^c \nu_R
             - \sum_\alpha \lambda_D^\alpha 
                           \bar L^\alpha \cdot \tilde H \nu_R  
             + \sum_{n=5}^{\infty} \frac{{\cal C}_n {\cal O}_n}{\Lambda^{n-4}_\text{NP}} + \text{h.c.}.
\end{align}
At the renormalizable level, it adds a kinetic term and a Majorana mass term for the RHN $\nu_R$ field as well as the usual Yukawa interactions $\lambda_D^\alpha$ to the SM lepton doublets. This gives rise to the usual seesaw mechanism, where we consider the single RHN $N \approx \nu_R$ with mass $m_N$ to generate a light neutrino mass scale $m_\nu = |V_{\ell N}|^2 m_N$ as described in Sec.~\ref{sec:model-neutrino}. Low energy effects of the heavy New Physics are then captured by effective operators $\mathcal{O}_n$ of increasing mass dimension $n$, constructed from SM fields and the RHN. They are accompanied by dimensionless Wilson coefficients ${\cal C}_n$ quantifying the individual operator strengths, usually expected to be $|{\cal C}_n| = \mathcal{O}(1)$ unless additional symmetry considerations lead to a suppression. The set of operators includes those of the SMEFT (without RHN) and additional operators incorporating the RHN. 

At the lowest order, $n=5$, there are two operators \footnote{There exists a third dimension-5 operator, ${\cal O}_{NB} = (\bar \nu_R^c \sigma^{\mu\nu} \nu_R)B_{\mu\nu} = 0$ that is usually considered but is absent for a single RHN, reflecting the fact that a Majorana fermion does not have a magnetic dipole moment.}, 
\begin{align}
    {\cal O}_W^{\alpha\beta} &= (\bar L^{\alpha,c}\cdot\tilde H^*)(\tilde H^\dag \cdot L^\beta),  \nonumber\\
    {\cal O}_{NH} &= (\bar \nu_R^c \nu_R)(H^\dag \cdot H).
\end{align}
The first, ${\cal O}_W^{\alpha\beta}$, is the well-known Weinberg operator inducing light neutrino masses after EW symmetry breaking. In the $\nu_R$SMEFT, it describes an additional contribution $\delta m_\nu^{\alpha\beta} = - \mathcal{C}_W^{\alpha\beta}v^2 / \Lambda_\text{NP}$, beyond that from the RHN-induced seesaw. This could, e.g., arise from additional RHN states heavier than the EW scale, or from other neutrino mass generation mechanisms in the ultraviolet. The second operator, ${\cal O}_{NH}$, is the RHN equivalent of the Weinberg operator. It induces an additional RHN Majorana mass contribution after EW symmetry breaking, $\delta m_N = - \mathcal{C}_{NH}v^2/\Lambda_\text{NP}$. More importantly in our context, it also generates the $hNN$ vertex with coupling strength $\mathcal{C}_{NH}v/\Lambda_\text{NP}$.

The operator ${\cal O}_{NH}$ will thus induce the Higgs branching ratio
\begin{align}
\label{eq:BrhnnEFT}
    \text{Br}(h\to NN) = 
    \frac{1}{4\pi} \frac{|\mathcal{C}_{NH}|^2 v^2}{\Lambda^2_\text{NP}} 
    \frac{m_h}{\Gamma_h^\text{SM}} 
    \left(1-\frac{4 m_N^2}{m_h^2}\right)^{3/2},
\end{align}
which is the equivalent of Eq.~(4) in the main text. With the RHN decaying via the active-sterile mixing $V_{\ell N}$, as in the $U(1)_{B-L}$ model, the constraints from the CMS search and future sensitivities can thus be reinterpreted in terms of the operator scale $\Lambda_\text{NP}/|\mathcal{C}_{NH}|$ in the $\nu_R$SMEFT. For this, we assume that other operators do not contribute significantly to the RHN decay width. 

\begin{figure}[t!]
    \centering
    \includegraphics[width=0.49\textwidth]{EFTcon}\\
    \includegraphics[width=0.49\columnwidth]{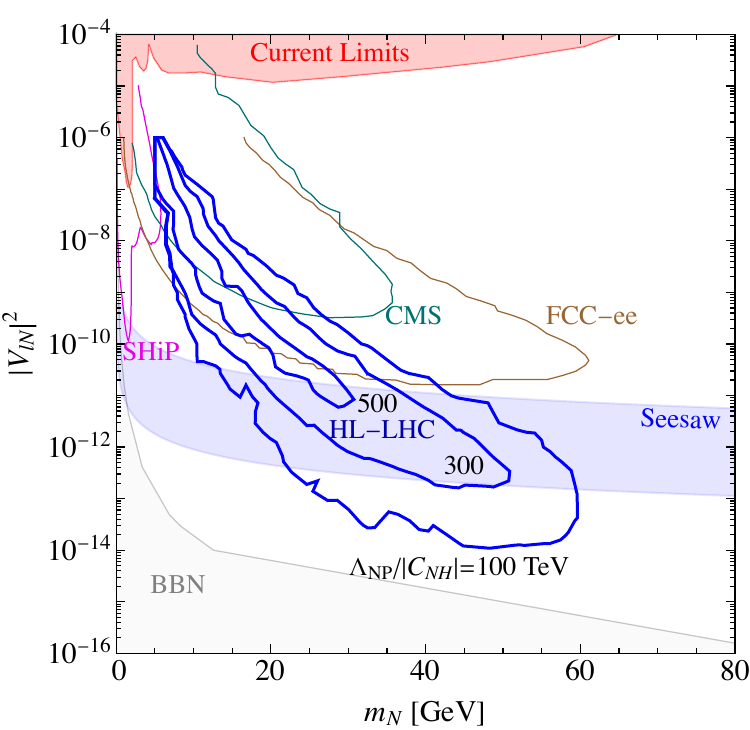}
    \includegraphics[width=0.49\columnwidth]{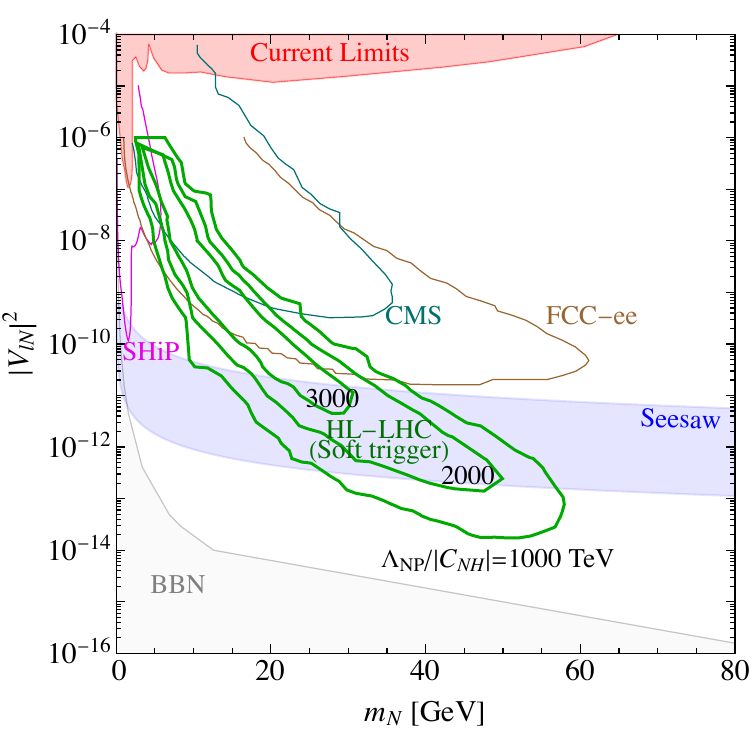}
    \caption{Lower limit on the $\nu_R$SMEFT operator scale $\Lambda_\text{NP}/|\mathcal{C}_{NH}|$ as a function of the RHN mass $m_N$ and the active-sterile mixing strength $|V_{\ell N}|^2$ using existing CMS data (top) as well as at the 14~TeV HL-LHC with 3000~fb$^{-1}$ using the CMS~(bottom left) and soft trigger strategies~(bottom right). The other elements in the plots are as described in Fig.~\ref{fig:higgslinearcombined}.}
\label{fig:EFTcon}
\end{figure}
For a given RHN mass $m_N$ and active-sterile mixing strength $|V_{\ell N}|^2$ we determine the smallest operator scale $\Lambda_\text{NP}/|\mathcal{C}_{NH}|$ that yields no significant signal in the CMS muon system at 95\% CL by reinterpreting it in terms of the Higgs branching ratio in Eq.~\eqref{eq:BrhnnEFT}. This is shown in Fig.~\ref{fig:EFTcon}, using the existing CMS data (top, identical to Fig.~3 in the main text) as well as at the 14~TeV HL-LHC with 3000~fb$^{-1}$ using the CMS~(bottom left) and soft trigger strategies~(bottom right). As in the $U(1)_{B-L}$ scenario, the search is sensitive as long as the RHN is sufficiently long-lived, $0.5~\text{m} \lesssim L_N \lesssim 6$~m, to be detected in the CMS muon system. Within this band, there is a soft dependence on the RHN mass, with the sensitivity decreasing with larger $m_N$ until nearing the threshold $m_N = m_h/2$.

\begin{figure}[t!]
    \centering
    \includegraphics[width=0.5\columnwidth]{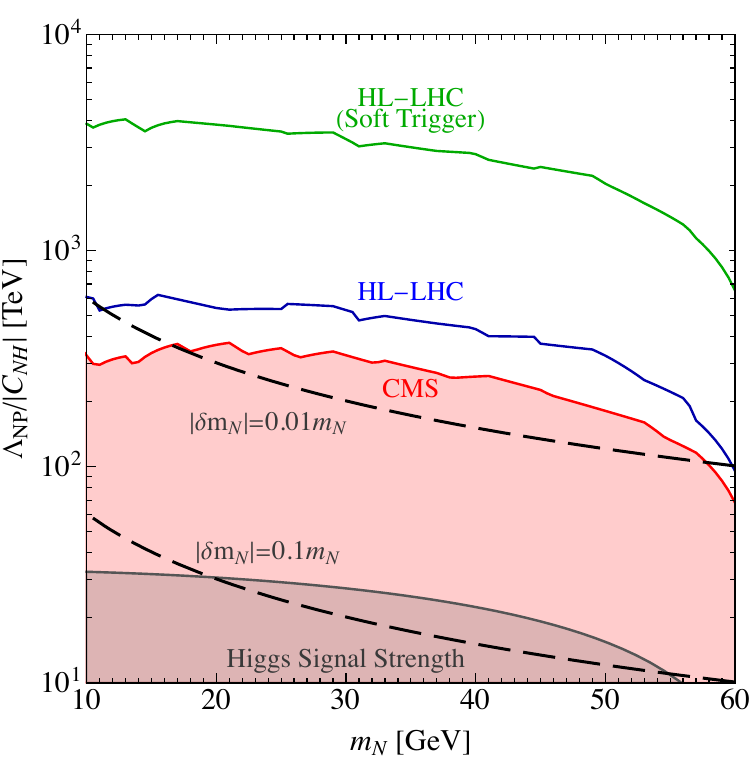}
    \caption{Best upper limit and projected sensitivity (maximized over $|V_{\ell N}|^2$) on the $\nu_R$SMEFT operator scale $\Lambda_\text{NP}/|\mathcal{C}_{NH}|$ as a function of the RHN mass $m_N$ from the existing CMS search~(red), at the HL-LHC with the CMS trigger~(blue) and at the HL-LHC with the soft trigger~(green). The current limit on the operator scale from the Higgs signal strength \cite{Fernandez-Martinez:2023phj} is shown as well. A contribution to the RHN mass from the operator $\mathcal{O}_{NH}$ at the level $|\delta m_N| = 0.1m_N$ and $0.01m_N$ is indicated by the dashed curves.}
\label{fig:EFT}
\end{figure}
In Fig.~\ref{fig:EFT}, we likewise show the best constraint on $\Lambda_\text{NP}/|\mathcal{C}_{NH}|$ for a given RHN mass $m_N$, comparing the existing CMS search with the future sensitivities at the HL-LHC with the CMS trigger and the soft trigger as described in the main text. The sensitivity peaks at $\Lambda_\text{NP}/|\mathcal{C}_{NH}| \gtrsim 300, 500, 3000$~TeV with CMS data, at the HL-LHC using the CMS trigger and at the HL-LHC using the soft trigger, respectively. We also show the limit from Higgs signal strength measurements, i.e., interpreting $h\to NN$ as an invisible decay, at $\Lambda_\text{NP}/|\mathcal{C}_{NH}| \gtrsim 33$~TeV~\cite{Fernandez-Martinez:2022stj, Fernandez-Martinez:2023phj}. The sensitivity of the existing CMS displaced shower search surpasses this limit by an order of magnitude. This is expected to improve by a hundred-fold at the HL-LHC using the soft trigger, which has already been installed since the start of Run-3~\cite{CMS-DP-2022-062}. As mentioned, the operator ${\cal O}_{NH}$ induces a correction to the bare mass of the RHN after EW symmetry breaking, $\delta m_N = -\mathcal{C}_{NH}v^2 / \Lambda_\text{NP}$. For consistency, it should be small compared to the bare mass and corrections at the 10\% and 1\% level are indicated by the dashed curves in Fig.~\ref{fig:EFT}.

As the main difference to the interpretation in the ultraviolet-complete $U(1)_{B-L}$ gauge model, the sensitivity to $\Lambda_{NP}$ does not decrease for small $m_N$. This is because the $hNN$ coupling in the $U(1)_{B-L}$ model arises from the breaking of lepton number, generating the RHN Majorana mass, and thus the $hNN$ vertex is proportional to $m_N$. This is expected to be a generic feature if the source of lepton number violation can be directly connected to the RHN mass. It can be incorporated within the $\nu_R$SMEFT via the inclusion of a spurion~\cite{Barducci:2020ncz}, associated with lepton number, to reflect the additional symmetry and its controlled breaking in the ultraviolet sector. As a consequence, effective operators violating lepton number vanish if lepton number is conserved in the renormalizable Lagrangian, i.e., as $m_N \to 0$. The relevant Wilson coefficients are thus suppressed as $\mathcal{C}_\text{LNV} = \mathcal{C}'_\text{LNV} m_N / \Lambda_\text{NP}$ with $|\mathcal{C}'_\text{LNV}| = \mathcal{O}(1)$. The $h\to NN$ branching ratio then scales as
\begin{align}
    \text{Br}(h\to NN) = 
    \frac{1}{4\pi} \frac{|\mathcal{C}'_{NH}|^2 m_N^2 v^2}{\Lambda^4_\text{NP}} 
    \frac{m_h}{\Gamma_h^\text{SM}} 
    \left(1-\frac{4 m_N^2}{m_h^2}\right)^{3/2}.
\end{align}
In this context, we can further reinterpret the sensitivity to $\Lambda_\text{NP}$ by making the substitution $|\mathcal{C}_{NH}|/\Lambda_\text{NP} \to |\mathcal{C}'_{NH}|m_N/\Lambda^2_\text{NP}$. The previous sensitivities $\Lambda_\text{NP}/|\mathcal{C}_{NH}| = 300, 500, 3000$~TeV for the CMS data, the HL-LHC (CMS trigger) and the HL-LHC (soft trigger) are then mapped to $\Lambda'_\text{NP}/|\mathcal{C}_{NH}| = 3.5, 4.5, 11$~TeV, respectively, for $m_N = 40$~GeV. This is compatible with the sensitivity with respect to the $B-L$-breaking scale and the $Z'$ mass in the $U(1)_{B-L}$ gauge model discussed in Sec.~\ref{sec:model-constraints}.

Other $\nu_R$SMEFT operators at higher dimensions \cite{Liao:2016qyd} may also contribute to the RHN and Higgs decay, and could be probed similarly. At dimension-6, operators coupling the RHN with the Higgs are \cite{Fernandez-Martinez:2023phj}
\begin{gather}
    {\cal O}_{LNH}^\alpha = (\bar L_\alpha \cdot \tilde H \nu_R) 
    (H^\dag \cdot H), \quad
    {\cal O}_{NHH} = (\bar\nu_R \gamma^\mu \nu_R) 
    (H^\dagger i \overleftrightarrow{D}_\mu H),
\end{gather}
and 
\begin{gather}
    {\cal O}_{N^2 H^4} = (\bar\nu_R \nu_R) (H^\dag \cdot H)^2, 
    \nonumber\\
    {\cal O}_{ND1} = (\bar\nu_R D_\mu \nu_R) 
    (H^\dagger \overleftrightarrow{D}^\mu H), \quad
    {\cal O}_{ND2} = (\bar\nu_R \nu_R) 
    \big((D_\mu H)^\dagger \cdot (D^\mu H)\big),
\end{gather}
at dimension-7 \cite{Liao:2016qyd}. Here, $D_\mu$ is the SM covariant derivative. Displaced shower searches can also be sensitive to $h \to N\nu$ which, in addition to the usual active-sterile neutrino mixing contribution, involves further operators.

\bibliography{submit}

\begin{thebibliography}{130}%
\makeatletter
\providecommand \@ifxundefined [1]{%
 \@ifx{#1\undefined}
}%
\providecommand \@ifnum [1]{%
 \ifnum #1\expandafter \@firstoftwo
 \else \expandafter \@secondoftwo
 \fi
}%
\providecommand \@ifx [1]{%
 \ifx #1\expandafter \@firstoftwo
 \else \expandafter \@secondoftwo
 \fi
}%
\providecommand \natexlab [1]{#1}%
\providecommand \enquote  [1]{``#1''}%
\providecommand \bibnamefont  [1]{#1}%
\providecommand \bibfnamefont [1]{#1}%
\providecommand \citenamefont [1]{#1}%
\providecommand \href@noop [0]{\@secondoftwo}%
\providecommand \href [0]{\begingroup \@sanitize@url \@href}%
\providecommand \@href[1]{\@@startlink{#1}\@@href}%
\providecommand \@@href[1]{\endgroup#1\@@endlink}%
\providecommand \@sanitize@url [0]{\catcode `\\12\catcode `\$12\catcode `\&12\catcode `\#12\catcode `\^12\catcode `\_12\catcode `\%12\relax}%
\providecommand \@@startlink[1]{}%
\providecommand \@@endlink[0]{}%
\providecommand \url  [0]{\begingroup\@sanitize@url \@url }%
\providecommand \@url [1]{\endgroup\@href {#1}{\urlprefix }}%
\providecommand \urlprefix  [0]{URL }%
\providecommand \Eprint [0]{\href }%
\providecommand \doibase [0]{https://doi.org/}%
\providecommand \selectlanguage [0]{\@gobble}%
\providecommand \bibinfo  [0]{\@secondoftwo}%
\providecommand \bibfield  [0]{\@secondoftwo}%
\providecommand \translation [1]{[#1]}%
\providecommand \BibitemOpen [0]{}%
\providecommand \bibitemStop [0]{}%
\providecommand \bibitemNoStop [0]{.\EOS\space}%
\providecommand \EOS [0]{\spacefactor3000\relax}%
\providecommand \BibitemShut  [1]{\csname bibitem#1\endcsname}%
\let\auto@bib@innerbib\@empty
\bibitem [{\citenamefont {Davis}(1994)}]{Davis:1994jw}%
  \BibitemOpen
  \bibfield  {author} {\bibinfo {author} {\bibfnamefont {R.}~\bibnamefont {Davis}},\ }\bibfield  {title} {\bibinfo {title} {{A review of the Homestake solar neutrino experiment}},\ }\href {https://doi.org/10.1016/0146-6410(94)90004-3} {\bibfield  {journal} {\bibinfo  {journal} {Prog. Part. Nucl. Phys.}\ }\textbf {\bibinfo {volume} {32}},\ \bibinfo {pages} {13} (\bibinfo {year} {1994})}\BibitemShut {NoStop}%
\bibitem [{\citenamefont {Fukuda}\ \emph {et~al.}(1998)\citenamefont {Fukuda} \emph {et~al.}}]{Super-Kamiokande:1998kpq}%
  \BibitemOpen
  \bibfield  {author} {\bibinfo {author} {\bibfnamefont {Y.}~\bibnamefont {Fukuda}} \emph {et~al.} (\bibinfo {collaboration} {Super-Kamiokande}),\ }\bibfield  {title} {\bibinfo {title} {{Evidence for oscillation of atmospheric neutrinos}},\ }\href {https://doi.org/10.1103/PhysRevLett.81.1562} {\bibfield  {journal} {\bibinfo  {journal} {Phys. Rev. Lett.}\ }\textbf {\bibinfo {volume} {81}},\ \bibinfo {pages} {1562} (\bibinfo {year} {1998})},\ \Eprint {https://arxiv.org/abs/hep-ex/9807003} {arXiv:hep-ex/9807003} \BibitemShut {NoStop}%
\bibitem [{\citenamefont {Eguchi}\ \emph {et~al.}(2003)\citenamefont {Eguchi} \emph {et~al.}}]{KamLAND:2002uet}%
  \BibitemOpen
  \bibfield  {author} {\bibinfo {author} {\bibfnamefont {K.}~\bibnamefont {Eguchi}} \emph {et~al.} (\bibinfo {collaboration} {KamLAND}),\ }\bibfield  {title} {\bibinfo {title} {{First results from KamLAND: Evidence for reactor anti-neutrino disappearance}},\ }\href {https://doi.org/10.1103/PhysRevLett.90.021802} {\bibfield  {journal} {\bibinfo  {journal} {Phys. Rev. Lett.}\ }\textbf {\bibinfo {volume} {90}},\ \bibinfo {pages} {021802} (\bibinfo {year} {2003})},\ \Eprint {https://arxiv.org/abs/hep-ex/0212021} {arXiv:hep-ex/0212021} \BibitemShut {NoStop}%
\bibitem [{\citenamefont {Ahmad}\ \emph {et~al.}(2002)\citenamefont {Ahmad} \emph {et~al.}}]{SNO:2002tuh}%
  \BibitemOpen
  \bibfield  {author} {\bibinfo {author} {\bibfnamefont {Q.~R.}\ \bibnamefont {Ahmad}} \emph {et~al.} (\bibinfo {collaboration} {SNO}),\ }\bibfield  {title} {\bibinfo {title} {{Direct evidence for neutrino flavor transformation from neutral current interactions in the Sudbury Neutrino Observatory}},\ }\href {https://doi.org/10.1103/PhysRevLett.89.011301} {\bibfield  {journal} {\bibinfo  {journal} {Phys. Rev. Lett.}\ }\textbf {\bibinfo {volume} {89}},\ \bibinfo {pages} {011301} (\bibinfo {year} {2002})},\ \Eprint {https://arxiv.org/abs/nucl-ex/0204008} {arXiv:nucl-ex/0204008} \BibitemShut {NoStop}%
\bibitem [{\citenamefont {Minkowski}(1977)}]{Minkowski:1977sc}%
  \BibitemOpen
  \bibfield  {author} {\bibinfo {author} {\bibfnamefont {P.}~\bibnamefont {Minkowski}},\ }\bibfield  {title} {\bibinfo {title} {{$\mu \to e\gamma$ at a Rate of One Out of $10^{9}$ Muon Decays?}},\ }\href {https://doi.org/10.1016/0370-2693(77)90435-X} {\bibfield  {journal} {\bibinfo  {journal} {Phys. Lett. B}\ }\textbf {\bibinfo {volume} {67}},\ \bibinfo {pages} {421} (\bibinfo {year} {1977})}\BibitemShut {NoStop}%
\bibitem [{\citenamefont {Aker}\ \emph {et~al.}(2024)\citenamefont {Aker} \emph {et~al.}}]{Katrin:2024tvg}%
  \BibitemOpen
  \bibfield  {author} {\bibinfo {author} {\bibfnamefont {M.}~\bibnamefont {Aker}} \emph {et~al.} (\bibinfo {collaboration} {Katrin}),\ }\bibfield  {title} {\bibinfo {title} {{Direct neutrino-mass measurement based on 259 days of KATRIN data}},\ }\href@noop {} {\bibfield  {journal} {\bibinfo  {journal} {arXiv}\ } (\bibinfo {year} {2024})},\ \Eprint {https://arxiv.org/abs/2406.13516} {arXiv:2406.13516 [nucl-ex]} \BibitemShut {NoStop}%
\bibitem [{\citenamefont {Esteban}\ \emph {et~al.}(2020)\citenamefont {Esteban}, \citenamefont {Gonzalez-Garcia}, \citenamefont {Maltoni}, \citenamefont {Schwetz},\ and\ \citenamefont {Zhou}}]{Esteban:2020cvm}%
  \BibitemOpen
  \bibfield  {author} {\bibinfo {author} {\bibfnamefont {I.}~\bibnamefont {Esteban}}, \bibinfo {author} {\bibfnamefont {M.~C.}\ \bibnamefont {Gonzalez-Garcia}}, \bibinfo {author} {\bibfnamefont {M.}~\bibnamefont {Maltoni}}, \bibinfo {author} {\bibfnamefont {T.}~\bibnamefont {Schwetz}},\ and\ \bibinfo {author} {\bibfnamefont {A.}~\bibnamefont {Zhou}},\ }\bibfield  {title} {\bibinfo {title} {{The fate of hints: updated global analysis of three-flavor neutrino oscillations}},\ }\href {https://doi.org/10.1007/JHEP09(2020)178} {\bibfield  {journal} {\bibinfo  {journal} {JHEP}\ }\textbf {\bibinfo {volume} {09}},\ \bibinfo {pages} {178}},\ \Eprint {https://arxiv.org/abs/2007.14792} {arXiv:2007.14792 [hep-ph]} \BibitemShut {NoStop}%
\bibitem [{\citenamefont {Kaneta}\ \emph {et~al.}(2017)\citenamefont {Kaneta}, \citenamefont {Kang},\ and\ \citenamefont {Lee}}]{Kaneta:2016vkq}%
  \BibitemOpen
  \bibfield  {author} {\bibinfo {author} {\bibfnamefont {K.}~\bibnamefont {Kaneta}}, \bibinfo {author} {\bibfnamefont {Z.}~\bibnamefont {Kang}},\ and\ \bibinfo {author} {\bibfnamefont {H.-S.}\ \bibnamefont {Lee}},\ }\bibfield  {title} {\bibinfo {title} {{Right-handed neutrino dark matter under the $B - L$ gauge interaction}},\ }\href {https://doi.org/10.1007/JHEP02(2017)031} {\bibfield  {journal} {\bibinfo  {journal} {JHEP}\ }\textbf {\bibinfo {volume} {02}},\ \bibinfo {pages} {031}},\ \Eprint {https://arxiv.org/abs/1606.09317} {arXiv:1606.09317 [hep-ph]} \BibitemShut {NoStop}%
\bibitem [{\citenamefont {Bondarenko}\ \emph {et~al.}(2018)\citenamefont {Bondarenko}, \citenamefont {Boyarsky}, \citenamefont {Gorbunov},\ and\ \citenamefont {Ruchayskiy}}]{Bondarenko:2018ptm}%
  \BibitemOpen
  \bibfield  {author} {\bibinfo {author} {\bibfnamefont {K.}~\bibnamefont {Bondarenko}}, \bibinfo {author} {\bibfnamefont {A.}~\bibnamefont {Boyarsky}}, \bibinfo {author} {\bibfnamefont {D.}~\bibnamefont {Gorbunov}},\ and\ \bibinfo {author} {\bibfnamefont {O.}~\bibnamefont {Ruchayskiy}},\ }\bibfield  {title} {\bibinfo {title} {{Phenomenology of GeV-scale Heavy Neutral Leptons}},\ }\href {https://doi.org/10.1007/JHEP11(2018)032} {\bibfield  {journal} {\bibinfo  {journal} {JHEP}\ }\textbf {\bibinfo {volume} {11}},\ \bibinfo {pages} {032}},\ \Eprint {https://arxiv.org/abs/1805.08567} {arXiv:1805.08567 [hep-ph]} \BibitemShut {NoStop}%
\bibitem [{\citenamefont {Bryman}\ and\ \citenamefont {Shrock}(2019)}]{Bryman:2019bjg}%
  \BibitemOpen
  \bibfield  {author} {\bibinfo {author} {\bibfnamefont {D.~A.}\ \bibnamefont {Bryman}}\ and\ \bibinfo {author} {\bibfnamefont {R.}~\bibnamefont {Shrock}},\ }\bibfield  {title} {\bibinfo {title} {{Constraints on Sterile Neutrinos in the MeV to GeV Mass Range}},\ }\href {https://doi.org/10.1103/PhysRevD.100.073011} {\bibfield  {journal} {\bibinfo  {journal} {Phys. Rev. D}\ }\textbf {\bibinfo {volume} {100}},\ \bibinfo {pages} {073011} (\bibinfo {year} {2019})},\ \Eprint {https://arxiv.org/abs/1909.11198} {arXiv:1909.11198 [hep-ph]} \BibitemShut {NoStop}%
\bibitem [{\citenamefont {Balaji}\ \emph {et~al.}(2020)\citenamefont {Balaji}, \citenamefont {Ramirez-Quezada},\ and\ \citenamefont {Zhou}}]{Balaji:2019fxd}%
  \BibitemOpen
  \bibfield  {author} {\bibinfo {author} {\bibfnamefont {S.}~\bibnamefont {Balaji}}, \bibinfo {author} {\bibfnamefont {M.}~\bibnamefont {Ramirez-Quezada}},\ and\ \bibinfo {author} {\bibfnamefont {Y.-L.}\ \bibnamefont {Zhou}},\ }\bibfield  {title} {\bibinfo {title} {{CP violation and circular polarisation in neutrino radiative decay}},\ }\href {https://doi.org/10.1007/JHEP04(2020)178} {\bibfield  {journal} {\bibinfo  {journal} {JHEP}\ }\textbf {\bibinfo {volume} {04}},\ \bibinfo {pages} {178}},\ \Eprint {https://arxiv.org/abs/1910.08558} {arXiv:1910.08558 [hep-ph]} \BibitemShut {NoStop}%
\bibitem [{\citenamefont {Bolton}\ \emph {et~al.}(2020)\citenamefont {Bolton}, \citenamefont {Deppisch},\ and\ \citenamefont {Bhupal~Dev}}]{Bolton:2019pcu}%
  \BibitemOpen
  \bibfield  {author} {\bibinfo {author} {\bibfnamefont {P.~D.}\ \bibnamefont {Bolton}}, \bibinfo {author} {\bibfnamefont {F.~F.}\ \bibnamefont {Deppisch}},\ and\ \bibinfo {author} {\bibfnamefont {P.~S.}\ \bibnamefont {Bhupal~Dev}},\ }\bibfield  {title} {\bibinfo {title} {{Neutrinoless double beta decay versus other probes of heavy sterile neutrinos}},\ }\href {https://doi.org/10.1007/JHEP03(2020)170} {\bibfield  {journal} {\bibinfo  {journal} {JHEP}\ }\textbf {\bibinfo {volume} {03}},\ \bibinfo {pages} {170}},\ \Eprint {https://arxiv.org/abs/1912.03058} {arXiv:1912.03058 [hep-ph]} \BibitemShut {NoStop}%
\bibitem [{\citenamefont {Liu}\ \emph {et~al.}(2022{\natexlab{a}})\citenamefont {Liu}, \citenamefont {Xie},\ and\ \citenamefont {Yi}}]{Liu:2021akf}%
  \BibitemOpen
  \bibfield  {author} {\bibinfo {author} {\bibfnamefont {W.}~\bibnamefont {Liu}}, \bibinfo {author} {\bibfnamefont {K.-P.}\ \bibnamefont {Xie}},\ and\ \bibinfo {author} {\bibfnamefont {Z.}~\bibnamefont {Yi}},\ }\bibfield  {title} {\bibinfo {title} {{Testing leptogenesis at the LHC and future muon colliders: A $Z^\prime$ scenario}},\ }\href {https://doi.org/10.1103/PhysRevD.105.095034} {\bibfield  {journal} {\bibinfo  {journal} {Phys. Rev. D}\ }\textbf {\bibinfo {volume} {105}},\ \bibinfo {pages} {095034} (\bibinfo {year} {2022}{\natexlab{a}})},\ \Eprint {https://arxiv.org/abs/2109.15087} {arXiv:2109.15087 [hep-ph]} \BibitemShut {NoStop}%
\bibitem [{\citenamefont {Abdullahi}\ \emph {et~al.}(2023)\citenamefont {Abdullahi} \emph {et~al.}}]{Abdullahi:2022jlv}%
  \BibitemOpen
  \bibfield  {author} {\bibinfo {author} {\bibfnamefont {A.~M.}\ \bibnamefont {Abdullahi}} \emph {et~al.},\ }\bibfield  {title} {\bibinfo {title} {{The present and future status of heavy neutral leptons}},\ }\href {https://doi.org/10.1088/1361-6471/ac98f9} {\bibfield  {journal} {\bibinfo  {journal} {J. Phys. G}\ }\textbf {\bibinfo {volume} {50}},\ \bibinfo {pages} {020501} (\bibinfo {year} {2023})},\ \Eprint {https://arxiv.org/abs/2203.08039} {arXiv:2203.08039 [hep-ph]} \BibitemShut {NoStop}%
\bibitem [{\citenamefont {Zhang}\ and\ \citenamefont {Liu}(2023)}]{Zhang:2023nxy}%
  \BibitemOpen
  \bibfield  {author} {\bibinfo {author} {\bibfnamefont {Y.}~\bibnamefont {Zhang}}\ and\ \bibinfo {author} {\bibfnamefont {W.}~\bibnamefont {Liu}},\ }\bibfield  {title} {\bibinfo {title} {{Probing active-sterile neutrino transition magnetic moments at LEP and CEPC}},\ }\href {https://doi.org/10.1103/PhysRevD.107.095031} {\bibfield  {journal} {\bibinfo  {journal} {Phys. Rev. D}\ }\textbf {\bibinfo {volume} {107}},\ \bibinfo {pages} {095031} (\bibinfo {year} {2023})},\ \Eprint {https://arxiv.org/abs/2301.06050} {arXiv:2301.06050 [hep-ph]} \BibitemShut {NoStop}%
\bibitem [{\citenamefont {Barducci}\ \emph {et~al.}(2023)\citenamefont {Barducci}, \citenamefont {Liu}, \citenamefont {Titov}, \citenamefont {Wang},\ and\ \citenamefont {Zhang}}]{Barducci:2023hzo}%
  \BibitemOpen
  \bibfield  {author} {\bibinfo {author} {\bibfnamefont {D.}~\bibnamefont {Barducci}}, \bibinfo {author} {\bibfnamefont {W.}~\bibnamefont {Liu}}, \bibinfo {author} {\bibfnamefont {A.}~\bibnamefont {Titov}}, \bibinfo {author} {\bibfnamefont {Z.~S.}\ \bibnamefont {Wang}},\ and\ \bibinfo {author} {\bibfnamefont {Y.}~\bibnamefont {Zhang}},\ }\bibfield  {title} {\bibinfo {title} {{Probing the dipole portal to heavy neutral leptons via meson decays at the high-luminosity LHC}},\ }\href {https://doi.org/10.1103/PhysRevD.108.115009} {\bibfield  {journal} {\bibinfo  {journal} {Phys. Rev. D}\ }\textbf {\bibinfo {volume} {108}},\ \bibinfo {pages} {115009} (\bibinfo {year} {2023})},\ \Eprint {https://arxiv.org/abs/2308.16608} {arXiv:2308.16608 [hep-ph]} \BibitemShut {NoStop}%
\bibitem [{\citenamefont {Liu}\ and\ \citenamefont {Deppisch}(2023)}]{Liu:2023klu}%
  \BibitemOpen
  \bibfield  {author} {\bibinfo {author} {\bibfnamefont {W.}~\bibnamefont {Liu}}\ and\ \bibinfo {author} {\bibfnamefont {F.~F.}\ \bibnamefont {Deppisch}},\ }\bibfield  {title} {\bibinfo {title} {{Testing Leptogenesis and Seesaw using Long-lived Particle Searches in the $B-L$ Model}},\ }\href@noop {} {\  (\bibinfo {year} {2023})},\ \Eprint {https://arxiv.org/abs/2312.11165} {arXiv:2312.11165 [hep-ph]} \BibitemShut {NoStop}%
\bibitem [{\citenamefont {Chatrchyan}\ \emph {et~al.}(2012)\citenamefont {Chatrchyan} \emph {et~al.}}]{CMS:2012wqj}%
  \BibitemOpen
  \bibfield  {author} {\bibinfo {author} {\bibfnamefont {S.}~\bibnamefont {Chatrchyan}} \emph {et~al.} (\bibinfo {collaboration} {CMS}),\ }\bibfield  {title} {\bibinfo {title} {{Search for heavy Majorana Neutrinos in $\mu^{\pm}\mu^{\pm} +$ Jets and $e^{\pm}e^{\pm} +$ Jets Events in pp Collisions at $\sqrt{s} =$ 7 TeV}},\ }\href {https://doi.org/10.1016/j.physletb.2012.09.012} {\bibfield  {journal} {\bibinfo  {journal} {Phys. Lett. B}\ }\textbf {\bibinfo {volume} {717}},\ \bibinfo {pages} {109} (\bibinfo {year} {2012})},\ \Eprint {https://arxiv.org/abs/1207.6079} {arXiv:1207.6079 [hep-ex]} \BibitemShut {NoStop}%
\bibitem [{\citenamefont {Aaij}\ \emph {et~al.}(2014)\citenamefont {Aaij} \emph {et~al.}}]{LHCb:2014osd}%
  \BibitemOpen
  \bibfield  {author} {\bibinfo {author} {\bibfnamefont {R.}~\bibnamefont {Aaij}} \emph {et~al.} (\bibinfo {collaboration} {LHCb}),\ }\bibfield  {title} {\bibinfo {title} {{Search for Majorana neutrinos in $B^- \to \pi^+\mu^-\mu^-$ decays}},\ }\href {https://doi.org/10.1103/PhysRevLett.112.131802} {\bibfield  {journal} {\bibinfo  {journal} {Phys. Rev. Lett.}\ }\textbf {\bibinfo {volume} {112}},\ \bibinfo {pages} {131802} (\bibinfo {year} {2014})},\ \Eprint {https://arxiv.org/abs/1401.5361} {arXiv:1401.5361 [hep-ex]} \BibitemShut {NoStop}%
\bibitem [{\citenamefont {Aad}\ \emph {et~al.}(2015)\citenamefont {Aad} \emph {et~al.}}]{ATLAS:2015gtp}%
  \BibitemOpen
  \bibfield  {author} {\bibinfo {author} {\bibfnamefont {G.}~\bibnamefont {Aad}} \emph {et~al.} (\bibinfo {collaboration} {ATLAS}),\ }\bibfield  {title} {\bibinfo {title} {{Search for heavy Majorana neutrinos with the ATLAS detector in pp collisions at $ \sqrt{s}=8 $ TeV}},\ }\href {https://doi.org/10.1007/JHEP07(2015)162} {\bibfield  {journal} {\bibinfo  {journal} {JHEP}\ }\textbf {\bibinfo {volume} {07}},\ \bibinfo {pages} {162}},\ \Eprint {https://arxiv.org/abs/1506.06020} {arXiv:1506.06020 [hep-ex]} \BibitemShut {NoStop}%
\bibitem [{\citenamefont {Khachatryan}\ \emph {et~al.}(2015)\citenamefont {Khachatryan} \emph {et~al.}}]{CMS:2015qur}%
  \BibitemOpen
  \bibfield  {author} {\bibinfo {author} {\bibfnamefont {V.}~\bibnamefont {Khachatryan}} \emph {et~al.} (\bibinfo {collaboration} {CMS}),\ }\bibfield  {title} {\bibinfo {title} {{Search for heavy Majorana neutrinos in $\mu^\pm \mu^\pm+$ jets events in proton-proton collisions at $\sqrt{s}$ = 8 TeV}},\ }\href {https://doi.org/10.1016/j.physletb.2015.06.070} {\bibfield  {journal} {\bibinfo  {journal} {Phys. Lett. B}\ }\textbf {\bibinfo {volume} {748}},\ \bibinfo {pages} {144} (\bibinfo {year} {2015})},\ \Eprint {https://arxiv.org/abs/1501.05566} {arXiv:1501.05566 [hep-ex]} \BibitemShut {NoStop}%
\bibitem [{\citenamefont {Khachatryan}\ \emph {et~al.}(2016)\citenamefont {Khachatryan} \emph {et~al.}}]{CMS:2016aro}%
  \BibitemOpen
  \bibfield  {author} {\bibinfo {author} {\bibfnamefont {V.}~\bibnamefont {Khachatryan}} \emph {et~al.} (\bibinfo {collaboration} {CMS}),\ }\bibfield  {title} {\bibinfo {title} {{Search for heavy Majorana neutrinos in e$^{±}$e$^{±}$+ jets and e$^{±}$ $\mu^{±}$+ jets events in proton-proton collisions at $ \sqrt{s}=8 $ TeV}},\ }\href {https://doi.org/10.1007/JHEP04(2016)169} {\bibfield  {journal} {\bibinfo  {journal} {JHEP}\ }\textbf {\bibinfo {volume} {04}},\ \bibinfo {pages} {169}},\ \Eprint {https://arxiv.org/abs/1603.02248} {arXiv:1603.02248 [hep-ex]} \BibitemShut {NoStop}%
\bibitem [{\citenamefont {Cortina~Gil}\ \emph {et~al.}(2018)\citenamefont {Cortina~Gil} \emph {et~al.}}]{NA62:2017qcd}%
  \BibitemOpen
  \bibfield  {author} {\bibinfo {author} {\bibfnamefont {E.}~\bibnamefont {Cortina~Gil}} \emph {et~al.} (\bibinfo {collaboration} {NA62}),\ }\bibfield  {title} {\bibinfo {title} {{Search for heavy neutral lepton production in $K^+$ decays}},\ }\href {https://doi.org/10.1016/j.physletb.2018.01.031} {\bibfield  {journal} {\bibinfo  {journal} {Phys. Lett. B}\ }\textbf {\bibinfo {volume} {778}},\ \bibinfo {pages} {137} (\bibinfo {year} {2018})},\ \Eprint {https://arxiv.org/abs/1712.00297} {arXiv:1712.00297 [hep-ex]} \BibitemShut {NoStop}%
\bibitem [{\citenamefont {Sirunyan}\ \emph {et~al.}(2018)\citenamefont {Sirunyan} \emph {et~al.}}]{CMS:2018iaf}%
  \BibitemOpen
  \bibfield  {author} {\bibinfo {author} {\bibfnamefont {A.~M.}\ \bibnamefont {Sirunyan}} \emph {et~al.} (\bibinfo {collaboration} {CMS}),\ }\bibfield  {title} {\bibinfo {title} {{Search for heavy neutral leptons in events with three charged leptons in proton-proton collisions at $\sqrt{s} =$ 13 TeV}},\ }\href {https://doi.org/10.1103/PhysRevLett.120.221801} {\bibfield  {journal} {\bibinfo  {journal} {Phys. Rev. Lett.}\ }\textbf {\bibinfo {volume} {120}},\ \bibinfo {pages} {221801} (\bibinfo {year} {2018})},\ \Eprint {https://arxiv.org/abs/1802.02965} {arXiv:1802.02965 [hep-ex]} \BibitemShut {NoStop}%
\bibitem [{\citenamefont {Sirunyan}\ \emph {et~al.}(2019)\citenamefont {Sirunyan} \emph {et~al.}}]{CMS:2018jxx}%
  \BibitemOpen
  \bibfield  {author} {\bibinfo {author} {\bibfnamefont {A.~M.}\ \bibnamefont {Sirunyan}} \emph {et~al.} (\bibinfo {collaboration} {CMS}),\ }\bibfield  {title} {\bibinfo {title} {{Search for heavy Majorana neutrinos in same-sign dilepton channels in proton-proton collisions at $ \sqrt{s}=13 $ TeV}},\ }\href {https://doi.org/10.1007/JHEP01(2019)122} {\bibfield  {journal} {\bibinfo  {journal} {JHEP}\ }\textbf {\bibinfo {volume} {01}},\ \bibinfo {pages} {122}},\ \Eprint {https://arxiv.org/abs/1806.10905} {arXiv:1806.10905 [hep-ex]} \BibitemShut {NoStop}%
\bibitem [{\citenamefont {Aad}\ \emph {et~al.}(2019{\natexlab{a}})\citenamefont {Aad} \emph {et~al.}}]{ATLAS:2019kpx}%
  \BibitemOpen
  \bibfield  {author} {\bibinfo {author} {\bibfnamefont {G.}~\bibnamefont {Aad}} \emph {et~al.} (\bibinfo {collaboration} {ATLAS}),\ }\bibfield  {title} {\bibinfo {title} {{Search for heavy neutral leptons in decays of $W$ bosons produced in 13 TeV $pp$ collisions using prompt and displaced signatures with the ATLAS detector}},\ }\href {https://doi.org/10.1007/JHEP10(2019)265} {\bibfield  {journal} {\bibinfo  {journal} {JHEP}\ }\textbf {\bibinfo {volume} {10}},\ \bibinfo {pages} {265}},\ \Eprint {https://arxiv.org/abs/1905.09787} {arXiv:1905.09787 [hep-ex]} \BibitemShut {NoStop}%
\bibitem [{\citenamefont {Aaij}\ \emph {et~al.}(2021)\citenamefont {Aaij} \emph {et~al.}}]{LHCb:2020wxx}%
  \BibitemOpen
  \bibfield  {author} {\bibinfo {author} {\bibfnamefont {R.}~\bibnamefont {Aaij}} \emph {et~al.} (\bibinfo {collaboration} {LHCb}),\ }\bibfield  {title} {\bibinfo {title} {{Search for heavy neutral leptons in $W^+\to\mu^{+}\mu^{\pm}\text{jet}$ decays}},\ }\href {https://doi.org/10.1140/epjc/s10052-021-08973-5} {\bibfield  {journal} {\bibinfo  {journal} {Eur. Phys. J. C}\ }\textbf {\bibinfo {volume} {81}},\ \bibinfo {pages} {248} (\bibinfo {year} {2021})},\ \Eprint {https://arxiv.org/abs/2011.05263} {arXiv:2011.05263 [hep-ex]} \BibitemShut {NoStop}%
\bibitem [{\citenamefont {Tumasyan}\ \emph {et~al.}(2022)\citenamefont {Tumasyan} \emph {et~al.}}]{CMS:2022fut}%
  \BibitemOpen
  \bibfield  {author} {\bibinfo {author} {\bibfnamefont {A.}~\bibnamefont {Tumasyan}} \emph {et~al.} (\bibinfo {collaboration} {CMS}),\ }\bibfield  {title} {\bibinfo {title} {{Search for long-lived heavy neutral leptons with displaced vertices in proton-proton collisions at $ \sqrt{\mathrm{s}} $ =13 TeV}},\ }\href {https://doi.org/10.1007/JHEP07(2022)081} {\bibfield  {journal} {\bibinfo  {journal} {JHEP}\ }\textbf {\bibinfo {volume} {07}},\ \bibinfo {pages} {081}},\ \Eprint {https://arxiv.org/abs/2201.05578} {arXiv:2201.05578 [hep-ex]} \BibitemShut {NoStop}%
\bibitem [{\citenamefont {Aad}\ \emph {et~al.}(2023{\natexlab{a}})\citenamefont {Aad} \emph {et~al.}}]{ATLAS:2023tkz}%
  \BibitemOpen
  \bibfield  {author} {\bibinfo {author} {\bibfnamefont {G.}~\bibnamefont {Aad}} \emph {et~al.} (\bibinfo {collaboration} {ATLAS}),\ }\bibfield  {title} {\bibinfo {title} {{Search for Majorana neutrinos in same-sign WW scattering events from pp collisions at $\sqrt{s}=13$~TeV}},\ }\href {https://doi.org/10.1140/epjc/s10052-023-11915-y} {\bibfield  {journal} {\bibinfo  {journal} {Eur. Phys. J. C}\ }\textbf {\bibinfo {volume} {83}},\ \bibinfo {pages} {824} (\bibinfo {year} {2023}{\natexlab{a}})},\ \Eprint {https://arxiv.org/abs/2305.14931} {arXiv:2305.14931 [hep-ex]} \BibitemShut {NoStop}%
\bibitem [{\citenamefont {Hayrapetyan}\ \emph {et~al.}(2024{\natexlab{a}})\citenamefont {Hayrapetyan} \emph {et~al.}}]{CMS:2024ake}%
  \BibitemOpen
  \bibfield  {author} {\bibinfo {author} {\bibfnamefont {A.}~\bibnamefont {Hayrapetyan}} \emph {et~al.} (\bibinfo {collaboration} {CMS}),\ }\bibfield  {title} {\bibinfo {title} {{Search for long-lived heavy neutral leptons decaying in the CMS muon detectors in proton-proton collisions at s=13\,\,TeV}},\ }\href {https://doi.org/10.1103/PhysRevD.110.012004} {\bibfield  {journal} {\bibinfo  {journal} {Phys. Rev. D}\ }\textbf {\bibinfo {volume} {110}},\ \bibinfo {pages} {012004} (\bibinfo {year} {2024}{\natexlab{a}})},\ \Eprint {https://arxiv.org/abs/2402.18658} {arXiv:2402.18658 [hep-ex]} \BibitemShut {NoStop}%
\bibitem [{\citenamefont {Davidson}(1979)}]{Davidson:1978pm}%
  \BibitemOpen
  \bibfield  {author} {\bibinfo {author} {\bibfnamefont {A.}~\bibnamefont {Davidson}},\ }\bibfield  {title} {\bibinfo {title} {{$B-L$ as the fourth color within an $\mathrm{SU}(2)_L \times \mathrm{U}(1)_R \times \mathrm{U}(1)$ model}},\ }\href {https://doi.org/10.1103/PhysRevD.20.776} {\bibfield  {journal} {\bibinfo  {journal} {Phys. Rev. D}\ }\textbf {\bibinfo {volume} {20}},\ \bibinfo {pages} {776} (\bibinfo {year} {1979})}\BibitemShut {NoStop}%
\bibitem [{\citenamefont {Mohapatra}\ and\ \citenamefont {Marshak}(1980)}]{Mohapatra:1980qe}%
  \BibitemOpen
  \bibfield  {author} {\bibinfo {author} {\bibfnamefont {R.~N.}\ \bibnamefont {Mohapatra}}\ and\ \bibinfo {author} {\bibfnamefont {R.~E.}\ \bibnamefont {Marshak}},\ }\bibfield  {title} {\bibinfo {title} {{Local B-L Symmetry of Electroweak Interactions, Majorana Neutrinos and Neutron Oscillations}},\ }\href {https://doi.org/10.1103/PhysRevLett.44.1316} {\bibfield  {journal} {\bibinfo  {journal} {Phys. Rev. Lett.}\ }\textbf {\bibinfo {volume} {44}},\ \bibinfo {pages} {1316} (\bibinfo {year} {1980})},\ \bibinfo {note} {[Erratum: Phys.Rev.Lett. 44, 1643 (1980)]}\BibitemShut {NoStop}%
\bibitem [{\citenamefont {Debnath}\ and\ \citenamefont {Fileviez~Perez}(2023)}]{Debnath:2023akj}%
  \BibitemOpen
  \bibfield  {author} {\bibinfo {author} {\bibfnamefont {H.}~\bibnamefont {Debnath}}\ and\ \bibinfo {author} {\bibfnamefont {P.}~\bibnamefont {Fileviez~Perez}},\ }\bibfield  {title} {\bibinfo {title} {{Low scale seesaw mechanism with local lepton number}},\ }\href {https://doi.org/10.1103/PhysRevD.108.075009} {\bibfield  {journal} {\bibinfo  {journal} {Phys. Rev. D}\ }\textbf {\bibinfo {volume} {108}},\ \bibinfo {pages} {075009} (\bibinfo {year} {2023})},\ \Eprint {https://arxiv.org/abs/2307.03646} {arXiv:2307.03646 [hep-ph]} \BibitemShut {NoStop}%
\bibitem [{\citenamefont {Fileviez~Perez}(2024)}]{FileviezPerez:2024fzc}%
  \BibitemOpen
  \bibfield  {author} {\bibinfo {author} {\bibfnamefont {P.}~\bibnamefont {Fileviez~Perez}},\ }\bibfield  {title} {\bibinfo {title} {{Lepton and baryon numbers as local gauge symmetries}},\ }\href {https://doi.org/10.1103/PhysRevD.110.035018} {\bibfield  {journal} {\bibinfo  {journal} {Phys. Rev. D}\ }\textbf {\bibinfo {volume} {110}},\ \bibinfo {pages} {035018} (\bibinfo {year} {2024})},\ \Eprint {https://arxiv.org/abs/2406.06866} {arXiv:2406.06866 [hep-ph]} \BibitemShut {NoStop}%
\bibitem [{\citenamefont {Mohapatra}\ and\ \citenamefont {Pati}(1975)}]{Mohapatra:1974gc}%
  \BibitemOpen
  \bibfield  {author} {\bibinfo {author} {\bibfnamefont {R.~N.}\ \bibnamefont {Mohapatra}}\ and\ \bibinfo {author} {\bibfnamefont {J.~C.}\ \bibnamefont {Pati}},\ }\bibfield  {title} {\bibinfo {title} {{A Natural Left-Right Symmetry}},\ }\href {https://doi.org/10.1103/PhysRevD.11.2558} {\bibfield  {journal} {\bibinfo  {journal} {Phys. Rev. D}\ }\textbf {\bibinfo {volume} {11}},\ \bibinfo {pages} {2558} (\bibinfo {year} {1975})}\BibitemShut {NoStop}%
\bibitem [{\citenamefont {Senjanovic}\ and\ \citenamefont {Mohapatra}(1975)}]{Senjanovic:1975rk}%
  \BibitemOpen
  \bibfield  {author} {\bibinfo {author} {\bibfnamefont {G.}~\bibnamefont {Senjanovic}}\ and\ \bibinfo {author} {\bibfnamefont {R.~N.}\ \bibnamefont {Mohapatra}},\ }\bibfield  {title} {\bibinfo {title} {{Exact Left-Right Symmetry and Spontaneous Violation of Parity}},\ }\href {https://doi.org/10.1103/PhysRevD.12.1502} {\bibfield  {journal} {\bibinfo  {journal} {Phys. Rev. D}\ }\textbf {\bibinfo {volume} {12}},\ \bibinfo {pages} {1502} (\bibinfo {year} {1975})}\BibitemShut {NoStop}%
\bibitem [{\citenamefont {Mohapatra}\ and\ \citenamefont {Senjanovic}(1980)}]{Mohapatra:1979ia}%
  \BibitemOpen
  \bibfield  {author} {\bibinfo {author} {\bibfnamefont {R.~N.}\ \bibnamefont {Mohapatra}}\ and\ \bibinfo {author} {\bibfnamefont {G.}~\bibnamefont {Senjanovic}},\ }\bibfield  {title} {\bibinfo {title} {{Neutrino Mass and Spontaneous Parity Nonconservation}},\ }\href {https://doi.org/10.1103/PhysRevLett.44.912} {\bibfield  {journal} {\bibinfo  {journal} {Phys. Rev. Lett.}\ }\textbf {\bibinfo {volume} {44}},\ \bibinfo {pages} {912} (\bibinfo {year} {1980})}\BibitemShut {NoStop}%
\bibitem [{\citenamefont {Pati}\ and\ \citenamefont {Salam}(1974)}]{Pati:1974yy}%
  \BibitemOpen
  \bibfield  {author} {\bibinfo {author} {\bibfnamefont {J.~C.}\ \bibnamefont {Pati}}\ and\ \bibinfo {author} {\bibfnamefont {A.}~\bibnamefont {Salam}},\ }\bibfield  {title} {\bibinfo {title} {{Lepton Number as the Fourth Color}},\ }\href {https://doi.org/10.1103/PhysRevD.10.275} {\bibfield  {journal} {\bibinfo  {journal} {Phys. Rev. D}\ }\textbf {\bibinfo {volume} {10}},\ \bibinfo {pages} {275} (\bibinfo {year} {1974})},\ \bibinfo {note} {[Erratum: Phys.Rev.D 11, 703--703 (1975)]}\BibitemShut {NoStop}%
\bibitem [{\citenamefont {Batell}\ \emph {et~al.}(2024)\citenamefont {Batell}, \citenamefont {Bhoonah},\ and\ \citenamefont {Huang}}]{Batell:2024ddo}%
  \BibitemOpen
  \bibfield  {author} {\bibinfo {author} {\bibfnamefont {B.}~\bibnamefont {Batell}}, \bibinfo {author} {\bibfnamefont {A.}~\bibnamefont {Bhoonah}},\ and\ \bibinfo {author} {\bibfnamefont {W.}~\bibnamefont {Huang}},\ }\bibfield  {title} {\bibinfo {title} {{Right-Handed Neutrino Masses from the Electroweak Scale}},\ }\href@noop {} {\bibfield  {journal} {\bibinfo  {journal} {arXiv}\ } (\bibinfo {year} {2024})},\ \Eprint {https://arxiv.org/abs/2411.07294} {arXiv:2411.07294 [hep-ph]} \BibitemShut {NoStop}%
\bibitem [{\citenamefont {Fukugita}\ and\ \citenamefont {Yanagida}(1986)}]{Fukugita:1986hr}%
  \BibitemOpen
  \bibfield  {author} {\bibinfo {author} {\bibfnamefont {M.}~\bibnamefont {Fukugita}}\ and\ \bibinfo {author} {\bibfnamefont {T.}~\bibnamefont {Yanagida}},\ }\bibfield  {title} {\bibinfo {title} {{Baryogenesis Without Grand Unification}},\ }\href {https://doi.org/10.1016/0370-2693(86)91126-3} {\bibfield  {journal} {\bibinfo  {journal} {Phys. Lett. B}\ }\textbf {\bibinfo {volume} {174}},\ \bibinfo {pages} {45} (\bibinfo {year} {1986})}\BibitemShut {NoStop}%
\bibitem [{\citenamefont {Luty}(1992)}]{Luty:1992un}%
  \BibitemOpen
  \bibfield  {author} {\bibinfo {author} {\bibfnamefont {M.~A.}\ \bibnamefont {Luty}},\ }\bibfield  {title} {\bibinfo {title} {{Baryogenesis via leptogenesis}},\ }\href {https://doi.org/10.1103/PhysRevD.45.455} {\bibfield  {journal} {\bibinfo  {journal} {Phys. Rev. D}\ }\textbf {\bibinfo {volume} {45}},\ \bibinfo {pages} {455} (\bibinfo {year} {1992})}\BibitemShut {NoStop}%
\bibitem [{\citenamefont {Davidson}\ \emph {et~al.}(2008)\citenamefont {Davidson}, \citenamefont {Nardi},\ and\ \citenamefont {Nir}}]{Davidson:2008bu}%
  \BibitemOpen
  \bibfield  {author} {\bibinfo {author} {\bibfnamefont {S.}~\bibnamefont {Davidson}}, \bibinfo {author} {\bibfnamefont {E.}~\bibnamefont {Nardi}},\ and\ \bibinfo {author} {\bibfnamefont {Y.}~\bibnamefont {Nir}},\ }\bibfield  {title} {\bibinfo {title} {{Leptogenesis}},\ }\href {https://doi.org/10.1016/j.physrep.2008.06.002} {\bibfield  {journal} {\bibinfo  {journal} {Phys. Rept.}\ }\textbf {\bibinfo {volume} {466}},\ \bibinfo {pages} {105} (\bibinfo {year} {2008})},\ \Eprint {https://arxiv.org/abs/0802.2962} {arXiv:0802.2962 [hep-ph]} \BibitemShut {NoStop}%
\bibitem [{\citenamefont {Iso}\ \emph {et~al.}(2009{\natexlab{a}})\citenamefont {Iso}, \citenamefont {Okada},\ and\ \citenamefont {Orikasa}}]{Iso:2009ss}%
  \BibitemOpen
  \bibfield  {author} {\bibinfo {author} {\bibfnamefont {S.}~\bibnamefont {Iso}}, \bibinfo {author} {\bibfnamefont {N.}~\bibnamefont {Okada}},\ and\ \bibinfo {author} {\bibfnamefont {Y.}~\bibnamefont {Orikasa}},\ }\bibfield  {title} {\bibinfo {title} {{Classically conformal $B^-$ L extended Standard Model}},\ }\href {https://doi.org/10.1016/j.physletb.2009.04.046} {\bibfield  {journal} {\bibinfo  {journal} {Phys. Lett. B}\ }\textbf {\bibinfo {volume} {676}},\ \bibinfo {pages} {81} (\bibinfo {year} {2009}{\natexlab{a}})},\ \Eprint {https://arxiv.org/abs/0902.4050} {arXiv:0902.4050 [hep-ph]} \BibitemShut {NoStop}%
\bibitem [{\citenamefont {Iso}\ \emph {et~al.}(2009{\natexlab{b}})\citenamefont {Iso}, \citenamefont {Okada},\ and\ \citenamefont {Orikasa}}]{Iso:2009nw}%
  \BibitemOpen
  \bibfield  {author} {\bibinfo {author} {\bibfnamefont {S.}~\bibnamefont {Iso}}, \bibinfo {author} {\bibfnamefont {N.}~\bibnamefont {Okada}},\ and\ \bibinfo {author} {\bibfnamefont {Y.}~\bibnamefont {Orikasa}},\ }\bibfield  {title} {\bibinfo {title} {{The minimal B-L model naturally realized at TeV scale}},\ }\href {https://doi.org/10.1103/PhysRevD.80.115007} {\bibfield  {journal} {\bibinfo  {journal} {Phys. Rev. D}\ }\textbf {\bibinfo {volume} {80}},\ \bibinfo {pages} {115007} (\bibinfo {year} {2009}{\natexlab{b}})},\ \Eprint {https://arxiv.org/abs/0909.0128} {arXiv:0909.0128 [hep-ph]} \BibitemShut {NoStop}%
\bibitem [{\citenamefont {Baldes}\ and\ \citenamefont {Olea-Romacho}(2024)}]{Baldes:2023rqv}%
  \BibitemOpen
  \bibfield  {author} {\bibinfo {author} {\bibfnamefont {I.}~\bibnamefont {Baldes}}\ and\ \bibinfo {author} {\bibfnamefont {M.~O.}\ \bibnamefont {Olea-Romacho}},\ }\bibfield  {title} {\bibinfo {title} {{Primordial black holes as dark matter: interferometric tests of phase transition origin}},\ }\href {https://doi.org/10.1007/JHEP01(2024)133} {\bibfield  {journal} {\bibinfo  {journal} {JHEP}\ }\textbf {\bibinfo {volume} {01}},\ \bibinfo {pages} {133}},\ \Eprint {https://arxiv.org/abs/2307.11639} {arXiv:2307.11639 [hep-ph]} \BibitemShut {NoStop}%
\bibitem [{\citenamefont {Graesser}(2007{\natexlab{a}})}]{Graesser:2007pc}%
  \BibitemOpen
  \bibfield  {author} {\bibinfo {author} {\bibfnamefont {M.~L.}\ \bibnamefont {Graesser}},\ }\href@noop {} {\bibinfo {title} {{Experimental Constraints on Higgs Boson Decays to TeV-scale Right-Handed Neutrinos}}} (\bibinfo {year} {2007}{\natexlab{a}}),\ \Eprint {https://arxiv.org/abs/0705.2190} {arXiv:0705.2190 [hep-ph]} \BibitemShut {NoStop}%
\bibitem [{\citenamefont {Graesser}(2007{\natexlab{b}})}]{Graesser:2007yj}%
  \BibitemOpen
  \bibfield  {author} {\bibinfo {author} {\bibfnamefont {M.~L.}\ \bibnamefont {Graesser}},\ }\bibfield  {title} {\bibinfo {title} {{Broadening the Higgs boson with right-handed neutrinos and a higher dimension operator at the electroweak scale}},\ }\href {https://doi.org/10.1103/PhysRevD.76.075006} {\bibfield  {journal} {\bibinfo  {journal} {Phys. Rev. D}\ }\textbf {\bibinfo {volume} {76}},\ \bibinfo {pages} {075006} (\bibinfo {year} {2007}{\natexlab{b}})},\ \Eprint {https://arxiv.org/abs/0704.0438} {arXiv:0704.0438 [hep-ph]} \BibitemShut {NoStop}%
\bibitem [{\citenamefont {del Aguila}\ \emph {et~al.}(2009)\citenamefont {del Aguila}, \citenamefont {Bar-Shalom}, \citenamefont {Soni},\ and\ \citenamefont {Wudka}}]{delAguila:2008ir}%
  \BibitemOpen
  \bibfield  {author} {\bibinfo {author} {\bibfnamefont {F.}~\bibnamefont {del Aguila}}, \bibinfo {author} {\bibfnamefont {S.}~\bibnamefont {Bar-Shalom}}, \bibinfo {author} {\bibfnamefont {A.}~\bibnamefont {Soni}},\ and\ \bibinfo {author} {\bibfnamefont {J.}~\bibnamefont {Wudka}},\ }\bibfield  {title} {\bibinfo {title} {{Heavy Majorana Neutrinos in the Effective Lagrangian Description: Application to Hadron Colliders}},\ }\href {https://doi.org/10.1016/j.physletb.2008.11.031} {\bibfield  {journal} {\bibinfo  {journal} {Phys. Lett. B}\ }\textbf {\bibinfo {volume} {670}},\ \bibinfo {pages} {399} (\bibinfo {year} {2009})},\ \Eprint {https://arxiv.org/abs/0806.0876} {arXiv:0806.0876 [hep-ph]} \BibitemShut {NoStop}%
\bibitem [{\citenamefont {Aparici}\ \emph {et~al.}(2009)\citenamefont {Aparici}, \citenamefont {Kim}, \citenamefont {Santamaria},\ and\ \citenamefont {Wudka}}]{Aparici:2009fh}%
  \BibitemOpen
  \bibfield  {author} {\bibinfo {author} {\bibfnamefont {A.}~\bibnamefont {Aparici}}, \bibinfo {author} {\bibfnamefont {K.}~\bibnamefont {Kim}}, \bibinfo {author} {\bibfnamefont {A.}~\bibnamefont {Santamaria}},\ and\ \bibinfo {author} {\bibfnamefont {J.}~\bibnamefont {Wudka}},\ }\bibfield  {title} {\bibinfo {title} {{Right-handed neutrino magnetic moments}},\ }\href {https://doi.org/10.1103/PhysRevD.80.013010} {\bibfield  {journal} {\bibinfo  {journal} {Phys. Rev. D}\ }\textbf {\bibinfo {volume} {80}},\ \bibinfo {pages} {013010} (\bibinfo {year} {2009})},\ \Eprint {https://arxiv.org/abs/0904.3244} {arXiv:0904.3244 [hep-ph]} \BibitemShut {NoStop}%
\bibitem [{\citenamefont {Liao}\ and\ \citenamefont {Ma}(2017)}]{Liao:2016qyd}%
  \BibitemOpen
  \bibfield  {author} {\bibinfo {author} {\bibfnamefont {Y.}~\bibnamefont {Liao}}\ and\ \bibinfo {author} {\bibfnamefont {X.-D.}\ \bibnamefont {Ma}},\ }\bibfield  {title} {\bibinfo {title} {{Operators up to Dimension Seven in Standard Model Effective Field Theory Extended with Sterile Neutrinos}},\ }\href {https://doi.org/10.1103/PhysRevD.96.015012} {\bibfield  {journal} {\bibinfo  {journal} {Phys. Rev. D}\ }\textbf {\bibinfo {volume} {96}},\ \bibinfo {pages} {015012} (\bibinfo {year} {2017})},\ \Eprint {https://arxiv.org/abs/1612.04527} {arXiv:1612.04527 [hep-ph]} \BibitemShut {NoStop}%
\bibitem [{\citenamefont {Deppisch}\ \emph {et~al.}(2014)\citenamefont {Deppisch}, \citenamefont {Desai},\ and\ \citenamefont {Valle}}]{Deppisch:2013cya}%
  \BibitemOpen
  \bibfield  {author} {\bibinfo {author} {\bibfnamefont {F.~F.}\ \bibnamefont {Deppisch}}, \bibinfo {author} {\bibfnamefont {N.}~\bibnamefont {Desai}},\ and\ \bibinfo {author} {\bibfnamefont {J.~W.~F.}\ \bibnamefont {Valle}},\ }\bibfield  {title} {\bibinfo {title} {{Is charged lepton flavor violation a high energy phenomenon?}},\ }\href {https://doi.org/10.1103/PhysRevD.89.051302} {\bibfield  {journal} {\bibinfo  {journal} {Phys. Rev. D}\ }\textbf {\bibinfo {volume} {89}},\ \bibinfo {pages} {051302} (\bibinfo {year} {2014})},\ \Eprint {https://arxiv.org/abs/1308.6789} {arXiv:1308.6789 [hep-ph]} \BibitemShut {NoStop}%
\bibitem [{\citenamefont {Batell}\ \emph {et~al.}(2016)\citenamefont {Batell}, \citenamefont {Pospelov},\ and\ \citenamefont {Shuve}}]{Batell:2016zod}%
  \BibitemOpen
  \bibfield  {author} {\bibinfo {author} {\bibfnamefont {B.}~\bibnamefont {Batell}}, \bibinfo {author} {\bibfnamefont {M.}~\bibnamefont {Pospelov}},\ and\ \bibinfo {author} {\bibfnamefont {B.}~\bibnamefont {Shuve}},\ }\bibfield  {title} {\bibinfo {title} {{Shedding Light on Neutrino Masses with Dark Forces}},\ }\href {https://doi.org/10.1007/JHEP08(2016)052} {\bibfield  {journal} {\bibinfo  {journal} {JHEP}\ }\textbf {\bibinfo {volume} {08}},\ \bibinfo {pages} {052}},\ \Eprint {https://arxiv.org/abs/1604.06099} {arXiv:1604.06099 [hep-ph]} \BibitemShut {NoStop}%
\bibitem [{\citenamefont {Deppisch}\ \emph {et~al.}(2019)\citenamefont {Deppisch}, \citenamefont {Kulkarni},\ and\ \citenamefont {Liu}}]{Deppisch:2019kvs}%
  \BibitemOpen
  \bibfield  {author} {\bibinfo {author} {\bibfnamefont {F.}~\bibnamefont {Deppisch}}, \bibinfo {author} {\bibfnamefont {S.}~\bibnamefont {Kulkarni}},\ and\ \bibinfo {author} {\bibfnamefont {W.}~\bibnamefont {Liu}},\ }\bibfield  {title} {\bibinfo {title} {{Heavy neutrino production via $Z^\prime$ at the lifetime frontier}},\ }\href {https://doi.org/10.1103/PhysRevD.100.035005} {\bibfield  {journal} {\bibinfo  {journal} {Phys. Rev. D}\ }\textbf {\bibinfo {volume} {100}},\ \bibinfo {pages} {035005} (\bibinfo {year} {2019})},\ \Eprint {https://arxiv.org/abs/1905.11889} {arXiv:1905.11889 [hep-ph]} \BibitemShut {NoStop}%
\bibitem [{\citenamefont {Accomando}\ \emph {et~al.}(2018)\citenamefont {Accomando}, \citenamefont {Delle~Rose}, \citenamefont {Moretti}, \citenamefont {Olaiya},\ and\ \citenamefont {Shepherd-Themistocleous}}]{Accomando:2017qcs}%
  \BibitemOpen
  \bibfield  {author} {\bibinfo {author} {\bibfnamefont {E.}~\bibnamefont {Accomando}}, \bibinfo {author} {\bibfnamefont {L.}~\bibnamefont {Delle~Rose}}, \bibinfo {author} {\bibfnamefont {S.}~\bibnamefont {Moretti}}, \bibinfo {author} {\bibfnamefont {E.}~\bibnamefont {Olaiya}},\ and\ \bibinfo {author} {\bibfnamefont {C.~H.}\ \bibnamefont {Shepherd-Themistocleous}},\ }\bibfield  {title} {\bibinfo {title} {{Extra Higgs boson and Z$^{\prime}$ as portals to signatures of heavy neutrinos at the LHC}},\ }\href {https://doi.org/10.1007/JHEP02(2018)109} {\bibfield  {journal} {\bibinfo  {journal} {JHEP}\ }\textbf {\bibinfo {volume} {02}},\ \bibinfo {pages} {109}},\ \Eprint {https://arxiv.org/abs/1708.03650} {arXiv:1708.03650 [hep-ph]} \BibitemShut {NoStop}%
\bibitem [{\citenamefont {Das}\ \emph {et~al.}(2019{\natexlab{a}})\citenamefont {Das}, \citenamefont {Dev},\ and\ \citenamefont {Okada}}]{Das:2019fee}%
  \BibitemOpen
  \bibfield  {author} {\bibinfo {author} {\bibfnamefont {A.}~\bibnamefont {Das}}, \bibinfo {author} {\bibfnamefont {P.~S.~B.}\ \bibnamefont {Dev}},\ and\ \bibinfo {author} {\bibfnamefont {N.}~\bibnamefont {Okada}},\ }\bibfield  {title} {\bibinfo {title} {{Long-lived TeV-scale right-handed neutrino production at the LHC in gauged $U(1)_X$ model}},\ }\href {https://doi.org/10.1016/j.physletb.2019.135052} {\bibfield  {journal} {\bibinfo  {journal} {Phys. Lett. B}\ }\textbf {\bibinfo {volume} {799}},\ \bibinfo {pages} {135052} (\bibinfo {year} {2019}{\natexlab{a}})},\ \Eprint {https://arxiv.org/abs/1906.04132} {arXiv:1906.04132 [hep-ph]} \BibitemShut {NoStop}%
\bibitem [{\citenamefont {Cheung}\ \emph {et~al.}(2021)\citenamefont {Cheung}, \citenamefont {Wang},\ and\ \citenamefont {Wang}}]{Cheung:2021utb}%
  \BibitemOpen
  \bibfield  {author} {\bibinfo {author} {\bibfnamefont {K.}~\bibnamefont {Cheung}}, \bibinfo {author} {\bibfnamefont {K.}~\bibnamefont {Wang}},\ and\ \bibinfo {author} {\bibfnamefont {Z.~S.}\ \bibnamefont {Wang}},\ }\bibfield  {title} {\bibinfo {title} {{Time-delayed electrons from neutral currents at the LHC}},\ }\href {https://doi.org/10.1007/JHEP09(2021)026} {\bibfield  {journal} {\bibinfo  {journal} {JHEP}\ }\textbf {\bibinfo {volume} {09}},\ \bibinfo {pages} {026}},\ \Eprint {https://arxiv.org/abs/2107.03203} {arXiv:2107.03203 [hep-ph]} \BibitemShut {NoStop}%
\bibitem [{\citenamefont {Chiang}\ \emph {et~al.}(2019)\citenamefont {Chiang}, \citenamefont {Cottin}, \citenamefont {Das},\ and\ \citenamefont {Mandal}}]{Chiang:2019ajm}%
  \BibitemOpen
  \bibfield  {author} {\bibinfo {author} {\bibfnamefont {C.-W.}\ \bibnamefont {Chiang}}, \bibinfo {author} {\bibfnamefont {G.}~\bibnamefont {Cottin}}, \bibinfo {author} {\bibfnamefont {A.}~\bibnamefont {Das}},\ and\ \bibinfo {author} {\bibfnamefont {S.}~\bibnamefont {Mandal}},\ }\bibfield  {title} {\bibinfo {title} {{Displaced heavy neutrinos from $Z^\prime$ decays at the LHC}},\ }\href {https://doi.org/10.1007/JHEP12(2019)070} {\bibfield  {journal} {\bibinfo  {journal} {JHEP}\ }\textbf {\bibinfo {volume} {12}},\ \bibinfo {pages} {070}},\ \Eprint {https://arxiv.org/abs/1908.09838} {arXiv:1908.09838 [hep-ph]} \BibitemShut {NoStop}%
\bibitem [{\citenamefont {Fileviez~P\'erez}\ and\ \citenamefont {Plascencia}(2020)}]{FileviezPerez:2020cgn}%
  \BibitemOpen
  \bibfield  {author} {\bibinfo {author} {\bibfnamefont {P.}~\bibnamefont {Fileviez~P\'erez}}\ and\ \bibinfo {author} {\bibfnamefont {A.~D.}\ \bibnamefont {Plascencia}},\ }\bibfield  {title} {\bibinfo {title} {{Probing the Nature of Neutrinos with a New Force}},\ }\href {https://doi.org/10.1103/PhysRevD.102.015010} {\bibfield  {journal} {\bibinfo  {journal} {Phys. Rev. D}\ }\textbf {\bibinfo {volume} {102}},\ \bibinfo {pages} {015010} (\bibinfo {year} {2020})},\ \Eprint {https://arxiv.org/abs/2005.04235} {arXiv:2005.04235 [hep-ph]} \BibitemShut {NoStop}%
\bibitem [{\citenamefont {Das}\ \emph {et~al.}(2019{\natexlab{b}})\citenamefont {Das}, \citenamefont {Okada}, \citenamefont {Okada},\ and\ \citenamefont {Raut}}]{Das:2018tbd}%
  \BibitemOpen
  \bibfield  {author} {\bibinfo {author} {\bibfnamefont {A.}~\bibnamefont {Das}}, \bibinfo {author} {\bibfnamefont {N.}~\bibnamefont {Okada}}, \bibinfo {author} {\bibfnamefont {S.}~\bibnamefont {Okada}},\ and\ \bibinfo {author} {\bibfnamefont {D.}~\bibnamefont {Raut}},\ }\bibfield  {title} {\bibinfo {title} {{Probing the seesaw mechanism at the 250 GeV ILC}},\ }\href {https://doi.org/10.1016/j.physletb.2019.134849} {\bibfield  {journal} {\bibinfo  {journal} {Phys. Lett. B}\ }\textbf {\bibinfo {volume} {797}},\ \bibinfo {pages} {134849} (\bibinfo {year} {2019}{\natexlab{b}})},\ \Eprint {https://arxiv.org/abs/1812.11931} {arXiv:1812.11931 [hep-ph]} \BibitemShut {NoStop}%
\bibitem [{\citenamefont {Liu}\ \emph {et~al.}(2022{\natexlab{b}})\citenamefont {Liu}, \citenamefont {Kulkarni},\ and\ \citenamefont {Deppisch}}]{Liu:2022kid}%
  \BibitemOpen
  \bibfield  {author} {\bibinfo {author} {\bibfnamefont {W.}~\bibnamefont {Liu}}, \bibinfo {author} {\bibfnamefont {S.}~\bibnamefont {Kulkarni}},\ and\ \bibinfo {author} {\bibfnamefont {F.~F.}\ \bibnamefont {Deppisch}},\ }\bibfield  {title} {\bibinfo {title} {{Heavy neutrinos at the FCC-hh in the U(1)B-L model}},\ }\href {https://doi.org/10.1103/PhysRevD.105.095043} {\bibfield  {journal} {\bibinfo  {journal} {Phys. Rev. D}\ }\textbf {\bibinfo {volume} {105}},\ \bibinfo {pages} {095043} (\bibinfo {year} {2022}{\natexlab{b}})},\ \Eprint {https://arxiv.org/abs/2202.07310} {arXiv:2202.07310 [hep-ph]} \BibitemShut {NoStop}%
\bibitem [{\citenamefont {Han}\ \emph {et~al.}(2021)\citenamefont {Han}, \citenamefont {Li},\ and\ \citenamefont {Yao}}]{Han:2021pun}%
  \BibitemOpen
  \bibfield  {author} {\bibinfo {author} {\bibfnamefont {C.}~\bibnamefont {Han}}, \bibinfo {author} {\bibfnamefont {T.}~\bibnamefont {Li}},\ and\ \bibinfo {author} {\bibfnamefont {C.-Y.}\ \bibnamefont {Yao}},\ }\bibfield  {title} {\bibinfo {title} {{Searching for heavy neutrino in terms of tau lepton at future hadron collider}},\ }\href {https://doi.org/10.1103/PhysRevD.104.015036} {\bibfield  {journal} {\bibinfo  {journal} {Phys. Rev. D}\ }\textbf {\bibinfo {volume} {104}},\ \bibinfo {pages} {015036} (\bibinfo {year} {2021})},\ \Eprint {https://arxiv.org/abs/2103.03548} {arXiv:2103.03548 [hep-ph]} \BibitemShut {NoStop}%
\bibitem [{\citenamefont {Das}\ and\ \citenamefont {Okada}(2017)}]{Das:2017nvm}%
  \BibitemOpen
  \bibfield  {author} {\bibinfo {author} {\bibfnamefont {A.}~\bibnamefont {Das}}\ and\ \bibinfo {author} {\bibfnamefont {N.}~\bibnamefont {Okada}},\ }\bibfield  {title} {\bibinfo {title} {{Bounds on heavy Majorana neutrinos in type-I seesaw and implications for collider searches}},\ }\href {https://doi.org/10.1016/j.physletb.2017.09.042} {\bibfield  {journal} {\bibinfo  {journal} {Phys. Lett. B}\ }\textbf {\bibinfo {volume} {774}},\ \bibinfo {pages} {32} (\bibinfo {year} {2017})},\ \Eprint {https://arxiv.org/abs/1702.04668} {arXiv:1702.04668 [hep-ph]} \BibitemShut {NoStop}%
\bibitem [{\citenamefont {Maiezza}\ \emph {et~al.}(2015)\citenamefont {Maiezza}, \citenamefont {Nemev\v{s}ek},\ and\ \citenamefont {Nesti}}]{Maiezza:2015lza}%
  \BibitemOpen
  \bibfield  {author} {\bibinfo {author} {\bibfnamefont {A.}~\bibnamefont {Maiezza}}, \bibinfo {author} {\bibfnamefont {M.}~\bibnamefont {Nemev\v{s}ek}},\ and\ \bibinfo {author} {\bibfnamefont {F.}~\bibnamefont {Nesti}},\ }\bibfield  {title} {\bibinfo {title} {{Lepton Number Violation in Higgs Decay at LHC}},\ }\href {https://doi.org/10.1103/PhysRevLett.115.081802} {\bibfield  {journal} {\bibinfo  {journal} {Phys. Rev. Lett.}\ }\textbf {\bibinfo {volume} {115}},\ \bibinfo {pages} {081802} (\bibinfo {year} {2015})},\ \Eprint {https://arxiv.org/abs/1503.06834} {arXiv:1503.06834 [hep-ph]} \BibitemShut {NoStop}%
\bibitem [{\citenamefont {Deppisch}\ \emph {et~al.}(2018)\citenamefont {Deppisch}, \citenamefont {Liu},\ and\ \citenamefont {Mitra}}]{Deppisch:2018eth}%
  \BibitemOpen
  \bibfield  {author} {\bibinfo {author} {\bibfnamefont {F.~F.}\ \bibnamefont {Deppisch}}, \bibinfo {author} {\bibfnamefont {W.}~\bibnamefont {Liu}},\ and\ \bibinfo {author} {\bibfnamefont {M.}~\bibnamefont {Mitra}},\ }\bibfield  {title} {\bibinfo {title} {{Long-lived Heavy Neutrinos from Higgs Decays}},\ }\href {https://doi.org/10.1007/JHEP08(2018)181} {\bibfield  {journal} {\bibinfo  {journal} {JHEP}\ }\textbf {\bibinfo {volume} {08}},\ \bibinfo {pages} {181}},\ \Eprint {https://arxiv.org/abs/1804.04075} {arXiv:1804.04075 [hep-ph]} \BibitemShut {NoStop}%
\bibitem [{\citenamefont {Mason}(2019)}]{Mason:2019okp}%
  \BibitemOpen
  \bibfield  {author} {\bibinfo {author} {\bibfnamefont {J.~D.}\ \bibnamefont {Mason}},\ }\bibfield  {title} {\bibinfo {title} {{Time-Delayed Electrons from Higgs Decays to Right-Handed Neutrinos}},\ }\href {https://doi.org/10.1007/JHEP07(2019)089} {\bibfield  {journal} {\bibinfo  {journal} {JHEP}\ }\textbf {\bibinfo {volume} {07}},\ \bibinfo {pages} {089}},\ \Eprint {https://arxiv.org/abs/1905.07772} {arXiv:1905.07772 [hep-ph]} \BibitemShut {NoStop}%
\bibitem [{\citenamefont {Accomando}\ \emph {et~al.}(2017)\citenamefont {Accomando}, \citenamefont {Delle~Rose}, \citenamefont {Moretti}, \citenamefont {Olaiya},\ and\ \citenamefont {Shepherd-Themistocleous}}]{Accomando:2016rpc}%
  \BibitemOpen
  \bibfield  {author} {\bibinfo {author} {\bibfnamefont {E.}~\bibnamefont {Accomando}}, \bibinfo {author} {\bibfnamefont {L.}~\bibnamefont {Delle~Rose}}, \bibinfo {author} {\bibfnamefont {S.}~\bibnamefont {Moretti}}, \bibinfo {author} {\bibfnamefont {E.}~\bibnamefont {Olaiya}},\ and\ \bibinfo {author} {\bibfnamefont {C.~H.}\ \bibnamefont {Shepherd-Themistocleous}},\ }\bibfield  {title} {\bibinfo {title} {{Novel SM-like Higgs decay into displaced heavy neutrino pairs in U(1)' models}},\ }\href {https://doi.org/10.1007/JHEP04(2017)081} {\bibfield  {journal} {\bibinfo  {journal} {JHEP}\ }\textbf {\bibinfo {volume} {04}},\ \bibinfo {pages} {081}},\ \Eprint {https://arxiv.org/abs/1612.05977} {arXiv:1612.05977 [hep-ph]} \BibitemShut {NoStop}%
\bibitem [{\citenamefont {Gao}\ \emph {et~al.}(2020)\citenamefont {Gao}, \citenamefont {Jin},\ and\ \citenamefont {Wang}}]{Gao:2019tio}%
  \BibitemOpen
  \bibfield  {author} {\bibinfo {author} {\bibfnamefont {Y.}~\bibnamefont {Gao}}, \bibinfo {author} {\bibfnamefont {M.}~\bibnamefont {Jin}},\ and\ \bibinfo {author} {\bibfnamefont {K.}~\bibnamefont {Wang}},\ }\bibfield  {title} {\bibinfo {title} {{Probing the Decoupled Seesaw Scalar in Rare Higgs Decay}},\ }\href {https://doi.org/10.1007/JHEP02(2020)101} {\bibfield  {journal} {\bibinfo  {journal} {JHEP}\ }\textbf {\bibinfo {volume} {02}},\ \bibinfo {pages} {101}},\ \Eprint {https://arxiv.org/abs/1904.12325} {arXiv:1904.12325 [hep-ph]} \BibitemShut {NoStop}%
\bibitem [{\citenamefont {Gago}\ \emph {et~al.}(2015)\citenamefont {Gago}, \citenamefont {Hern\'andez}, \citenamefont {Jones-P\'erez}, \citenamefont {Losada},\ and\ \citenamefont {Moreno Brice\~no}}]{Gago:2015vma}%
  \BibitemOpen
  \bibfield  {author} {\bibinfo {author} {\bibfnamefont {A.~M.}\ \bibnamefont {Gago}}, \bibinfo {author} {\bibfnamefont {P.}~\bibnamefont {Hern\'andez}}, \bibinfo {author} {\bibfnamefont {J.}~\bibnamefont {Jones-P\'erez}}, \bibinfo {author} {\bibfnamefont {M.}~\bibnamefont {Losada}},\ and\ \bibinfo {author} {\bibfnamefont {A.}~\bibnamefont {Moreno Brice\~no}},\ }\bibfield  {title} {\bibinfo {title} {{Probing the Type I Seesaw Mechanism with Displaced Vertices at the LHC}},\ }\href {https://doi.org/10.1140/epjc/s10052-015-3693-1} {\bibfield  {journal} {\bibinfo  {journal} {Eur. Phys. J. C}\ }\textbf {\bibinfo {volume} {75}},\ \bibinfo {pages} {470} (\bibinfo {year} {2015})},\ \Eprint {https://arxiv.org/abs/1505.05880} {arXiv:1505.05880 [hep-ph]} \BibitemShut {NoStop}%
\bibitem [{\citenamefont {Jones-P\'erez}\ \emph {et~al.}(2020)\citenamefont {Jones-P\'erez}, \citenamefont {Masias},\ and\ \citenamefont {Ruiz-\'Alvarez}}]{Jones-Perez:2019plk}%
  \BibitemOpen
  \bibfield  {author} {\bibinfo {author} {\bibfnamefont {J.}~\bibnamefont {Jones-P\'erez}}, \bibinfo {author} {\bibfnamefont {J.}~\bibnamefont {Masias}},\ and\ \bibinfo {author} {\bibfnamefont {J.~D.}\ \bibnamefont {Ruiz-\'Alvarez}},\ }\bibfield  {title} {\bibinfo {title} {{Search for Long-Lived Heavy Neutrinos at the LHC with a VBF Trigger}},\ }\href {https://doi.org/10.1140/epjc/s10052-020-8188-z} {\bibfield  {journal} {\bibinfo  {journal} {Eur. Phys. J. C}\ }\textbf {\bibinfo {volume} {80}},\ \bibinfo {pages} {642} (\bibinfo {year} {2020})},\ \Eprint {https://arxiv.org/abs/1912.08206} {arXiv:1912.08206 [hep-ph]} \BibitemShut {NoStop}%
\bibitem [{\citenamefont {Liu}\ \emph {et~al.}(2022{\natexlab{c}})\citenamefont {Liu}, \citenamefont {Li}, \citenamefont {Li},\ and\ \citenamefont {Sun}}]{Liu:2022ugx}%
  \BibitemOpen
  \bibfield  {author} {\bibinfo {author} {\bibfnamefont {W.}~\bibnamefont {Liu}}, \bibinfo {author} {\bibfnamefont {J.}~\bibnamefont {Li}}, \bibinfo {author} {\bibfnamefont {J.}~\bibnamefont {Li}},\ and\ \bibinfo {author} {\bibfnamefont {H.}~\bibnamefont {Sun}},\ }\bibfield  {title} {\bibinfo {title} {{Testing the seesaw mechanisms via displaced right-handed neutrinos from a light scalar at the HL-LHC}},\ }\href {https://doi.org/10.1103/PhysRevD.106.015019} {\bibfield  {journal} {\bibinfo  {journal} {Phys. Rev. D}\ }\textbf {\bibinfo {volume} {106}},\ \bibinfo {pages} {015019} (\bibinfo {year} {2022}{\natexlab{c}})},\ \Eprint {https://arxiv.org/abs/2204.03819} {arXiv:2204.03819 [hep-ph]} \BibitemShut {NoStop}%
\bibitem [{\citenamefont {Li}\ \emph {et~al.}(2024)\citenamefont {Li}, \citenamefont {Liu},\ and\ \citenamefont {Sun}}]{Li:2023dbs}%
  \BibitemOpen
  \bibfield  {author} {\bibinfo {author} {\bibfnamefont {J.}~\bibnamefont {Li}}, \bibinfo {author} {\bibfnamefont {W.}~\bibnamefont {Liu}},\ and\ \bibinfo {author} {\bibfnamefont {H.}~\bibnamefont {Sun}},\ }\bibfield  {title} {\bibinfo {title} {{Z' mediated right-handed neutrinos from meson decays at the FASER}},\ }\href {https://doi.org/10.1103/PhysRevD.109.035022} {\bibfield  {journal} {\bibinfo  {journal} {Phys. Rev. D}\ }\textbf {\bibinfo {volume} {109}},\ \bibinfo {pages} {035022} (\bibinfo {year} {2024})},\ \Eprint {https://arxiv.org/abs/2309.05020} {arXiv:2309.05020 [hep-ph]} \BibitemShut {NoStop}%
\bibitem [{\citenamefont {Deppisch}\ \emph {et~al.}(2023)\citenamefont {Deppisch}, \citenamefont {Kulkarni},\ and\ \citenamefont {Liu}}]{Deppisch:2023sga}%
  \BibitemOpen
  \bibfield  {author} {\bibinfo {author} {\bibfnamefont {F.~F.}\ \bibnamefont {Deppisch}}, \bibinfo {author} {\bibfnamefont {S.}~\bibnamefont {Kulkarni}},\ and\ \bibinfo {author} {\bibfnamefont {W.}~\bibnamefont {Liu}},\ }\bibfield  {title} {\bibinfo {title} {{Sterile Neutrinos at MAPP in the B-L Model}},\ }\href@noop {} {\  (\bibinfo {year} {2023})},\ \Eprint {https://arxiv.org/abs/2311.01719} {arXiv:2311.01719 [hep-ph]} \BibitemShut {NoStop}%
\bibitem [{\citenamefont {Bernal}\ \emph {et~al.}(2023)\citenamefont {Bernal}, \citenamefont {Deka},\ and\ \citenamefont {Losada}}]{Bernal:2023coo}%
  \BibitemOpen
  \bibfield  {author} {\bibinfo {author} {\bibfnamefont {N.}~\bibnamefont {Bernal}}, \bibinfo {author} {\bibfnamefont {K.}~\bibnamefont {Deka}},\ and\ \bibinfo {author} {\bibfnamefont {M.}~\bibnamefont {Losada}},\ }\bibfield  {title} {\bibinfo {title} {{Discovering Heavy Neutral Leptons with the Higgs Boson}},\ }\href@noop {} {\  (\bibinfo {year} {2023})},\ \Eprint {https://arxiv.org/abs/2311.18033} {arXiv:2311.18033 [hep-ph]} \BibitemShut {NoStop}%
\bibitem [{\citenamefont {Atre}\ \emph {et~al.}(2009)\citenamefont {Atre}, \citenamefont {Han}, \citenamefont {Pascoli},\ and\ \citenamefont {Zhang}}]{Atre:2009rg}%
  \BibitemOpen
  \bibfield  {author} {\bibinfo {author} {\bibfnamefont {A.}~\bibnamefont {Atre}}, \bibinfo {author} {\bibfnamefont {T.}~\bibnamefont {Han}}, \bibinfo {author} {\bibfnamefont {S.}~\bibnamefont {Pascoli}},\ and\ \bibinfo {author} {\bibfnamefont {B.}~\bibnamefont {Zhang}},\ }\bibfield  {title} {\bibinfo {title} {{The Search for Heavy Majorana Neutrinos}},\ }\href {https://doi.org/10.1088/1126-6708/2009/05/030} {\bibfield  {journal} {\bibinfo  {journal} {JHEP}\ }\textbf {\bibinfo {volume} {05}},\ \bibinfo {pages} {030}},\ \Eprint {https://arxiv.org/abs/0901.3589} {arXiv:0901.3589 [hep-ph]} \BibitemShut {NoStop}%
\bibitem [{\citenamefont {Tumasyan}\ \emph {et~al.}(2021)\citenamefont {Tumasyan} \emph {et~al.}}]{CMS:2021juv}%
  \BibitemOpen
  \bibfield  {author} {\bibinfo {author} {\bibfnamefont {A.}~\bibnamefont {Tumasyan}} \emph {et~al.} (\bibinfo {collaboration} {CMS}),\ }\bibfield  {title} {\bibinfo {title} {{Search for Long-Lived Particles Decaying in the CMS End Cap Muon Detectors in Proton-Proton Collisions at $\sqrt s$ =13\,\,TeV}},\ }\href {https://doi.org/10.1103/PhysRevLett.127.261804} {\bibfield  {journal} {\bibinfo  {journal} {Phys. Rev. Lett.}\ }\textbf {\bibinfo {volume} {127}},\ \bibinfo {pages} {261804} (\bibinfo {year} {2021})},\ \Eprint {https://arxiv.org/abs/2107.04838} {arXiv:2107.04838 [hep-ex]} \BibitemShut {NoStop}%
\bibitem [{\citenamefont {Aaboud}\ \emph {et~al.}(2019)\citenamefont {Aaboud} \emph {et~al.}}]{ATLAS:2018tup}%
  \BibitemOpen
  \bibfield  {author} {\bibinfo {author} {\bibfnamefont {M.}~\bibnamefont {Aaboud}} \emph {et~al.} (\bibinfo {collaboration} {ATLAS}),\ }\bibfield  {title} {\bibinfo {title} {{Search for long-lived particles produced in $pp$ collisions at $\sqrt{s}=13$ TeV that decay into displaced hadronic jets in the ATLAS muon spectrometer}},\ }\href {https://doi.org/10.1103/PhysRevD.99.052005} {\bibfield  {journal} {\bibinfo  {journal} {Phys. Rev. D}\ }\textbf {\bibinfo {volume} {99}},\ \bibinfo {pages} {052005} (\bibinfo {year} {2019})},\ \Eprint {https://arxiv.org/abs/1811.07370} {arXiv:1811.07370 [hep-ex]} \BibitemShut {NoStop}%
\bibitem [{\citenamefont {Aad}\ \emph {et~al.}(2020)\citenamefont {Aad} \emph {et~al.}}]{ATLAS:2019jcm}%
  \BibitemOpen
  \bibfield  {author} {\bibinfo {author} {\bibfnamefont {G.}~\bibnamefont {Aad}} \emph {et~al.} (\bibinfo {collaboration} {ATLAS}),\ }\bibfield  {title} {\bibinfo {title} {{Search for long-lived neutral particles produced in $pp$ collisions at $\sqrt{s} = 13$ TeV decaying into displaced hadronic jets in the ATLAS inner detector and muon spectrometer}},\ }\href {https://doi.org/10.1103/PhysRevD.101.052013} {\bibfield  {journal} {\bibinfo  {journal} {Phys. Rev. D}\ }\textbf {\bibinfo {volume} {101}},\ \bibinfo {pages} {052013} (\bibinfo {year} {2020})},\ \Eprint {https://arxiv.org/abs/1911.12575} {arXiv:1911.12575 [hep-ex]} \BibitemShut {NoStop}%
\bibitem [{\citenamefont {Aad}\ \emph {et~al.}(2023{\natexlab{b}})\citenamefont {Aad} \emph {et~al.}}]{ATLAS:2023oti}%
  \BibitemOpen
  \bibfield  {author} {\bibinfo {author} {\bibfnamefont {G.}~\bibnamefont {Aad}} \emph {et~al.} (\bibinfo {collaboration} {ATLAS}),\ }\bibfield  {title} {\bibinfo {title} {{Search for long-lived, massive particles in events with displaced vertices and multiple jets in pp collisions at $ \sqrt{s} $ = 13 TeV with the ATLAS detector}},\ }\href {https://doi.org/10.1007/JHEP06(2023)200} {\bibfield  {journal} {\bibinfo  {journal} {JHEP}\ }\textbf {\bibinfo {volume} {2306}},\ \bibinfo {pages} {200}},\ \Eprint {https://arxiv.org/abs/2301.13866} {arXiv:2301.13866 [hep-ex]} \BibitemShut {NoStop}%
\bibitem [{\citenamefont {Cottin}\ \emph {et~al.}(2023)\citenamefont {Cottin}, \citenamefont {Helo}, \citenamefont {Hirsch}, \citenamefont {Pe\~na}, \citenamefont {Wang},\ and\ \citenamefont {Xie}}]{Cottin:2022nwp}%
  \BibitemOpen
  \bibfield  {author} {\bibinfo {author} {\bibfnamefont {G.}~\bibnamefont {Cottin}}, \bibinfo {author} {\bibfnamefont {J.~C.}\ \bibnamefont {Helo}}, \bibinfo {author} {\bibfnamefont {M.}~\bibnamefont {Hirsch}}, \bibinfo {author} {\bibfnamefont {C.}~\bibnamefont {Pe\~na}}, \bibinfo {author} {\bibfnamefont {C.}~\bibnamefont {Wang}},\ and\ \bibinfo {author} {\bibfnamefont {S.}~\bibnamefont {Xie}},\ }\bibfield  {title} {\bibinfo {title} {{Long-lived heavy neutral leptons with a displaced shower signature at CMS}},\ }\href {https://doi.org/10.1007/JHEP02(2023)011} {\bibfield  {journal} {\bibinfo  {journal} {JHEP}\ }\textbf {\bibinfo {volume} {02}},\ \bibinfo {pages} {011}},\ \Eprint {https://arxiv.org/abs/2210.17446} {arXiv:2210.17446 [hep-ph]} \BibitemShut {NoStop}%
\bibitem [{\citenamefont {Aad}\ \emph {et~al.}(2019{\natexlab{b}})\citenamefont {Aad} \emph {et~al.}}]{ATLAS:2019erb}%
  \BibitemOpen
  \bibfield  {author} {\bibinfo {author} {\bibfnamefont {G.}~\bibnamefont {Aad}} \emph {et~al.} (\bibinfo {collaboration} {ATLAS}),\ }\bibfield  {title} {\bibinfo {title} {{Search for high-mass dilepton resonances using 139 fb$^{-1}$ of $pp$ collision data collected at $\sqrt{s}=$13 TeV with the ATLAS detector}},\ }\href {https://doi.org/10.1016/j.physletb.2019.07.016} {\bibfield  {journal} {\bibinfo  {journal} {Phys. Lett. B}\ }\textbf {\bibinfo {volume} {796}},\ \bibinfo {pages} {68} (\bibinfo {year} {2019}{\natexlab{b}})},\ \Eprint {https://arxiv.org/abs/1903.06248} {arXiv:1903.06248 [hep-ex]} \BibitemShut {NoStop}%
\bibitem [{\citenamefont {Aaboud}\ \emph {et~al.}(2017)\citenamefont {Aaboud} \emph {et~al.}}]{ATLAS:2016oxs}%
  \BibitemOpen
  \bibfield  {author} {\bibinfo {author} {\bibfnamefont {M.}~\bibnamefont {Aaboud}} \emph {et~al.} (\bibinfo {collaboration} {ATLAS}),\ }\bibfield  {title} {\bibinfo {title} {{Measurements of top-quark pair to $Z$-boson cross-section ratios at $\sqrt s = 13, 8, 7$ TeV with the ATLAS detector}},\ }\href {https://doi.org/10.1007/JHEP02(2017)117} {\bibfield  {journal} {\bibinfo  {journal} {JHEP}\ }\textbf {\bibinfo {volume} {02}},\ \bibinfo {pages} {117}},\ \Eprint {https://arxiv.org/abs/1612.03636} {arXiv:1612.03636 [hep-ex]} \BibitemShut {NoStop}%
\bibitem [{\citenamefont {Dev}\ \emph {et~al.}(2021)\citenamefont {Dev}, \citenamefont {Rodejohann}, \citenamefont {Xu},\ and\ \citenamefont {Zhang}}]{Dev:2021otb}%
  \BibitemOpen
  \bibfield  {author} {\bibinfo {author} {\bibfnamefont {P.~S.~B.}\ \bibnamefont {Dev}}, \bibinfo {author} {\bibfnamefont {W.}~\bibnamefont {Rodejohann}}, \bibinfo {author} {\bibfnamefont {X.-J.}\ \bibnamefont {Xu}},\ and\ \bibinfo {author} {\bibfnamefont {Y.}~\bibnamefont {Zhang}},\ }\bibfield  {title} {\bibinfo {title} {{Searching for $Z^\prime$ bosons at the P2 experiment}},\ }\href {https://doi.org/10.1007/JHEP06(2021)039} {\bibfield  {journal} {\bibinfo  {journal} {JHEP}\ }\textbf {\bibinfo {volume} {06}},\ \bibinfo {pages} {039}},\ \Eprint {https://arxiv.org/abs/2103.09067} {arXiv:2103.09067 [hep-ph]} \BibitemShut {NoStop}%
\bibitem [{\citenamefont {CERN}()}]{CERNYellowReportPageAt13TeV}%
  \BibitemOpen
  \bibfield  {author} {\bibinfo {author} {\bibnamefont {CERN}},\ }\href@noop {} {\bibinfo {title} {Cern yellow reports 13 tev}},\ \bibinfo {howpublished} {\url{https://twiki.cern.ch/twiki/bin/view/LHCPhysics/CERNYellowReportPageAt13TeV}}\BibitemShut {NoStop}%
\bibitem [{LHC()}]{LHCHiggsWG}%
  \BibitemOpen
  \href@noop {} {\bibinfo {title} {Lhc higgs cross section working group}},\ \bibinfo {howpublished} {\url{https://twiki.cern.ch/twiki/bin/view/LHCPhysics/HiggsEuropeanStrategy}}\BibitemShut {NoStop}%
\bibitem [{\citenamefont {de~Florian}\ \emph {et~al.}(2016)\citenamefont {de~Florian} \emph {et~al.}}]{LHCHiggsCrossSectionWorkingGroup:2016ypw}%
  \BibitemOpen
  \bibfield  {author} {\bibinfo {author} {\bibfnamefont {D.}~\bibnamefont {de~Florian}} \emph {et~al.} (\bibinfo {collaboration} {LHC Higgs Cross Section Working Group}),\ }\bibfield  {title} {\bibinfo {title} {{Handbook of LHC Higgs Cross Sections: 4. Deciphering the Nature of the Higgs Sector}}\ }\textbf {\bibinfo {volume} {2/2017}},\ \href {https://doi.org/10.23731/CYRM-2017-002} {10.23731/CYRM-2017-002} (\bibinfo {year} {2016}),\ \Eprint {https://arxiv.org/abs/1610.07922} {arXiv:1610.07922 [hep-ph]} \BibitemShut {NoStop}%
\bibitem [{\citenamefont {Cacciapaglia}\ \emph {et~al.}(2006)\citenamefont {Cacciapaglia}, \citenamefont {Csaki}, \citenamefont {Marandella},\ and\ \citenamefont {Strumia}}]{Cacciapaglia:2006pk}%
  \BibitemOpen
  \bibfield  {author} {\bibinfo {author} {\bibfnamefont {G.}~\bibnamefont {Cacciapaglia}}, \bibinfo {author} {\bibfnamefont {C.}~\bibnamefont {Csaki}}, \bibinfo {author} {\bibfnamefont {G.}~\bibnamefont {Marandella}},\ and\ \bibinfo {author} {\bibfnamefont {A.}~\bibnamefont {Strumia}},\ }\bibfield  {title} {\bibinfo {title} {{The Minimal Set of Electroweak Precision Parameters}},\ }\href {https://doi.org/10.1103/PhysRevD.74.033011} {\bibfield  {journal} {\bibinfo  {journal} {Phys. Rev. D}\ }\textbf {\bibinfo {volume} {74}},\ \bibinfo {pages} {033011} (\bibinfo {year} {2006})},\ \Eprint {https://arxiv.org/abs/hep-ph/0604111} {arXiv:hep-ph/0604111} \BibitemShut {NoStop}%
\bibitem [{\citenamefont {Alcaraz}\ \emph {et~al.}(2006)\citenamefont {Alcaraz} \emph {et~al.}}]{ALEPH:2006bhb}%
  \BibitemOpen
  \bibfield  {author} {\bibinfo {author} {\bibfnamefont {J.}~\bibnamefont {Alcaraz}} \emph {et~al.} (\bibinfo {collaboration} {ALEPH, DELPHI, L3, OPAL, LEP Electroweak Working Group}),\ }\bibfield  {title} {\bibinfo {title} {{A Combination of preliminary electroweak measurements and constraints on the standard model}},\ }\href@noop {} {\  (\bibinfo {year} {2006})},\ \Eprint {https://arxiv.org/abs/hep-ex/0612034} {arXiv:hep-ex/0612034} \BibitemShut {NoStop}%
\bibitem [{\citenamefont {Degrande}\ \emph {et~al.}(2012)\citenamefont {Degrande}, \citenamefont {Duhr}, \citenamefont {Fuks}, \citenamefont {Grellscheid}, \citenamefont {Mattelaer},\ and\ \citenamefont {Reiter}}]{Degrande:2011ua}%
  \BibitemOpen
  \bibfield  {author} {\bibinfo {author} {\bibfnamefont {C.}~\bibnamefont {Degrande}}, \bibinfo {author} {\bibfnamefont {C.}~\bibnamefont {Duhr}}, \bibinfo {author} {\bibfnamefont {B.}~\bibnamefont {Fuks}}, \bibinfo {author} {\bibfnamefont {D.}~\bibnamefont {Grellscheid}}, \bibinfo {author} {\bibfnamefont {O.}~\bibnamefont {Mattelaer}},\ and\ \bibinfo {author} {\bibfnamefont {T.}~\bibnamefont {Reiter}},\ }\bibfield  {title} {\bibinfo {title} {{UFO - The Universal FeynRules Output}},\ }\href {https://doi.org/10.1016/j.cpc.2012.01.022} {\bibfield  {journal} {\bibinfo  {journal} {Comput. Phys. Commun.}\ }\textbf {\bibinfo {volume} {183}},\ \bibinfo {pages} {1201} (\bibinfo {year} {2012})},\ \Eprint {https://arxiv.org/abs/1108.2040} {arXiv:1108.2040 [hep-ph]} \BibitemShut {NoStop}%
\bibitem [{\citenamefont {Alwall}\ \emph {et~al.}(2014)\citenamefont {Alwall}, \citenamefont {Frederix}, \citenamefont {Frixione}, \citenamefont {Hirschi}, \citenamefont {Maltoni}, \citenamefont {Mattelaer}, \citenamefont {Shao}, \citenamefont {Stelzer}, \citenamefont {Torrielli},\ and\ \citenamefont {Zaro}}]{Alwall:2014hca}%
  \BibitemOpen
  \bibfield  {author} {\bibinfo {author} {\bibfnamefont {J.}~\bibnamefont {Alwall}}, \bibinfo {author} {\bibfnamefont {R.}~\bibnamefont {Frederix}}, \bibinfo {author} {\bibfnamefont {S.}~\bibnamefont {Frixione}}, \bibinfo {author} {\bibfnamefont {V.}~\bibnamefont {Hirschi}}, \bibinfo {author} {\bibfnamefont {F.}~\bibnamefont {Maltoni}}, \bibinfo {author} {\bibfnamefont {O.}~\bibnamefont {Mattelaer}}, \bibinfo {author} {\bibfnamefont {H.~S.}\ \bibnamefont {Shao}}, \bibinfo {author} {\bibfnamefont {T.}~\bibnamefont {Stelzer}}, \bibinfo {author} {\bibfnamefont {P.}~\bibnamefont {Torrielli}},\ and\ \bibinfo {author} {\bibfnamefont {M.}~\bibnamefont {Zaro}},\ }\bibfield  {title} {\bibinfo {title} {{The automated computation of tree-level and next-to-leading order differential cross sections, and their matching to parton shower simulations}},\ }\href {https://doi.org/10.1007/JHEP07(2014)079} {\bibfield  {journal} {\bibinfo  {journal} {JHEP}\ }\textbf {\bibinfo {volume} {07}},\ \bibinfo {pages} {079}},\ \Eprint
  {https://arxiv.org/abs/1405.0301} {arXiv:1405.0301 [hep-ph]} \BibitemShut {NoStop}%
\bibitem [{\citenamefont {Mangano}\ \emph {et~al.}(2007)\citenamefont {Mangano}, \citenamefont {Moretti}, \citenamefont {Piccinini},\ and\ \citenamefont {Treccani}}]{Mangano:2006rw}%
  \BibitemOpen
  \bibfield  {author} {\bibinfo {author} {\bibfnamefont {M.~L.}\ \bibnamefont {Mangano}}, \bibinfo {author} {\bibfnamefont {M.}~\bibnamefont {Moretti}}, \bibinfo {author} {\bibfnamefont {F.}~\bibnamefont {Piccinini}},\ and\ \bibinfo {author} {\bibfnamefont {M.}~\bibnamefont {Treccani}},\ }\bibfield  {title} {\bibinfo {title} {{Matching matrix elements and shower evolution for top-quark production in hadronic collisions}},\ }\href {https://doi.org/10.1088/1126-6708/2007/01/013} {\bibfield  {journal} {\bibinfo  {journal} {JHEP}\ }\textbf {\bibinfo {volume} {01}},\ \bibinfo {pages} {013}},\ \Eprint {https://arxiv.org/abs/hep-ph/0611129} {arXiv:hep-ph/0611129} \BibitemShut {NoStop}%
\bibitem [{\citenamefont {Hoeche}\ \emph {et~al.}(2005)\citenamefont {Hoeche}, \citenamefont {Krauss}, \citenamefont {Lavesson}, \citenamefont {Lonnblad}, \citenamefont {Mangano}, \citenamefont {Schalicke},\ and\ \citenamefont {Schumann}}]{Hoeche:2005vzu}%
  \BibitemOpen
  \bibfield  {author} {\bibinfo {author} {\bibfnamefont {S.}~\bibnamefont {Hoeche}}, \bibinfo {author} {\bibfnamefont {F.}~\bibnamefont {Krauss}}, \bibinfo {author} {\bibfnamefont {N.}~\bibnamefont {Lavesson}}, \bibinfo {author} {\bibfnamefont {L.}~\bibnamefont {Lonnblad}}, \bibinfo {author} {\bibfnamefont {M.}~\bibnamefont {Mangano}}, \bibinfo {author} {\bibfnamefont {A.}~\bibnamefont {Schalicke}},\ and\ \bibinfo {author} {\bibfnamefont {S.}~\bibnamefont {Schumann}},\ }\bibfield  {title} {\bibinfo {title} {{Matching parton showers and matrix elements}},\ }in\ \href {https://doi.org/10.5170/CERN-2005-014.288} {\emph {\bibinfo {booktitle} {{HERA and the LHC: A Workshop on the Implications of HERA for LHC Physics: CERN - DESY Workshop 2004/2005 (Midterm Meeting, CERN, 11-13 October 2004; Final Meeting, DESY, 17-21 January 2005)}}}}\ (\bibinfo {year} {2005})\ pp.\ \bibinfo {pages} {288--289},\ \Eprint {https://arxiv.org/abs/hep-ph/0602031} {arXiv:hep-ph/0602031} \BibitemShut {NoStop}%
\bibitem [{\citenamefont {Sj$\ddot{\text{o}}$strand}\ \emph {et~al.}(2015)\citenamefont {Sj$\ddot{\text{o}}$strand}, \citenamefont {Ask}, \citenamefont {Christiansen}, \citenamefont {Corke}, \citenamefont {Desai}, \citenamefont {Ilten}, \citenamefont {Mrenna}, \citenamefont {Prestel}, \citenamefont {Rasmussen},\ and\ \citenamefont {Skands}}]{Sjostrand:2014zea}%
  \BibitemOpen
  \bibfield  {author} {\bibinfo {author} {\bibfnamefont {T.}~\bibnamefont {Sj$\ddot{\text{o}}$strand}}, \bibinfo {author} {\bibfnamefont {S.}~\bibnamefont {Ask}}, \bibinfo {author} {\bibfnamefont {J.~R.}\ \bibnamefont {Christiansen}}, \bibinfo {author} {\bibfnamefont {R.}~\bibnamefont {Corke}}, \bibinfo {author} {\bibfnamefont {N.}~\bibnamefont {Desai}}, \bibinfo {author} {\bibfnamefont {P.}~\bibnamefont {Ilten}}, \bibinfo {author} {\bibfnamefont {S.}~\bibnamefont {Mrenna}}, \bibinfo {author} {\bibfnamefont {S.}~\bibnamefont {Prestel}}, \bibinfo {author} {\bibfnamefont {C.~O.}\ \bibnamefont {Rasmussen}},\ and\ \bibinfo {author} {\bibfnamefont {P.~Z.}\ \bibnamefont {Skands}},\ }\bibfield  {title} {\bibinfo {title} {{An Introduction to PYTHIA 8.2}},\ }\href {https://doi.org/10.1016/j.cpc.2015.01.024} {\bibfield  {journal} {\bibinfo  {journal} {Comput. Phys. Commun.}\ }\textbf {\bibinfo {volume} {191}},\ \bibinfo {pages} {159} (\bibinfo {year} {2015})},\ \Eprint {https://arxiv.org/abs/1410.3012} {arXiv:1410.3012
  [hep-ph]} \BibitemShut {NoStop}%
\bibitem [{\citenamefont {Cacciari}\ \emph {et~al.}(2012)\citenamefont {Cacciari}, \citenamefont {Salam},\ and\ \citenamefont {Soyez}}]{Cacciari:2011ma}%
  \BibitemOpen
  \bibfield  {author} {\bibinfo {author} {\bibfnamefont {M.}~\bibnamefont {Cacciari}}, \bibinfo {author} {\bibfnamefont {G.~P.}\ \bibnamefont {Salam}},\ and\ \bibinfo {author} {\bibfnamefont {G.}~\bibnamefont {Soyez}},\ }\bibfield  {title} {\bibinfo {title} {{FastJet User Manual}},\ }\href {https://doi.org/10.1140/epjc/s10052-012-1896-2} {\bibfield  {journal} {\bibinfo  {journal} {Eur. Phys. J. C}\ }\textbf {\bibinfo {volume} {72}},\ \bibinfo {pages} {1896} (\bibinfo {year} {2012})},\ \Eprint {https://arxiv.org/abs/1111.6097} {arXiv:1111.6097 [hep-ph]} \BibitemShut {NoStop}%
\bibitem [{\citenamefont {de~Favereau}\ \emph {et~al.}(2014)\citenamefont {de~Favereau}, \citenamefont {Delaere}, \citenamefont {Demin}, \citenamefont {Giammanco}, \citenamefont {Lema\^\i{}tre}, \citenamefont {Mertens},\ and\ \citenamefont {Selvaggi}}]{deFavereau:2013fsa}%
  \BibitemOpen
  \bibfield  {author} {\bibinfo {author} {\bibfnamefont {J.}~\bibnamefont {de~Favereau}}, \bibinfo {author} {\bibfnamefont {C.}~\bibnamefont {Delaere}}, \bibinfo {author} {\bibfnamefont {P.}~\bibnamefont {Demin}}, \bibinfo {author} {\bibfnamefont {A.}~\bibnamefont {Giammanco}}, \bibinfo {author} {\bibfnamefont {V.}~\bibnamefont {Lema\^\i{}tre}}, \bibinfo {author} {\bibfnamefont {A.}~\bibnamefont {Mertens}},\ and\ \bibinfo {author} {\bibfnamefont {M.}~\bibnamefont {Selvaggi}} (\bibinfo {collaboration} {DELPHES 3}),\ }\bibfield  {title} {\bibinfo {title} {{DELPHES 3, A modular framework for fast simulation of a generic collider experiment}},\ }\href {https://doi.org/10.1007/JHEP02(2014)057} {\bibfield  {journal} {\bibinfo  {journal} {JHEP}\ }\textbf {\bibinfo {volume} {02}},\ \bibinfo {pages} {057}},\ \Eprint {https://arxiv.org/abs/1307.6346} {arXiv:1307.6346 [hep-ex]} \BibitemShut {NoStop}%
\bibitem [{\citenamefont {Hayrapetyan}\ \emph {et~al.}(2024{\natexlab{b}})\citenamefont {Hayrapetyan} \emph {et~al.}}]{CMS:2024bvl}%
  \BibitemOpen
  \bibfield  {author} {\bibinfo {author} {\bibfnamefont {A.}~\bibnamefont {Hayrapetyan}} \emph {et~al.} (\bibinfo {collaboration} {CMS}),\ }\bibfield  {title} {\bibinfo {title} {{Search for long-lived particles decaying in the CMS muon detectors in proton-proton collisions at s=13\,\,TeV}},\ }\href {https://doi.org/10.1103/PhysRevD.110.032007} {\bibfield  {journal} {\bibinfo  {journal} {Phys. Rev. D}\ }\textbf {\bibinfo {volume} {110}},\ \bibinfo {pages} {032007} (\bibinfo {year} {2024}{\natexlab{b}})},\ \Eprint {https://arxiv.org/abs/2402.01898} {arXiv:2402.01898 [hep-ex]} \BibitemShut {NoStop}%
\bibitem [{\citenamefont {CMS}(2024)}]{CMS:2024nua}%
  \BibitemOpen
  \bibfield  {author} {\bibinfo {author} {\bibnamefont {CMS}},\ }\href@noop {} {\bibinfo {title} {{Search for vector-like leptons with long-lived particle decays in the CMS muon system}}} (\bibinfo {year} {2024})\BibitemShut {NoStop}%
\bibitem [{\citenamefont {Wang}(2022)}]{delphes_pr}%
  \BibitemOpen
  \bibfield  {author} {\bibinfo {author} {\bibfnamefont {C.}~\bibnamefont {Wang}},\ }\href@noop {} {\bibinfo {title} {{Dedicated Delphes Module}}},\ \bibinfo {howpublished} {{\url{https://github.com/delphes/delphes/pull/103}}} (\bibinfo {year} {2022})\BibitemShut {NoStop}%
\bibitem [{\citenamefont {{CMS Collaboration}}(2021)}]{hepdata.104408.v2}%
  \BibitemOpen
  \bibfield  {author} {\bibinfo {author} {\bibnamefont {{CMS Collaboration}}},\ }\href@noop {} {\bibinfo {title} {{Search for long-lived particles decaying in the CMS endcap muon detectors in proton-proton collisions at $\sqrt{s} = $ 13 TeV (Version 2)}}},\ \bibinfo {howpublished} {{\href{https://doi.org/10.17182/hepdata.104408.v2}{HEPData (collection)}}} (\bibinfo {year} {2021})\BibitemShut {NoStop}%
\bibitem [{\citenamefont {Alwall}\ \emph {et~al.}(2011)\citenamefont {Alwall}, \citenamefont {Herquet}, \citenamefont {Maltoni}, \citenamefont {Mattelaer},\ and\ \citenamefont {Stelzer}}]{Alwall:2011uj}%
  \BibitemOpen
  \bibfield  {author} {\bibinfo {author} {\bibfnamefont {J.}~\bibnamefont {Alwall}}, \bibinfo {author} {\bibfnamefont {M.}~\bibnamefont {Herquet}}, \bibinfo {author} {\bibfnamefont {F.}~\bibnamefont {Maltoni}}, \bibinfo {author} {\bibfnamefont {O.}~\bibnamefont {Mattelaer}},\ and\ \bibinfo {author} {\bibfnamefont {T.}~\bibnamefont {Stelzer}},\ }\bibfield  {title} {\bibinfo {title} {{MadGraph 5 : Going Beyond}},\ }\href {https://doi.org/10.1007/JHEP06(2011)128} {\bibfield  {journal} {\bibinfo  {journal} {JHEP}\ }\textbf {\bibinfo {volume} {06}},\ \bibinfo {pages} {128}},\ \Eprint {https://arxiv.org/abs/1106.0522} {arXiv:1106.0522 [hep-ph]} \BibitemShut {NoStop}%
\bibitem [{\citenamefont {Jones}\ \emph {et~al.}(2018)\citenamefont {Jones}, \citenamefont {Kerner},\ and\ \citenamefont {Luisoni}}]{Jones:2018hbb}%
  \BibitemOpen
  \bibfield  {author} {\bibinfo {author} {\bibfnamefont {S.~P.}\ \bibnamefont {Jones}}, \bibinfo {author} {\bibfnamefont {M.}~\bibnamefont {Kerner}},\ and\ \bibinfo {author} {\bibfnamefont {G.}~\bibnamefont {Luisoni}},\ }\bibfield  {title} {\bibinfo {title} {{Next-to-Leading-Order QCD Corrections to Higgs Boson Plus Jet Production with Full Top-Quark Mass Dependence}},\ }\href {https://doi.org/10.1103/PhysRevLett.120.162001} {\bibfield  {journal} {\bibinfo  {journal} {Phys. Rev. Lett.}\ }\textbf {\bibinfo {volume} {120}},\ \bibinfo {pages} {162001} (\bibinfo {year} {2018})},\ \bibinfo {note} {[Erratum: Phys.Rev.Lett. 128, 059901 (2022)]},\ \Eprint {https://arxiv.org/abs/1802.00349} {arXiv:1802.00349 [hep-ph]} \BibitemShut {NoStop}%
\bibitem [{\citenamefont {Becker}\ \emph {et~al.}(2024)\citenamefont {Becker} \emph {et~al.}}]{Becker:2020rjp}%
  \BibitemOpen
  \bibfield  {author} {\bibinfo {author} {\bibfnamefont {K.}~\bibnamefont {Becker}} \emph {et~al.},\ }\bibfield  {title} {\bibinfo {title} {{Precise predictions for boosted Higgs production}},\ }\href {https://doi.org/10.21468/SciPostPhysCore.7.1.001} {\bibfield  {journal} {\bibinfo  {journal} {SciPost Phys. Core}\ }\textbf {\bibinfo {volume} {7}},\ \bibinfo {pages} {001} (\bibinfo {year} {2024})},\ \Eprint {https://arxiv.org/abs/2005.07762} {arXiv:2005.07762 [hep-ph]} \BibitemShut {NoStop}%
\bibitem [{\citenamefont {Maltoni}\ \emph {et~al.}(2014)\citenamefont {Maltoni}, \citenamefont {Vryonidou},\ and\ \citenamefont {Zaro}}]{Maltoni:2014eza}%
  \BibitemOpen
  \bibfield  {author} {\bibinfo {author} {\bibfnamefont {F.}~\bibnamefont {Maltoni}}, \bibinfo {author} {\bibfnamefont {E.}~\bibnamefont {Vryonidou}},\ and\ \bibinfo {author} {\bibfnamefont {M.}~\bibnamefont {Zaro}},\ }\bibfield  {title} {\bibinfo {title} {{Top-quark mass effects in double and triple Higgs production in gluon-gluon fusion at NLO}},\ }\href {https://doi.org/10.1007/JHEP11(2014)079} {\bibfield  {journal} {\bibinfo  {journal} {JHEP}\ }\textbf {\bibinfo {volume} {11}},\ \bibinfo {pages} {079}},\ \Eprint {https://arxiv.org/abs/1408.6542} {arXiv:1408.6542 [hep-ph]} \BibitemShut {NoStop}%
\bibitem [{CMS(2022)}]{CMS-DP-2022-062}%
  \BibitemOpen
  \href {https://cds.cern.ch/record/2842376} {\bibinfo {title} {{CSC High Multiplicity Trigger in Run 3}}} (\bibinfo {year} {2022})\BibitemShut {NoStop}%
\bibitem [{\citenamefont {Bolton}\ \emph {et~al.}(2022)\citenamefont {Bolton}, \citenamefont {Deppisch},\ and\ \citenamefont {Dev}}]{Bolton:2022pyf}%
  \BibitemOpen
  \bibfield  {author} {\bibinfo {author} {\bibfnamefont {P.~D.}\ \bibnamefont {Bolton}}, \bibinfo {author} {\bibfnamefont {F.~F.}\ \bibnamefont {Deppisch}},\ and\ \bibinfo {author} {\bibfnamefont {P.~S.~B.}\ \bibnamefont {Dev}},\ }\bibfield  {title} {\bibinfo {title} {{Probes of Heavy Sterile Neutrinos}},\ }in\ \href@noop {} {\emph {\bibinfo {booktitle} {{56th Rencontres de Moriond on Electroweak Interactions and Unified Theories}}}}\ (\bibinfo {year} {2022})\ \Eprint {https://arxiv.org/abs/2206.01140} {arXiv:2206.01140 [hep-ph]} \BibitemShut {NoStop}%
\bibitem [{\citenamefont {Cepeda}\ \emph {et~al.}(2019)\citenamefont {Cepeda} \emph {et~al.}}]{Cepeda:2019klc}%
  \BibitemOpen
  \bibfield  {author} {\bibinfo {author} {\bibfnamefont {M.}~\bibnamefont {Cepeda}} \emph {et~al.},\ }\bibfield  {title} {\bibinfo {title} {{Report from Working Group 2}: {Higgs Physics at the HL-LHC and HE-LHC}},\ }\href {https://doi.org/10.23731/CYRM-2019-007.221} {\bibfield  {journal} {\bibinfo  {journal} {CERN Yellow Rep. Monogr.}\ }\textbf {\bibinfo {volume} {7}},\ \bibinfo {pages} {221} (\bibinfo {year} {2019})},\ \Eprint {https://arxiv.org/abs/1902.00134} {arXiv:1902.00134 [hep-ph]} \BibitemShut {NoStop}%
\bibitem [{\citenamefont {ATLAS}\ and\ \citenamefont {CMS}(2019)}]{ATLAS:2019mfr}%
  \BibitemOpen
  \bibfield  {author} {\bibinfo {author} {\bibnamefont {ATLAS}}\ and\ \bibinfo {author} {\bibnamefont {CMS}},\ }\bibfield  {title} {\bibinfo {title} {{Addendum to the report on the physics at the HL-LHC, and perspectives for the HE-LHC: Collection of notes from ATLAS and CMS}},\ }\href {https://doi.org/10.23731/CYRM-2019-007.Addendum} {\bibfield  {journal} {\bibinfo  {journal} {CERN Yellow Rep. Monogr.}\ }\textbf {\bibinfo {volume} {7}},\ \bibinfo {pages} {Addendum} (\bibinfo {year} {2019})},\ \Eprint {https://arxiv.org/abs/1902.10229} {arXiv:1902.10229 [hep-ex]} \BibitemShut {NoStop}%
\bibitem [{\citenamefont {Barducci}\ \emph {et~al.}(2021)\citenamefont {Barducci}, \citenamefont {Bertuzzo}, \citenamefont {Caputo}, \citenamefont {Hernandez},\ and\ \citenamefont {Mele}}]{Barducci:2020icf}%
  \BibitemOpen
  \bibfield  {author} {\bibinfo {author} {\bibfnamefont {D.}~\bibnamefont {Barducci}}, \bibinfo {author} {\bibfnamefont {E.}~\bibnamefont {Bertuzzo}}, \bibinfo {author} {\bibfnamefont {A.}~\bibnamefont {Caputo}}, \bibinfo {author} {\bibfnamefont {P.}~\bibnamefont {Hernandez}},\ and\ \bibinfo {author} {\bibfnamefont {B.}~\bibnamefont {Mele}},\ }\bibfield  {title} {\bibinfo {title} {{The see-saw portal at future Higgs Factories}},\ }\href {https://doi.org/10.1007/JHEP03(2021)117} {\bibfield  {journal} {\bibinfo  {journal} {JHEP}\ }\textbf {\bibinfo {volume} {03}},\ \bibinfo {pages} {117}},\ \Eprint {https://arxiv.org/abs/2011.04725} {arXiv:2011.04725 [hep-ph]} \BibitemShut {NoStop}%
\bibitem [{\citenamefont {Fern\'andez-Mart\'\i{}nez}\ \emph {et~al.}(2023{\natexlab{a}})\citenamefont {Fern\'andez-Mart\'\i{}nez}, \citenamefont {L\'opez-Pav\'on}, \citenamefont {No}, \citenamefont {Ota},\ and\ \citenamefont {Rosauro-Alcaraz}}]{Fernandez-Martinez:2022stj}%
  \BibitemOpen
  \bibfield  {author} {\bibinfo {author} {\bibfnamefont {E.}~\bibnamefont {Fern\'andez-Mart\'\i{}nez}}, \bibinfo {author} {\bibfnamefont {J.}~\bibnamefont {L\'opez-Pav\'on}}, \bibinfo {author} {\bibfnamefont {J.~M.}\ \bibnamefont {No}}, \bibinfo {author} {\bibfnamefont {T.}~\bibnamefont {Ota}},\ and\ \bibinfo {author} {\bibfnamefont {S.}~\bibnamefont {Rosauro-Alcaraz}},\ }\bibfield  {title} {\bibinfo {title} {{$\nu $ Electroweak baryogenesis: the scalar singlet strikes back}},\ }\href {https://doi.org/10.1140/epjc/s10052-023-11887-z} {\bibfield  {journal} {\bibinfo  {journal} {Eur. Phys. J. C}\ }\textbf {\bibinfo {volume} {83}},\ \bibinfo {pages} {715} (\bibinfo {year} {2023}{\natexlab{a}})},\ \Eprint {https://arxiv.org/abs/2210.16279} {arXiv:2210.16279 [hep-ph]} \BibitemShut {NoStop}%
\bibitem [{\citenamefont {Fern\'andez-Mart\'\i{}nez}\ \emph {et~al.}(2023{\natexlab{b}})\citenamefont {Fern\'andez-Mart\'\i{}nez}, \citenamefont {Gonz\'alez-L\'opez}, \citenamefont {Hern\'andez-Garc\'\i{}a}, \citenamefont {Hostert},\ and\ \citenamefont {L\'opez-Pav\'on}}]{Fernandez-Martinez:2023phj}%
  \BibitemOpen
  \bibfield  {author} {\bibinfo {author} {\bibfnamefont {E.}~\bibnamefont {Fern\'andez-Mart\'\i{}nez}}, \bibinfo {author} {\bibfnamefont {M.}~\bibnamefont {Gonz\'alez-L\'opez}}, \bibinfo {author} {\bibfnamefont {J.}~\bibnamefont {Hern\'andez-Garc\'\i{}a}}, \bibinfo {author} {\bibfnamefont {M.}~\bibnamefont {Hostert}},\ and\ \bibinfo {author} {\bibfnamefont {J.}~\bibnamefont {L\'opez-Pav\'on}},\ }\bibfield  {title} {\bibinfo {title} {{Effective portals to heavy neutral leptons}},\ }\href {https://doi.org/10.1007/JHEP09(2023)001} {\bibfield  {journal} {\bibinfo  {journal} {JHEP}\ }\textbf {\bibinfo {volume} {09}},\ \bibinfo {pages} {001}},\ \Eprint {https://arxiv.org/abs/2304.06772} {arXiv:2304.06772 [hep-ph]} \BibitemShut {NoStop}%
\bibitem [{\citenamefont {Robens}\ and\ \citenamefont {Stefaniak}(2015)}]{Robens:2015gla}%
  \BibitemOpen
  \bibfield  {author} {\bibinfo {author} {\bibfnamefont {T.}~\bibnamefont {Robens}}\ and\ \bibinfo {author} {\bibfnamefont {T.}~\bibnamefont {Stefaniak}},\ }\bibfield  {title} {\bibinfo {title} {{Status of the Higgs Singlet Extension of the Standard Model after LHC Run 1}},\ }\href {https://doi.org/10.1140/epjc/s10052-015-3323-y} {\bibfield  {journal} {\bibinfo  {journal} {Eur. Phys. J. C}\ }\textbf {\bibinfo {volume} {75}},\ \bibinfo {pages} {104} (\bibinfo {year} {2015})},\ \Eprint {https://arxiv.org/abs/1501.02234} {arXiv:1501.02234 [hep-ph]} \BibitemShut {NoStop}%
\bibitem [{Note1()}]{Note1}%
  \BibitemOpen
  \bibinfo {note} {The case where $m_\Phi < m_h$ is considered in \cite {Liu:2022ugx}.}\BibitemShut {Stop}%
\bibitem [{\citenamefont {Robens}(2023)}]{Robens:2022cun}%
  \BibitemOpen
  \bibfield  {author} {\bibinfo {author} {\bibfnamefont {T.}~\bibnamefont {Robens}},\ }\bibfield  {title} {\bibinfo {title} {{Constraining Extended Scalar Sectors at~Current and~Future Colliders\textemdash{}An Update}},\ }\href {https://doi.org/10.1007/978-3-031-30459-0_13} {\bibfield  {journal} {\bibinfo  {journal} {Springer Proc. Phys.}\ }\textbf {\bibinfo {volume} {292}},\ \bibinfo {pages} {141} (\bibinfo {year} {2023})},\ \Eprint {https://arxiv.org/abs/2209.15544} {arXiv:2209.15544 [hep-ph]} \BibitemShut {NoStop}%
\bibitem [{\citenamefont {Papaefstathiou}\ \emph {et~al.}(2022)\citenamefont {Papaefstathiou}, \citenamefont {Robens},\ and\ \citenamefont {White}}]{Papaefstathiou:2022oyi}%
  \BibitemOpen
  \bibfield  {author} {\bibinfo {author} {\bibfnamefont {A.}~\bibnamefont {Papaefstathiou}}, \bibinfo {author} {\bibfnamefont {T.}~\bibnamefont {Robens}},\ and\ \bibinfo {author} {\bibfnamefont {G.}~\bibnamefont {White}},\ }\bibfield  {title} {\bibinfo {title} {{Signal strength and W-boson mass measurements as a probe of the electro-weak phase transition at colliders - Snowmass White Paper}},\ }in\ \href@noop {} {\emph {\bibinfo {booktitle} {{Snowmass 2021}}}}\ (\bibinfo {year} {2022})\ \Eprint {https://arxiv.org/abs/2205.14379} {arXiv:2205.14379 [hep-ph]} \BibitemShut {NoStop}%
\bibitem [{\citenamefont {ATLAS}(2021)}]{ATLAS:2021vrm}%
  \BibitemOpen
  \bibfield  {author} {\bibinfo {author} {\bibnamefont {ATLAS}},\ }\href@noop {} {\bibinfo {title} {{Combined measurements of Higgs boson production and decay using up to $139$ fb$^{-1}$ of proton-proton collision data at $\sqrt{s}= 13$ TeV collected with the ATLAS experiment}}} (\bibinfo {year} {2021})\BibitemShut {NoStop}%
\bibitem [{\citenamefont {Robens}(2021)}]{Robens:2021rkl}%
  \BibitemOpen
  \bibfield  {author} {\bibinfo {author} {\bibfnamefont {T.}~\bibnamefont {Robens}},\ }\bibfield  {title} {\bibinfo {title} {{Extended scalar sectors at current and future colliders}},\ }in\ \href@noop {} {\emph {\bibinfo {booktitle} {{55th Rencontres de Moriond on QCD and High Energy Interactions}}}}\ (\bibinfo {year} {2021})\ \Eprint {https://arxiv.org/abs/2105.07719} {arXiv:2105.07719 [hep-ph]} \BibitemShut {NoStop}%
\bibitem [{\citenamefont {L\'opez-Val}\ and\ \citenamefont {Robens}(2014)}]{Lopez-Val:2014jva}%
  \BibitemOpen
  \bibfield  {author} {\bibinfo {author} {\bibfnamefont {D.}~\bibnamefont {L\'opez-Val}}\ and\ \bibinfo {author} {\bibfnamefont {T.}~\bibnamefont {Robens}},\ }\bibfield  {title} {\bibinfo {title} {{\ensuremath{\Delta}r and the W-boson mass in the singlet extension of the standard model}},\ }\href {https://doi.org/10.1103/PhysRevD.90.114018} {\bibfield  {journal} {\bibinfo  {journal} {Phys. Rev. D}\ }\textbf {\bibinfo {volume} {90}},\ \bibinfo {pages} {114018} (\bibinfo {year} {2014})},\ \Eprint {https://arxiv.org/abs/1406.1043} {arXiv:1406.1043 [hep-ph]} \BibitemShut {NoStop}%
\bibitem [{\citenamefont {Workman}\ \emph {et~al.}(2022)\citenamefont {Workman} \emph {et~al.}}]{ParticleDataGroup:2022pth}%
  \BibitemOpen
  \bibfield  {author} {\bibinfo {author} {\bibfnamefont {R.~L.}\ \bibnamefont {Workman}} \emph {et~al.} (\bibinfo {collaboration} {Particle Data Group}),\ }\bibfield  {title} {\bibinfo {title} {{Review of Particle Physics}},\ }\href {https://doi.org/10.1093/ptep/ptac097} {\bibfield  {journal} {\bibinfo  {journal} {PTEP}\ }\textbf {\bibinfo {volume} {2022}},\ \bibinfo {pages} {083C01} (\bibinfo {year} {2022})}\BibitemShut {NoStop}%
\bibitem [{\citenamefont {Sirunyan}\ \emph {et~al.}(2020)\citenamefont {Sirunyan} \emph {et~al.}}]{CMS:2019buh}%
  \BibitemOpen
  \bibfield  {author} {\bibinfo {author} {\bibfnamefont {A.~M.}\ \bibnamefont {Sirunyan}} \emph {et~al.} (\bibinfo {collaboration} {CMS}),\ }\bibfield  {title} {\bibinfo {title} {{Search for a Narrow Resonance Lighter than 200 GeV Decaying to a Pair of Muons in Proton-Proton Collisions at $\sqrt{s} =$ TeV}},\ }\href {https://doi.org/10.1103/PhysRevLett.124.131802} {\bibfield  {journal} {\bibinfo  {journal} {Phys. Rev. Lett.}\ }\textbf {\bibinfo {volume} {124}},\ \bibinfo {pages} {131802} (\bibinfo {year} {2020})},\ \Eprint {https://arxiv.org/abs/1912.04776} {arXiv:1912.04776 [hep-ex]} \BibitemShut {NoStop}%
\bibitem [{\citenamefont {Sirunyan}\ \emph {et~al.}(2021)\citenamefont {Sirunyan} \emph {et~al.}}]{CMS:2021ctt}%
  \BibitemOpen
  \bibfield  {author} {\bibinfo {author} {\bibfnamefont {A.~M.}\ \bibnamefont {Sirunyan}} \emph {et~al.} (\bibinfo {collaboration} {CMS}),\ }\bibfield  {title} {\bibinfo {title} {{Search for resonant and nonresonant new phenomena in high-mass dilepton final states at $ \sqrt{s} $ = 13 TeV}},\ }\href {https://doi.org/10.1007/JHEP07(2021)208} {\bibfield  {journal} {\bibinfo  {journal} {JHEP}\ }\textbf {\bibinfo {volume} {07}},\ \bibinfo {pages} {208}},\ \Eprint {https://arxiv.org/abs/2103.02708} {arXiv:2103.02708 [hep-ex]} \BibitemShut {NoStop}%
\bibitem [{\citenamefont {Aaij}\ \emph {et~al.}(2020)\citenamefont {Aaij} \emph {et~al.}}]{LHCb:2019vmc}%
  \BibitemOpen
  \bibfield  {author} {\bibinfo {author} {\bibfnamefont {R.}~\bibnamefont {Aaij}} \emph {et~al.} (\bibinfo {collaboration} {LHCb}),\ }\bibfield  {title} {\bibinfo {title} {{Search for $A'\to\mu^+\mu^-$ Decays}},\ }\href {https://doi.org/10.1103/PhysRevLett.124.041801} {\bibfield  {journal} {\bibinfo  {journal} {Phys. Rev. Lett.}\ }\textbf {\bibinfo {volume} {124}},\ \bibinfo {pages} {041801} (\bibinfo {year} {2020})},\ \Eprint {https://arxiv.org/abs/1910.06926} {arXiv:1910.06926 [hep-ex]} \BibitemShut {NoStop}%
\bibitem [{\citenamefont {Fox}\ \emph {et~al.}(2011)\citenamefont {Fox}, \citenamefont {Harnik}, \citenamefont {Kopp},\ and\ \citenamefont {Tsai}}]{Fox:2011fx}%
  \BibitemOpen
  \bibfield  {author} {\bibinfo {author} {\bibfnamefont {P.~J.}\ \bibnamefont {Fox}}, \bibinfo {author} {\bibfnamefont {R.}~\bibnamefont {Harnik}}, \bibinfo {author} {\bibfnamefont {J.}~\bibnamefont {Kopp}},\ and\ \bibinfo {author} {\bibfnamefont {Y.}~\bibnamefont {Tsai}},\ }\bibfield  {title} {\bibinfo {title} {{LEP Shines Light on Dark Matter}},\ }\href {https://doi.org/10.1103/PhysRevD.84.014028} {\bibfield  {journal} {\bibinfo  {journal} {Phys. Rev. D}\ }\textbf {\bibinfo {volume} {84}},\ \bibinfo {pages} {014028} (\bibinfo {year} {2011})},\ \Eprint {https://arxiv.org/abs/1103.0240} {arXiv:1103.0240 [hep-ph]} \BibitemShut {NoStop}%
\bibitem [{\citenamefont {Ilten}\ \emph {et~al.}(2018)\citenamefont {Ilten}, \citenamefont {Soreq}, \citenamefont {Williams},\ and\ \citenamefont {Xue}}]{Ilten:2018crw}%
  \BibitemOpen
  \bibfield  {author} {\bibinfo {author} {\bibfnamefont {P.}~\bibnamefont {Ilten}}, \bibinfo {author} {\bibfnamefont {Y.}~\bibnamefont {Soreq}}, \bibinfo {author} {\bibfnamefont {M.}~\bibnamefont {Williams}},\ and\ \bibinfo {author} {\bibfnamefont {W.}~\bibnamefont {Xue}},\ }\bibfield  {title} {\bibinfo {title} {{Serendipity in dark photon searches}},\ }\href {https://doi.org/10.1007/JHEP06(2018)004} {\bibfield  {journal} {\bibinfo  {journal} {JHEP}\ }\textbf {\bibinfo {volume} {06}},\ \bibinfo {pages} {004}},\ \Eprint {https://arxiv.org/abs/1801.04847} {arXiv:1801.04847 [hep-ph]} \BibitemShut {NoStop}%
\bibitem [{\citenamefont {Lindner}\ \emph {et~al.}(2018)\citenamefont {Lindner}, \citenamefont {Queiroz}, \citenamefont {Rodejohann},\ and\ \citenamefont {Xu}}]{Lindner:2018kjo}%
  \BibitemOpen
  \bibfield  {author} {\bibinfo {author} {\bibfnamefont {M.}~\bibnamefont {Lindner}}, \bibinfo {author} {\bibfnamefont {F.~S.}\ \bibnamefont {Queiroz}}, \bibinfo {author} {\bibfnamefont {W.}~\bibnamefont {Rodejohann}},\ and\ \bibinfo {author} {\bibfnamefont {X.-J.}\ \bibnamefont {Xu}},\ }\bibfield  {title} {\bibinfo {title} {{Neutrino-electron scattering: general constraints on Z$^{\prime}$ and dark photon models}},\ }\href {https://doi.org/10.1007/JHEP05(2018)098} {\bibfield  {journal} {\bibinfo  {journal} {JHEP}\ }\textbf {\bibinfo {volume} {05}},\ \bibinfo {pages} {098}},\ \Eprint {https://arxiv.org/abs/1803.00060} {arXiv:1803.00060 [hep-ph]} \BibitemShut {NoStop}%
\bibitem [{\citenamefont {Aad}\ \emph {et~al.}(2016)\citenamefont {Aad} \emph {et~al.}}]{ATLAS:2016fij}%
  \BibitemOpen
  \bibfield  {author} {\bibinfo {author} {\bibfnamefont {G.}~\bibnamefont {Aad}} \emph {et~al.} (\bibinfo {collaboration} {ATLAS}),\ }\bibfield  {title} {\bibinfo {title} {{Measurement of $W^{\pm}$ and $Z$-boson production cross sections in $pp$ collisions at $\sqrt{s}=13$ TeV with the ATLAS detector}},\ }\href {https://doi.org/10.1016/j.physletb.2016.06.023} {\bibfield  {journal} {\bibinfo  {journal} {Phys. Lett. B}\ }\textbf {\bibinfo {volume} {759}},\ \bibinfo {pages} {601} (\bibinfo {year} {2016})},\ \Eprint {https://arxiv.org/abs/1603.09222} {arXiv:1603.09222 [hep-ex]} \BibitemShut {NoStop}%
\bibitem [{\citenamefont {ATLAS}(2023)}]{ATLAS:2023sjw}%
  \BibitemOpen
  \bibfield  {author} {\bibinfo {author} {\bibnamefont {ATLAS}},\ }\href@noop {} {\bibinfo {title} {{Measurement of $t\bar{t}$ and $Z$-boson cross sections and their ratio using $pp$ collisions at $\sqrt{s} = 13.6$ TeV with the ATLAS detector}}} (\bibinfo {year} {2023})\BibitemShut {NoStop}%
\bibitem [{\citenamefont {Ahdida}\ \emph {et~al.}(2019)\citenamefont {Ahdida} \emph {et~al.}}]{SHiP:2018xqw}%
  \BibitemOpen
  \bibfield  {author} {\bibinfo {author} {\bibfnamefont {C.}~\bibnamefont {Ahdida}} \emph {et~al.} (\bibinfo {collaboration} {SHiP}),\ }\bibfield  {title} {\bibinfo {title} {{Sensitivity of the SHiP experiment to Heavy Neutral Leptons}},\ }\href {https://doi.org/10.1007/JHEP04(2019)077} {\bibfield  {journal} {\bibinfo  {journal} {JHEP}\ }\textbf {\bibinfo {volume} {04}},\ \bibinfo {pages} {077}},\ \Eprint {https://arxiv.org/abs/1811.00930} {arXiv:1811.00930 [hep-ph]} \BibitemShut {NoStop}%
\bibitem [{\citenamefont {Drewes}\ and\ \citenamefont {Hajer}(2020)}]{Drewes:2019fou}%
  \BibitemOpen
  \bibfield  {author} {\bibinfo {author} {\bibfnamefont {M.}~\bibnamefont {Drewes}}\ and\ \bibinfo {author} {\bibfnamefont {J.}~\bibnamefont {Hajer}},\ }\bibfield  {title} {\bibinfo {title} {{Heavy Neutrinos in displaced vertex searches at the LHC and HL-LHC}},\ }\href {https://doi.org/10.1007/JHEP02(2020)070} {\bibfield  {journal} {\bibinfo  {journal} {JHEP}\ }\textbf {\bibinfo {volume} {02}},\ \bibinfo {pages} {070}},\ \Eprint {https://arxiv.org/abs/1903.06100} {arXiv:1903.06100 [hep-ph]} \BibitemShut {NoStop}%
\bibitem [{\citenamefont {Blondel}\ \emph {et~al.}(2022)\citenamefont {Blondel} \emph {et~al.}}]{Blondel:2022qqo}%
  \BibitemOpen
  \bibfield  {author} {\bibinfo {author} {\bibfnamefont {A.}~\bibnamefont {Blondel}} \emph {et~al.},\ }\bibfield  {title} {\bibinfo {title} {{Searches for long-lived particles at the future FCC-ee}},\ }\href {https://doi.org/10.3389/fphy.2022.967881} {\bibfield  {journal} {\bibinfo  {journal} {Front. in Phys.}\ }\textbf {\bibinfo {volume} {10}},\ \bibinfo {pages} {967881} (\bibinfo {year} {2022})},\ \Eprint {https://arxiv.org/abs/2203.05502} {arXiv:2203.05502 [hep-ex]} \BibitemShut {NoStop}%
\bibitem [{Note2()}]{Note2}%
  \BibitemOpen
  \bibinfo {note} {There exists a third dimension-5 operator, ${\protect \cal O}_{NB} = (\protect \bar \nu _R^c \sigma ^{\mu \nu } \nu _R)B_{\mu \nu } = 0$ that is usually considered but is absent for a single RHN, reflecting the fact that a Majorana fermion does not have a magnetic dipole moment.}\BibitemShut {Stop}%
\bibitem [{\citenamefont {Barducci}\ \emph {et~al.}(2020)\citenamefont {Barducci}, \citenamefont {Bertuzzo}, \citenamefont {Caputo},\ and\ \citenamefont {Hernandez}}]{Barducci:2020ncz}%
  \BibitemOpen
  \bibfield  {author} {\bibinfo {author} {\bibfnamefont {D.}~\bibnamefont {Barducci}}, \bibinfo {author} {\bibfnamefont {E.}~\bibnamefont {Bertuzzo}}, \bibinfo {author} {\bibfnamefont {A.}~\bibnamefont {Caputo}},\ and\ \bibinfo {author} {\bibfnamefont {P.}~\bibnamefont {Hernandez}},\ }\bibfield  {title} {\bibinfo {title} {{Minimal flavor violation in the see-saw portal}},\ }\href {https://doi.org/10.1007/JHEP06(2020)185} {\bibfield  {journal} {\bibinfo  {journal} {JHEP}\ }\textbf {\bibinfo {volume} {06}},\ \bibinfo {pages} {185}},\ \Eprint {https://arxiv.org/abs/2003.08391} {arXiv:2003.08391 [hep-ph]} \BibitemShut {NoStop}%
\end{thebibliography}%
\end{document}